\documentclass[a4paper,11pt]{article}
\pdfoutput=1 

\usepackage[usenames,dvipsnames,svgnames,table]{xcolor} 
\usepackage{jcapmod}
\usepackage{verbatim}

\usepackage[T1]{fontenc} 
\usepackage[latin9]{inputenc}
\definecolor{lightgray}{gray}{0.9}
\definecolor{Amber}{rgb}{1.0, 0.75, 0.0}
\definecolor{blizzardblue}{rgb}{0.67, 0.9, 0.93}
\definecolor{burningsand}{RGB}{220, 148, 129}
\definecolor{burgundy}{rgb}{0.5, 0.0, 0.13}

\usepackage{float}
\usepackage{graphicx}
\usepackage{hyperref}
\usepackage{amsmath}
\usepackage{amssymb}
\usepackage{microtype}
\usepackage[shortlabels]{enumitem}
\usepackage{cancel}
\usepackage{amsbsy}
\usepackage{color}
\usepackage[capitalise]{cleveref}
\usepackage{breakurl}
\usepackage[normalem]{ulem}
\usepackage{mathtools}
\usepackage{csquotes}
\usepackage{lmodern}
\usepackage{bm}
\usepackage{dsfont}
\usepackage{bbold}
\usepackage{empheq}
\usepackage{caption}
\usepackage{subcaption}

\pdfpageheight\paperheight
\pdfpagewidth\paperwidth

\usepackage[textwidth=0.71\paperwidth, textheight=0.8\paperheight]{geometry}

\renewcommand*{\vec}[1]{\bm{#1}}
\newcommand*{\unitvec}[1]{\vec{\hat{#1}}}
\newcommand*{\mat}[1]{\bm{\mathsf{#1}}}
\DeclareMathOperator{\diag}{diag}

\newcommand{\identity}{\ensuremath{\mathds{1}}}
\newcommand*{\transpose}[1]{\ensuremath{{#1}^{T}}}
\newcommand*{\inverse}[1]{\ensuremath{{#1}^{-1}}}
\newcommand{\integers}{\mathbb{Z}}

\newcommand*{\E}[1]{\texorpdfstring{\ensuremath{E_{#1}}}{E#1}}
\newcommand*{\Espace}{\texorpdfstring{\ensuremath{E^3}}{E(3)}}
\newcommand*{\A}[1]{\texorpdfstring{\ensuremath{A_{#1}}}{A#1}}
\newcommand*{\B}[1]{\texorpdfstring{\ensuremath{B_{#1}}}{B#1}}
\newcommand{\slabh}{\texorpdfstring{\ensuremath{\E{16}^{(\mathrm{h})}}}{E16h}}
\newcommand{\slabi}{\texorpdfstring{\ensuremath{\E{16}^{(\mathrm{i})}}}{E16i}}%
\newcommand{\Kdelta}{\delta^{(\textrm{K})}}
\newcommand{\Ddelta}{\delta^{(\textrm{D})}}
\newcommand{\setN}{\mathcal{N}}
\newcommand*{\Mod}[1]{\,(\mathrm{mod}\ #1)}
\newcommand{\DeltaYstarforell}{\DeltaYstar}
\newcommand{\DeltaYstar}{\Delta^{Y*}}
\newcommand{\vecnp}{\smash{\vec{n}'}}

\newcommand{\LLSS}{L_{\mathrm{LSS}}}
\newcommand{\Lcircle}{L_{\mathrm{circle}}}
\newcommand{\eexp}{\mathrm{e}}
\newcommand{\dderiv}{\mathrm{d}}

\usepackage[table]{xcolor}

\allowdisplaybreaks

\makeatletter
\DeclareRobustCommand{\rcite}[1]{%
  \rcite@aux#1,\@nil{#1}%
}
\def\rcite@aux#1,#2\@nil#3{%
  \if\relax#2\relax
    Ref.~\cite{#3}%
  \else
    Refs.~\cite{#3}%
  \fi
}
\makeatother

\subheader{IFT-UAM/CSIC-25-107}

\title{Cosmic topology. Part IIb. Eigenmodes, correlation matrices, and detectability of non-orientable Euclidean manifolds}

\author[a]{Craig J. Copi,}
\author[a]{Amirhossein Samandar,}
\author[a]{Glenn D. Starkman,}
\author[b]{Javier Carr\'on Duque,}
\author[b,a,c]{Yashar Akrami,}
\author[d,e,f]{Stefano Anselmi,}
\author[c]{Andrew H. Jaffe,}
\author[g]{Arthur Kosowsky,}
\author[a,h]{Fernando Cornet-Gomez,}
\author[i]{Johannes R. Eskilt,}
\author[b,j]{Mikel Martin Barandiaran,}
\author[a]{Deyan P. Mihaylov,}
\author[a]{Anna Negro,}           
\author[b,k]{Joline Noltmann,}
\author[l]{Thiago S. Pereira,}
\author[b,a]{Andrius Tamosiunas}

\collaboration{(COMPACT Collaboration)}
\collaborationImg{\includegraphics[width = 0.1\textwidth]{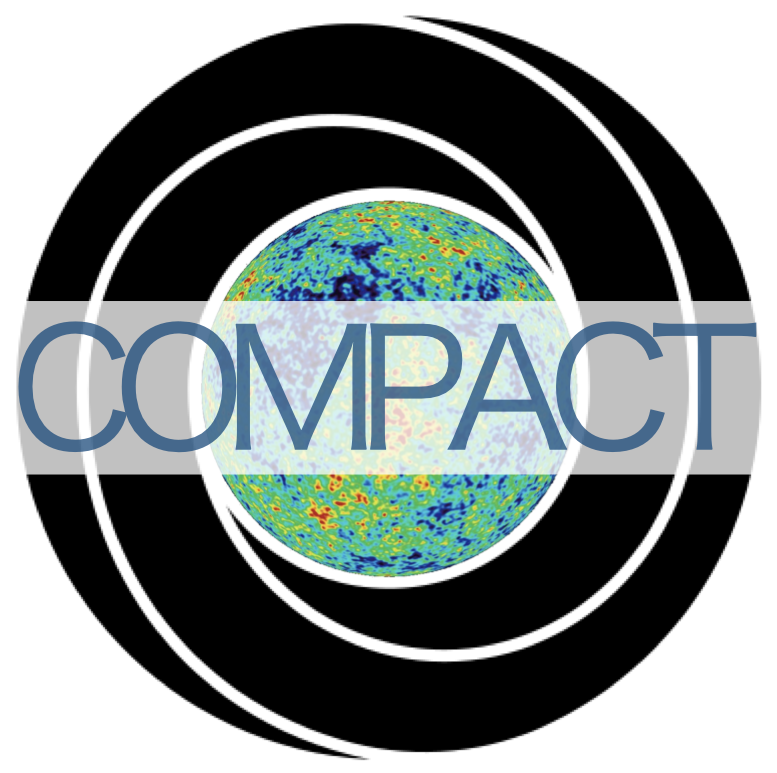}}

\affiliation[a]{CERCA/ISO, Department of Physics, Case Western Reserve University, 10900 Euclid Avenue, Cleveland, OH 44106, USA}
\affiliation[b]{Instituto de F\'isica Te\'orica (IFT) UAM-CSIC, C/ Nicol\'as Cabrera 13-15, Campus de Cantoblanco UAM, 28049 Madrid, Spain}
\affiliation[c]{Astrophysics Group \& Imperial Centre for Inference and Cosmology, Department of Physics, Imperial College London, Blackett Laboratory, Prince Consort Road, London SW7 2AZ, United Kingdom}
\affiliation[d]{INFN, Sezione di Padova, via Marzolo 8, I-35131 Padova, Italy}
\affiliation[e]{Dipartimento di Fisica e Astronomia ``G. Galilei'', Universit\`a degli Studi di Padova, via Marzolo 8, I-35131 Padova, Italy}
\affiliation[f]{Laboratoire Univers et Th\'eories, Observatoire de Paris, Universit\'e PSL, Universit\'e Paris Cit\'e, CNRS, F-92190 Meudon, France}
\affiliation[g]{Department of Physics and Astronomy, University of Pittsburgh, Pittsburgh, PA 15260, USA}
\affiliation[h]{Departamento de F\'isica, Universidad de C\'ordoba, Campus Universitario de Rabanales, Ctra. N-IV Km. 396, E-14071 C\'ordoba, Spain}
\affiliation[i]{Institute of Theoretical Astrophysics, University of Oslo, P.O. Box 1029 Blindern, N-0315 Oslo, Norway}
\affiliation[j]{Departamento de F\'isica Te\'orica, Universidad Aut\'onoma de Madrid, 28049 Madrid, Spain}
\affiliation[k]{Institute for Theoretical Particle Physics and Cosmology, RWTH Aachen University, Templergraben 55, 52062 Aachen, Germany}
\affiliation[l]{Departamento de F\'{i}sica, Universidade Estadual de Londrina, Rod. Celso Garcia Cid, Km 380, 86057-970, Londrina, Paran\'{a}, Brazil}

\emailAdd{craig.copi@case.edu}
\emailAdd{amirhossein.samandar@case.edu}
\emailAdd{glenn.starkman@case.edu}
\emailAdd{javier.carron@csic.es}
\emailAdd{yashar.akrami@csic.es}
\emailAdd{stefano.anselmi@pd.infn.it}
\emailAdd{a.jaffe@imperial.ac.uk}
\emailAdd{kosowsky@pitt.edu}
\emailAdd{fernando.cornetgomez@case.edu}
\emailAdd{j.r.eskilt@astro.uio.no}
\emailAdd{mikel.martin@uam.es}
\emailAdd{deyan.mihaylov@case.edu}
\emailAdd{anna.negro@case.edu}
\emailAdd{joline.noltmann@rwth-aachen.de}
\emailAdd{tspereira@uel.br}
\emailAdd{andrius.tamosiunas@case.edu}

\abstract{
    If the Universe has non-trivial spatial topology, 
    observables depend on both the parameters of the spatial manifold and the position and orientation of the observer.
    In infinite Euclidean space, most cosmological observables arise from the amplitudes of Fourier modes of primordial scalar curvature perturbations.
    Topological boundary conditions replace the full set of Fourier modes with specific linear combinations of selected Fourier modes as the  eigenmodes of the scalar Laplacian.
    In an  earlier work we provided a comprehensive treatment of orientable Euclidean three-manifolds;
    but a thorough exploration of cosmic topology must include non-orientable three-manifolds as candidates for the geometry of space.
    In this paper we consider the non-orientable Euclidean topologies \E{7}--\E{10}, \E{13}--\E{15}, and \E{17}, 
    encompassing the full range of manifold parameters and observer positions, generalizing previous treatments.
    Under the assumption that the amplitudes of primordial scalar curvature eigenmodes are independent random variables, for each topology we obtain the correlation matrices of Fourier-mode amplitudes (of scalar fields linearly related to the scalar curvature) and the correlation matrices of spherical-harmonic coefficients of such fields sampled on a sphere, such as the temperature of the cosmic microwave background (CMB).
    We evaluate the detectability of these correlations given the cosmic variance of the  CMB sky. 
    As for orientable three-manifolds, we find that in manifolds where the distance to our nearest clone is less than about $1.2$ times the diameter of the last scattering surface
    of the CMB, we expect a correlation signal that is larger than cosmic variance noise in the CMB.
    The parameter space of the non-orientable Euclidean manifolds is quite rich, supporting, for example, complex dependencies of clone distances on those parameters.
    Our limited selection of manifold parameters -- both the values of those we fix, and the choices of which to vary -- are therefore exemplary of interesting behaviors (e.g., of how well the manifold can be distinguished from the covering space), but not necessarily representative.  
    Future searches for topology will certainly require a much more thorough exploration of the parameter space to determine what values of the parameters predict statistical correlations that are convincingly attributable to topology.
}


\begin{document}
\maketitle
\flushbottom

\section{Introduction}
\label{secn:intro}

In a non-orientable space, there is no well-defined notion of clockwise and counter-clockwise.
Place a clockface on a M\"obius strip.
Standing at the center of the face and watching the hands move around you, you find they move clockwise, as expected.
Have a  friend slide the clockface (but not you) around the strip back to your position.
Now the hands are moving in the opposite direction around you.

Locally, we have not observed any such phenomenon -- left-handed astronauts do not return right-handed, their pocket watch hands move clockwise.
Socks may disappear in dryers, but left socks do not reappear as right socks.
Apparently, locally space is orientable -- there are no little analogs of M\"obius strips lurking in our neighborhood.
But is this a global feature of space?  
If you accidentally bought two left shoes, could you ship one on  a closed loop around the Universe so that when it arrived it would be the matching right shoe?
Or, as some have argued, as investigations of the topology of the Universe move forward, should we simply not bother considering the possibility that space is a non-orientable 3-manifold?\footnote{
    The standard lore has argued \cite{Hawking:1973uf} that we must live in an orientable space because of the structure of the Standard Model (SM) of particle physics, which contains ``undoubled chiral fermions.''
    The electron, for example, is a negatively charged spin-$1/2$ particle, that  comes in two varieties -- left-handed $e_L^-$ and right-handed $E_R^-$.
    However, these are fundamentally two entirely different particles that have the same name for historic reasons.
    The $e_L^-$ and $E_R^-$ couple identically to the photon -- the gauge boson of the electromagnetic interaction -- but couple differently to the $W$ and $Z$ bosons of the weak interaction -- the $E_R^-$ doesn't couple at all, it is a singlet representation of the $SU(2)_L$ symmetry of the SM\@.
    The $e_L^-$ on the other hand is, together with the left-handed electron neutrino $\nu_{eL}$, a member of a doublet representation of the $SU(2)_L$ symmetry.
    This confusing name-sharing comes about because the Higgs-boson doublet couples to the left-handed electron doublet and the right-handed-electron singlet.
    This coupling is responsible for ``giving the electron mass.''
    Electron mass is a property that the $e_L^-$ and the $E_R^-$ share.
    So two completely different particles with the same charge and the same mass -- but completely different couplings to weak interactions -- share a single name.
    
    Why is this a problem for orientability?  If one carried a $E_R^-$ around an appropriate loop of a non-orientable space then one would return with a left-handed particle that has the charge and mass of the electron but does not couple to weak interactions. This is not the $e_L^-$; it is a particle that does not exist in the SM\@.
    We say that the SM is a chiral theory.}

The topology of the Universe is as-yet undetermined (e.g., see \rcite{COMPACT:2022gbl}).
General relativity is largely agnostic to topology -- the Einstein Field Equations are systems of non-linear but local second-order partial differential equations for the components of the metric tensor, describing the local geometry, sourced by the stress-energy density field; they are not directly sensitive to the topology of the manifold.
Observations of the Cosmic Microwave Background (CMB) have allowed us to infer that the shortest closed loop around the Universe \rcite{Cornish:2003db,Cornish:2011ys} through us must be longer than $98.5\%$ of the diameter of the last scattering surface of the CMB.
Nobody, not even Mother Nature, has carried anything around any closed loops, at least since the recombination of the cosmological plasma $13.8$ billion years ago.
Claims that the Universe must be orientable therefore rest on a certain confidence that the Standard Model of particle physics can be globally defined.\footnote{
    If there is a SM field that defines the $e_L^-$ as part of a SU(2) doublet, then there must also be a $e_{R}^-$ field that is part of another SU(2) doublet; similarly, since there is an $E_R^-$ field that is an SU(2) singlet, there must also be an $E_L^-$ that is an SU(2) singlet field.
    The excitations of these fields -- i.e., all four varieties of electrons -- would then necessarily appear in our laboratories.
    But only the two with which we are already familiar -- the $e_L^-$ and the $e_R^-$ -- ever do.}

We argue that, as part of the broader program of searching for the spatial topology of the Universe, it is worth including non-orientable 3-manifolds.
No-go theorems are always dangerous -- they often have underlying assumptions that become more apparent on closer inspection, or discovery of a counter example.  
It has already been argued \cite{Brian_Pitts_2012} that the prohibition on undoubled chiral fermions in non-orientable space is inaccurate.
Others have suggested that it merely requires that every closed loop that explores the non-orientability pass through some chirality-flipping surface.

This paper therefore extends to non-orientable Euclidean manifolds the work that was done by the COMPACT collaboration in \cite{COMPACT:2023rkp} for orientable Euclidean manifolds.
In \cite{COMPACT:2023rkp}, we provided motivation for the search for cosmic topology and a computational and notational framework for making statistical predictions of cosmological observables (especially the CMB) in the case that we live in a three-dimensional spatial manifold with non-trivial spatial topology.
This framework is essential for any search for the topology of the Universe in cosmological observables.
The key ingredients are the eigenmodes of the Laplacian on each relevant manifold, and the correlations they imply between the amplitudes  of covering space eigenmodes -- i.e., Fourier modes, or spherical harmonics.

For each of the four non-orientable compact Euclidean manifolds 
    (\E{7}, Klein space; 
    \E{8}, Klein space with  horizontal flip; 
    \E{9}, Klein space with  vertical flip; and
    \E{10}, Klein space with  half turn)
and each of the four non-orientable non-compact Euclidean manifolds 
    (\E{13}, Chimney space with  vertical flip; 
    \E{14}, Chimney space with  horizontal flip; 
    \E{15}, Chimney Klein with  half-turn and flip; and
    \E{17}, slab space with flip)
we characterize the space (i.e., in section \ref{secn:topologiesmanifolds} we present a set of group generators of complete generality), and present (in section \ref{secn:eigenmodes}) the eigenmodes of the scalar Laplacian, the resulting correlation matrices of Fourier modes, and the correlation matrices of spherical-harmomic coefficients of the scalar contributions to CMB Temperature and $E$-mode polarization.\footnote{
    Here scalar, vector, and tensor refer to the transformation properties of the field, operator, etc.\ under rotations.
} 
This work generalizes an earlier study of the CMB on non-orientable Euclidean 3-manifolds \cite{COMPACT:2023rkp}.
However, in that work, specific choices were made for the representations of the transformation groups that underlie these manifolds.
Some of these choices were specific, not general.
Elsewhere, we have \cite{Samandar:2025kuf} considered the tensor eigenmodes and their contributions to CMB Temperature and both $E$-mode and $B$-mode polarizations for orientable Euclidean manifolds and will do so in \cite{COMPACTnon-orientablespin2} for the non-orientable manifolds addressed in this work.
We have also explored machine learning techniques to classify manifolds with trivial and non-trivial topologies in \cite{COMPACT:2024dqe}.
These are all essential ingredients for making statistical predictions for cosmological observables in such spaces, and thus for determining whether we likely inhabit one.
In future works we will endeavor to do both those things.

\section{Topologies and manifolds of \Espace: general considerations}
\label{secn:topologiesmanifolds-general}

In \cite{COMPACT:2023rkp}, we presented in detail the 10 orientable topologies of \Espace: \E{1}--\E{6}, \E{11}, \E{12}, \E{16}, and \E{18}.
However, we began in section 2 of that paper with a general presentation of the properties of Euclidean manifolds.
For the reader's convenience, we repeat that general presentation here in somewhat abbreviated form, but refer them to \cite{COMPACT:2023rkp} for a more complete treatment.
 
The isometry group $E(3)$ of Euclidean three-space \Espace, consists of arbitrary translations, rotations, and reflections, and all products of these.
We are interested in the freely acting discrete subgroups $\Gamma^{\E{i}}$ of $E(3)$, i.e., consisting of transformations that, except for the identify transformation, take no point of \Espace\ to itself.
These include translations, ``corkscrew motions'' (rotations about arbitrary axes followed by translations with a component parallel to that axis), ``glide reflections'' (reflections across planes followed by translations with components parallel to the plane), and certain products of these.
The non-trivial \Espace\ topologies \E{i} are formed by modding out $E(3)$ by such $\Gamma^{\E{i}}$, i.e., $E(3)\to E(3)/\Gamma^{\E{i}}$ (see, e.g., \rcite{hempel20043, hirsch2012differential}).

There are 18 distinct  $\Gamma^{\E{i}}$ (including the trivial group), one for each of the 18 distinct topologies \E{i} for $i\in\{1,\ldots,18\}$ \cite{Lachieze-Rey1995pr, Luminet1999, hitchman2009geometry, Riazuelo2004:prd}.
Each  $\Gamma^{\E{i}}$ allows for certain continuous parameters that characterize their translations and the translation components of their corkscrew motions and glide reflections.
These parameters are physical -- their values affect the statistical (or, under certain circumstances, deterministic) predictions for observables.
Also, since the topology boundary conditions break isotropy, and in most cases homogeneity, to fully describe the statistical properties of observables one must typically specify the position and orientation of the observer. This introduces up to 6 more real physical parameters.
Throughout this paper, we assume that cosmic topology is the only source of statistical isotropy or homogeneity violation.

Consider a simple three-torus, \E{1}.
Its symmetry group $\Gamma^{\E{1}}$ is generated by three pure translations, 
\begin{equation}
    \label{eqn:E1generator}
    g^{\E{1}}_i: \vec{x} \to \vec{x} + \vec{T}^{\E{1}}_i,
\end{equation}
for any three linearly independent vectors $\vec{T}^{\E{1}}_i$, $i=1,2,3$.
The simplest and most familiar special case is the cubic three-torus where the translations are orthogonal and of equal length, $\vec{T}^{\E{1}}_i = L \unitvec{e}_i$, for an orthogonal set of 3 unit vectors $\unitvec{e}_i$.
A general element of $\Gamma^{\E{1}}$ is a product of integer powers of these $g^{\E{1}}_i$, i.e., it is a translation by an integer linear combination of these three translations,
\begin{equation}
  \vec{T}^{\E{1}}_{\vec{n}}=n_1\vec{T}^{\E{1}}_1 + n_2\vec{T}^{\E{1}}_2 + n_3\vec{T}^{\E{1}}_3.
\end{equation}
If the speed of light were infinite, 
an observer in \E{1} would perceive themselves to have a lattice of ``clones'' displaced from themselves by these vectors $\vec{T}^{\E{1}}_{\vec{n}}$, for all sets of integers $\{n_1,n_2,n_3\}$ -- as they look out in the Universe in directions parallel to $\vec{T}^{\E{1}}_{\vec{n}}$, the paths trace out closed curves that return to themselves.
They would also perceive any object they see around them to also have clones displaced from its closest instance by these same vectors.
Figure 1 of \cite{COMPACT:2023rkp} illustrates the actions of the generators $g_i^{\E{1}}$, and in that paper we give a detailed description of what an observer would see depending on the specific choice of $\vec{T}^{\E{1}}_i$.
Of course, light and other signals travel at a finite speed, so this lattice of clones is observable only if the clones are close enough (see, e.g., \rcite{Sokolov:1974,Fang:1983,Fagundes:1987,Lehoucq:1996qe,Roukema:1996cu,Weatherley:2003,Fujii:2011ga,Fujii:2013xsa}).

The set of points in the space closer to the observer than to any of their clones is that observer's Dirichlet domain.
They might call this region their ``fundamental domain'' (FD) or unit cell \cite{hitchman2009geometry} and imagine tiling all of \Espace (i.e., \E{18}) with such FDs -- whatever happens in the observer's FD happens simultaneously and in precisely the same way in every other ``tile.''
These are equivalent descriptions of the same reality, and we often use them interchangeably in describing topologically non-trivial universes.

The shape of a FD is not itself an observable,
and is certainly not a physical property of the manifold.
In two dimensions this was perhaps most famously made evident by the many interesting FD shapes represented by the Dutch artist M.C.~Escher -- containing birds, fish, \textit{etc}.\ \cite{arthive}.
What is physical is the set of group elements in $\Gamma^{\E{1}}$, or, equivalently, the relative locations (and orientations) of the ``clones'' of a given point in space -- i.e., its images under elements of $\Gamma^{\E{1}}$.
Connected with this ambiguity of the shape of a FD, we noted in \cite{COMPACT:2023rkp} that there are many distinct choices of $\vec{T}^{\E{1}}_i$ that lead to the exact same lattice of clones.

Although the $\vec{T}^{\E{1}}_i$ have 9 degrees of freedom, we have chosen to associate just 6 of them with the manifold -- we can think of this as choosing the lengths of the three $T^{\E{1}}_i$ and the angles between them.
The remaining three degrees of freedom can then be taken to be the Euler angles describing the orientation of an observer's coordinate system relative to the $T^{\E{1}}_i$.

Since the three-torus \E{1} (along with \E{11} and \slabh) breaks the rotational invariance of \E{18}, the covering space, but preserves homogeneity, only the orientation of the observer needs to be specified.
The other Euclidean manifolds also break at least some of the translation invariance by establishing one or more preferred axes of rotation or planes of reflection.
Observables, such as the locations of clones relative to the observer, will depend on the observer's location.
In those spaces, we will need to specify both the position of the observer relative to those axes or planes and the orientation of the observer relative to some directions fixed by the topology.

This separation of parameters into those associated with the manifold and those specific to the observer is not unique: there is some freedom to make the generators simpler at the expense of moving the origin of the observer's coordinate system and vice-versa.
Where to assign the parameters for the simplest description is problem dependent.
It will be important for us in choosing how to describe the statistical properties of observables in candidate manifolds and how to compare them to  observations.
This freedom is discussed for each of the manifolds below.

Returning to our illustrative example of \E{1},
the group $\Gamma^{\E{1}}$ associated with \E{1} has 6 real parameters and all allowed choices of these parameters result in the same \E{1} topology, but generically they result in different lattices of clones and so are physically distinct (and distinguishable) manifolds.
However, there are equivalence classes, each with a countably infinite number of members, in which we replace the three vectors $\vec{T}^{\E{1}}_i$ by three linearly independent integer linear combinations of these three vectors chosen to not change the lattice of clones.
Thus if we are trying to characterize the allowed possibilities for $\Gamma$ without double counting we must take care in choosing the ranges of the parameters.\footnote{
    We try to be careful to use the word 
    ``manifold'' to refer to each equivalence class of physically indistinguishable spaces, and the word ``topology'' to refer collectively to all the equivalence classes with the same isometry group $\Gamma^{\E{1}}$. In other words, when we specify the vectors $\vec{T}^{\E{1}}_i$ we have a particular manifold, when we leave them unspecified we have a particular topology.  }
As in \cite{COMPACT:2023rkp},  for each of the topologies under consideration we will carefully lay out a particular choice of parameter spaces that avoids any double counting.

For \E{1}, the actions of the generators of $\Gamma^{\E{1}}$, given by \eqref{eqn:E1generator} were particularly simple.
As described in \cite{COMPACT:2023rkp}, for all \E{i}, each generator $g^{\E{i}}_{a_j}$ of $\Gamma^{\E{i}}$ acts on a point $\vec{x}$ in the manifold as
\begin{equation}
    \label{eqn:actionofgenerator}
    g_{a_j}^{\E{i}}: \vec{x} \to \mat{M}^{\E{i}}_a (\vec{x}-\vec{x}_0^{\E{i}}) + \vec{T}^{\E{i}}_{a_j} + \vec{x}^{\E{i}}_0 ,
\end{equation}
where $\mat{M}^{\E{i}}_a \in O(3)$ and $\vec{T}^{\E{i}}_{a_j}$ is a translation vector appropriate for the given topology \E{i}.
We use the index $a$ to distinguish among the (up to three) distinct $\mat{M}^{\E{i}}_a \in O(3)$ and the index $j$ to label the distinct vectors $\vec{T}^{\E{i}}_{a_j}$ for a given $\mat{M}^{\E{i}}_a$.
The vector $\vec{x}^{\E{i}}_0$ is the position (relative to some arbitrary coordinate origin) of a point on the axis about which $\mat{M}^{\E{i}}_a$ rotates or on the plane across which it reflects.
Thus when $\vec{x}=\vec{x}^{\E{i}}_0$, $g_{a_j}^{\E{i}}$ is a pure translation by $\vec{T}^{\E{i}}_{a_j}$.
Since the $\mat{M}^{\E{i}}_a$ are such that the axes about which they rotate or the normals to the planes across which they reflect are orthogonal \cite{Thurston1982ThreeDM}, we can choose a single $\vec{x}_0^{\E{i}}$ for all the generators.
(This is why $\vec{x}_0^{\E{i}}$ needs neither $a$ nor $j$ labels.)

In the case of \E{1}, $\mat{M}^{\E{i}}_a$ is the identity for all three generators.
More generally the generators can be chosen so that each $\mat{M}^{\E{i}}_a$ is one of: the identity,  a rotation about a coordinate axis, or the reflection of a single coordinate.
The generators are referred to as translations if $\mat{M}^{\E{i}}_a=\identity$, corkscrew motions if $\mat{M}^{\E{i}}_a$ is a proper rotation, and glide reflections if $\mat{M}^{\E{i}}_a$ is a reflection.
$\vec{T}^{\E{i}}_{a_j}$ can never be $\vec{0}$; if it were, then $g_{a_j}^{\E{i}}$ would not be freely acting since it would take $\vec{x}_0^{\E{i}}$ to itself.

If one or more of the $\mat{M}^{\E{i}}_a$ is not the identity, the manifold is not homogeneous, i.e., the lattice of clones of an observer depends on the location of the invariant axis/plane of $\mat{M}^{\E{i}}_a$ relative to the observer. 
For that reason, the shape of the Dirichlet domain also depends on the observer's location.
We might have been tempted to interpret the unit cell of the clone lattice or the observer's Dirichlet domain as ``the shape of the Universe'', but clearly by that definition the ``the shape of the Universe'' is in the eye of the beholder.
For a detailed explanation see \cite{COMPACT:2023rkp}.

For certain purposes, it might be useful to use the shift in origin  to ``simplify'' the set of $\vec{T}^{\E{i}}_{a_j}$, for example, to set certain components to zero, or to equate certain components to one another.
As explained in detail in \cite{COMPACT:2023rkp}:
\begin{itemize}
    \item If $\mat{M}^{\E{i}}_a$ is a proper rotation about an axis, 
    then only the components of $\vec{T}^{\E{i}}_{a_j}$ in the plane of rotation  can be altered by shifting $\vec{x}^{\E{i}}_0$.
    \item If $\mat{M}^{\E{i}}_a$ is a reflection across a plane, then the component of $\vec{T}^{\E{i}}_{a_j}$ normal to the plane of reflection can be altered.
    \item If $\mat{M}^{\E{i}}_a=\identity$, then none of the components of $\vec{T}^{\E{i}}_{a_j}$ can be altered; $\vec{T}^{\E{i}}_{a_j}$ remains an arbitrary vector.
\end{itemize}

If more than one of the $\mat{M}^{\E{i}}_a$ is not the identity, then their axes/planes must be orthogonal to one another (see, for example, \rcite{Thurston1982ThreeDM}).
Since there are at most three distinct $\mat{M}^{\E{i}}_a$, the associated axes/planes can always be taken to be parallel to coordinate axes/planes.

From a mathematical point of view, we could use our freedom to choose the origin to eliminate or relate as many as three of the components of  $\vec{T}^{\E{i}}_{a_j}$.
While this ability to simplify the $\vec{T}^{\E{i}}_{a_j}$ may prove useful for enumerating manifolds or for simulating cosmological observables,
for an observer, the most sensible choice of origin is  likely to be their own position, which may be very far from the point one would choose to yield a simplified set of generators.
We therefore preserve both $\vec{x}^{\E{i}}_0$ and $\vec{T}^{\E{i}}_{a_j}$ in our expressions for eigenmodes, and comment appropriately.

\begin{figure}[t]
     \centering
     \begin{subfigure}[b]{0.5\textwidth}
         \centering
         \includegraphics[scale=1.0]{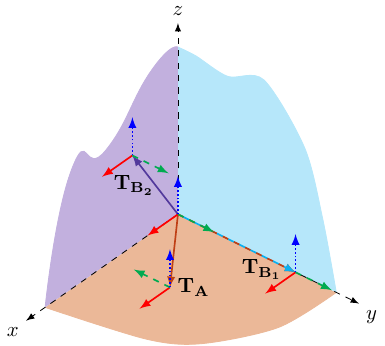}
         \caption{\E{7}}
         \label{fig:E7}
     \end{subfigure}%
     \begin{subfigure}[b]{0.5\textwidth}
         \centering
         \includegraphics[scale=1.0]{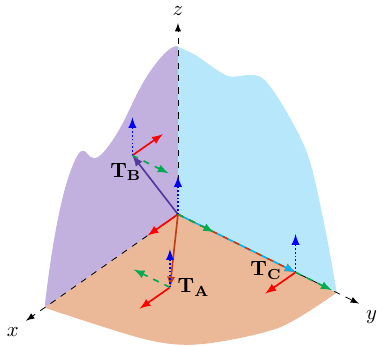}
         \caption{\E{8}}
         \label{fig:E8}
     \end{subfigure}
     \newline
     \begin{subfigure}[b]{0.5\textwidth}
         \centering
         \includegraphics[scale=1.0]{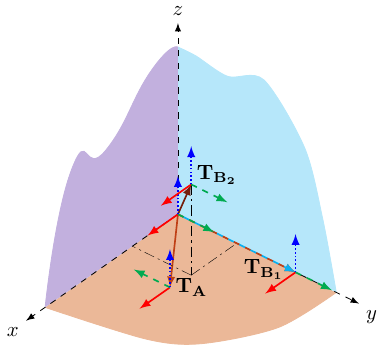}
         \caption{\E{9}}
         \label{fig:E9}
     \end{subfigure}%
     \begin{subfigure}[b]{0.5\textwidth}
         \centering
         \includegraphics[scale=1.0]{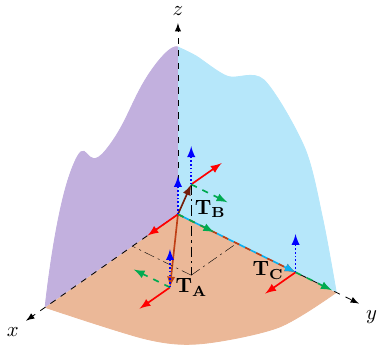}
         \caption{\E{10}}
         \label{fig:E10}
     \end{subfigure}
    \caption{Diagrams showing the actions of the generators for the topologies \E{7}--\E{10}.
    In each subdiagram, an observer at the origin is represented by an orthogonal triad $\unitvec{x}$, $\unitvec{y}$, $\unitvec{z}$ shown as  short red, green, and blue arrows, respectively.
    Translation vectors for each of the three generators are rooted at the origin, and may be labeled $\vec{T}_A$, $\vec{T}_{B_i}$, $\vec{T}_{B}$, and $\vec{T}_{C}$ depending on the details of the topology.
    At the head of each of those translation vectors is another red-green-blue triad representing one of the observer's topological clones, showing how they have been reflected or rotated compared to the observer at the origin.
    The colored coordinate planes are provided only as visual aids.
    }
    \label{fig:topologyplots}
\end{figure}

\begin{figure}[t]
     \centering
     \begin{subfigure}[b]{0.5\textwidth}
         \centering
         \includegraphics[scale=1.0]{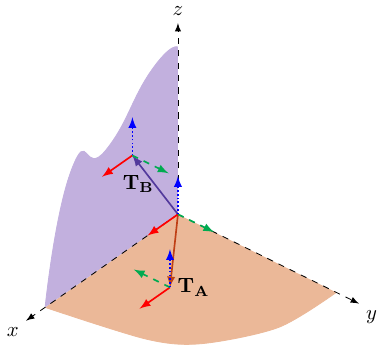}
         \caption{\E{13}}
         \label{fig:E13}
     \end{subfigure}%
     \begin{subfigure}[b]{0.5\textwidth}
         \centering
         \includegraphics[scale=1.0]{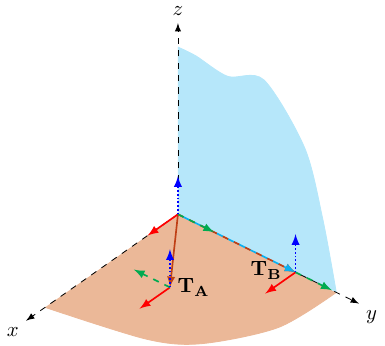}
         \caption{\E{14}}
         \label{fig:E14}
     \end{subfigure}
     \newline
     \begin{subfigure}[b]{0.5\textwidth}
         \centering
         \includegraphics[scale=1.0]{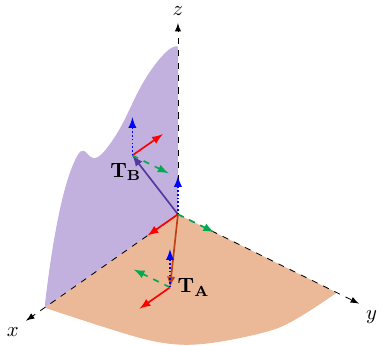}
         \caption{\E{15}}
         \label{fig:E15}
     \end{subfigure}%
     \begin{subfigure}[b]{0.5\textwidth}
         \centering
         \includegraphics[scale=1.0]{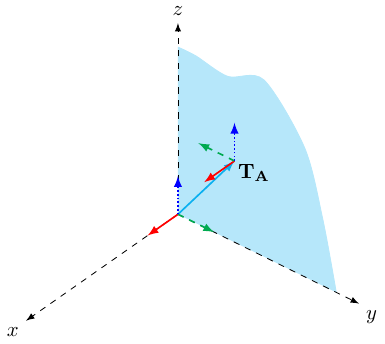}
         \caption{\E{17}}
         \label{fig:E17}
     \end{subfigure}
    \caption{Diagrams as in \cref{fig:topologyplots}, but showing the actions of the generators for the non-compact topologies \E{13}--\E{15}, and \E{17}.
    }
    \label{fig:topologyplotsnoncompact}
\end{figure}

\cref{fig:topologyplots,fig:topologyplotsnoncompact} illustrate the actions of the generators for the eight non-orientable Euclidean topologies.
All elements of $\Gamma^{\E{i}}$ can be obtained by successive actions of these generators and their inverses.
For \E{7}--\E{10} three generators are required (one can choose to include extras though we refrain from doing so in this work).
These are the compact Euclidean topologies.
For \E{13}--\E{15} two generators are required; for  \E{17} one generator is required.

\newgeometry{textwidth=0.8\paperwidth, textheight=0.85\paperheight}

\restoregeometry

\section{Properties of non-orientable Euclidean topologies}
\label{secn:topologiesmanifolds}

\begin{table}
    \newcommand{\highlightcolor}{yellow}
    \begin{tabular}{clcccc} \hline
        \textbf{Symbol} & \textbf{Name} &  \textbf{Compact} & \textbf{Orientable} & \textbf{Homogeneous} & \textbf{Isotropic} \\ 
        & & \textbf{Dimensions} & & & \\ \hline
         \E{1} & 3-torus & 3 & Yes & Yes & No \\
         \E{2} & Half-turn& 3 & Yes & No & No \\
         \E{3} & Quarter-turn & 3 & Yes & No & No \\
         \E{4} & Third-turn & 3 & Yes & No & No \\
         \E{5} & Sixth-turn & 3 & Yes & No & No \\
         \E{6} & Hantzsche-Wendt & 3 & Yes & No & No \\ \hline
         \rowcolor{\highlightcolor} \E{7} & Klein space & 3 & No & No & No \\
         \rowcolor{\highlightcolor} \E{8} & --- (horizontal flip) & 3 & No & No & No \\
         \rowcolor{\highlightcolor} \E{9} & --- (vertical flip) & 3 & No & No & No \\
         \rowcolor{\highlightcolor} \E{10} & --- (half-turn) & 3 & No & No & No \\ \hline
         \E{11} & Chimney space & 2 & Yes & Yes & No \\
         \E{12} & --- (half-turn) & 2 & Yes & No & No \\ \hline
         \rowcolor{\highlightcolor} \E{13} & --- (vertical flip) & 2 & No & No & No \\
         \rowcolor{\highlightcolor} \E{14} & --- (horizontal flip) & 2 & No & No & No \\
         \rowcolor{\highlightcolor} \E{15} & --- (half-turn + flip) & 2 & No & No & No \\ \hline
         \slabh & Slab (unrotated) & 1 & Yes & Yes & No \\
         \slabi & Slab (rotated) & 1 & Yes & No & No \\ \hline
         \rowcolor{\highlightcolor} \E{17} & Slab (flip) & 1 & No & No & No \\ \hline
         \E{18} & Covering space & 0 & Yes & Yes & Yes \\ \hline
    \end{tabular}
    \caption{Properties of the 18 three-dimensional Euclidean topologies.
    The non-orientable topologies, the focus of this work, are highlighted.
    }
    \label{tab:properties}
\end{table}

The 18 Euclidean topologies can be categorized according to their number of compact dimensions and whether or not they are orientable, homogeneous, and/or isotropic.
The topologies, with their names, symbols, and properties, are listed in \cref{tab:properties}.
In a previous paper \cite{COMPACT:2023rkp} we presented the generators, scalar-Laplacian eigenmodes, and various cosmological correlation matrices in orientable Euclidean three-manifolds.
The balance of this paper concerns the remaining Euclidean topologies, i.e., those with non-orientable manifolds: the fully compact \E{7}--\E{10} (illustrated in \cref{fig:topologyplots}), those with compact-cross-sectional area \E{13}--\E{15} (\cref{fig:topologyplotsnoncompact}), and one that is compact in one dimension  \E{17} (\cref{fig:topologyplotsnoncompact}),  as highlighted in Table~\ref{tab:properties}.
Topologies of three-manifolds that admit non-Euclidean homogeneous local geometries will be addressed in future papers.

We structure this paper to closely match that of \cite{COMPACT:2023rkp}.
In this section we summarize the important features of the manifolds of each  non-orientable Euclidean topology.
First, we list its important properties as summarized in Table~\ref{tab:properties}.
Next, we provide an action of the generators $g_{a_j}^{\E{i}}$ of its associated discrete subgroup  $\Gamma^{\E{i}}$ of $E(3)$.
In other words, we specify a set of matrices $\mat{M}^{\E{i}}_a$ and associated non-zero translation vectors $\vec{T}^{\E{i}}_{a_j}$ that characterize a manifold of each topology.

For orientable manifolds \cite{COMPACT:2023rkp}, all the matrices $\mat{M}^{\E{i}}_a$ were elements of $SO(3)$, because every element of $\Gamma^{\E{i}}$ was either a pure translation or a ``corkscrew motion'' (e.g., see \rcite{Thurston1982ThreeDM}).
For non-orientable manifolds, at least one $\mat{M}^{\E{i}}_a$ must be an element of $O(3)$ that is {\em not} in $SO(3)$, because $\Gamma^{\E{i}}$ necessarily includes glide reflections.
However, the choices of $\mat{M}^{\E{i}}_a$ are not unique even for physically indistinguishable manifolds.\footnote{
    It is clear that for a specific set of $\mat{M}^{\E{i}}_a$ the associated translation vectors are not unique, since they are specified by several real parameters, as for orientiable manifolds \cite{COMPACT:2023rkp}.
    However, the point here is that the exact same group action can be generated by different choices of group elements  $g_{a_j}^{\E{i}}$ that can have different $\mat{M}^{\E{i}}_a$.
    This point was not emphasized in \cite{COMPACT:2023rkp}. The derivations and multiple forms of the generators are provided in \cref{app:derivations}.
} 

Within each set of topologies with the same number of compact dimensions, there is exactly one for which all of its generators, and thus all the elements of $\Gamma^{\E{i}}$, are pure translations.
These are the 3-torus, \E{1}, with three compact dimensions; the chimney space, \E{11}, with two compact dimensions; the unrotated slab space, \slabh, with one compact dimension; and, trivially, the covering space (i.e., the full Euclidean space), \E{18}, with no compact dimensions.
The other topologies in each set, whether orientable or non-orientable, can be viewed as ``roots'' of these homogeneous manifolds,\footnote{%
    The one exception is \slabi\ -- the rotated slab space -- if the rotation about the axis is chosen to be by an irrational multiple of $\pi$.
    This is a rather pathological case in which all eigenmodes of the Laplacian that satisfy the boundary conditions are axially symmetric about the rotation axis of the manifold -- i.e., in cylindrical coordinates they are independent of the azimuthal angle.
    We do not consider this case further.
} as follows.
As noted above, for each group $\Gamma^{\E{i}}$ and for each $O(3)$ matrix $\mat{M}^{\E{i}}_a$ of its group elements, in particular of its generators, either $\mat{M}^{\E{i}}_a$ is itself the identity, or there is a positive integer $N\geq 2$ such that $(\mat{M}^{\E{i}}_a)^N=\identity$.
Thus the generator applied $N$ times is a pure translation, and
we are always able to construct a subgroup of $\Gamma^{\E{i}}$ of the same rank composed of such pure translations.
For \E{7}--\E{10} this subgroup is rank 3, for \E{13}--\E{15} it is rank 2, and for \E{17} it is rank 1.
Those integer-powers of generators generate an associated homogeneous manifold (AHM):
    for \E{7}--\E{10} we call this \E{1} the ``associated \E{1}'' of this manifold; 
    for \E{13}--\E{15} it is called the associated \E{11};
    for \E{17} it is the associated \slabh.

The Dirichlet domain of an observer tiles a fundamental domain of their AHM, and it will be convenient at times to think of the tiling of the covering space \E{18} by the Dirichlet domain hierarchically, i.e., the Dirichlet domain tiles the AHM, which in turn tiles the covering space.
For example, as detailed in \rcite{COMPACT:2023rkp}, the Laplacian eigenmodes can be represented as sums over Fourier modes -- one Fourier mode for each Dirichlet domain in the AHM.
Similarly, for a rank $n$ AHM, if we consider an $n\times n$ block of AHMs formed by applying the $n$ pure translations and their inverses to an AHM containing the Dirichlet domain of a point, then the nearest clone to that point will always be located within that $n\times n$ block.

As remarked above, the action of the generators is affected by the choice of orientation and origin of the coordinate system.
The choices made are contained in the description of each manifold and fall into two broad categories.
\begin{itemize}
    \item The orientation of the coordinate system used in the action of the generators allows for the simplification of the translation vectors $\vec{T}^{\E{i}}_{a_j}$ and/or to fix the ratios of some of their parameters.
    In particular, we will first use the rotational freedom to fix the normal to any plane of reflection and the axes associated with any corkscrew motions to be along a coordinate axis.
    When additional rotational freedom remains, we will use it to fix one or more of the components of a translation vector.
    \item Shifting the origin of the coordinate system, $\vec{x}_0$, allows us to freely adjust the two components of $\vec{T}^{\E{i}}_{a_j}$ perpendicular to the axis of any corkscrew motion, and the one component of $\vec{T}^{\E{i}}_{a_j}$ perpendicular to the plane of any (glide) reflection.
\end{itemize}
We describe how these are implemented and how components could be adjusted by the freedom to shift the origin.

Care must be taken when varying the parameters in generators to ensure that choices are not redundant, i.e., that choices of parameters that appear different actually generate a different lattice of clones.
A list of conditions is provided to allow one to  vary the parameters over all allowed values without ``double-counting.''

As noted above, a fundamental domain is commonly used as a tool to describe the three homogeneous spaces, but it is always observer-dependent in inhomogeneous ones.
Due to this, we do not provide fundamental domains for the non-orientable manifolds.

In the remainder of this section we provide the generators and important properties of each of the non-orientable Euclidean manifolds.
For consistency and simplicity, we deviate from the choices made elsewhere (for example, in  \cite{Riazuelo2004:prd}).
For the $O(3)$ structure we always choose $\mat{M}_A$ to be a flip across the $xz$-plane (taking $y\to -y$).
We then choose $\mat{M}_B$ to be a flip across an orthogonal plane, a rotation, or the identity, as appropriate.
Finally, we choose $\mat{M}_C \equiv \identity$ when required.
For simplicity, we choose a set of generators with the maximum number of allowed pure translations.
While these choices break with previous conventions in some cases, they lead to consistent choices of generators that make it natural to see how the partially compact spaces (\E{13}, \E{14}, \E{15}, \E{17}) follow as limits of the fully compact spaces (in particular, \E{7} and \E{8}).
They also simplify the expressions for some important quantities, like the correlation matrices, and simplify some calculations, like of the fraction of observers whose nearest clone is larger than some distance.
These choices may not make everything simpler.
For example, they make the connection to the geometry-inspired conventional names of the spaces less obvious.
A derivation of the general form of the generators, the relationship among the Klein spaces, and alternative choices for the $O(3)$ structure and the resulting generators is provided in \cref{app:derivations}.

\subsection{\E{7}: Klein space}
\label{secn:topologyE7}

\noindent \textit{Properties}: As listed in \cref{tab:properties}, manifolds of this topology are compact, non-orientable, inhomogeneous, and anisotropic.

\noindent
\textit{Generators}: A simple choice for the generators of \E{7} can be written as
\begin{align} 
    \label{eqn:E7generalT}
    & \mat{M}^{\E{7}}_A = \diag(1, -1, 1), \quad \mat{M}^{\E{7}}_B = \identity, \quad \mbox{with} \nonumber \\
    & \vec{T}^{\E{7}}_A = \begin{pmatrix} L_{Ax} \\ L_{Ay} \\ 0 \end{pmatrix}, \quad    
    \vec{T}^{\E{7}}_{\B{1}} = \begin{pmatrix} 0 \\ L_{\B{1}y} \\ 0 \end{pmatrix} \equiv \vec{T}^{\E{7}}_2, \quad    
    \vec{T}^{\E{7}}_{\B{2}} = \begin{pmatrix} L_{\B{2}x} \\ 0 \\ L_{\B{2}z}\end{pmatrix} \equiv \vec{T}^{\E{7}}_3,\,
\end{align}
where $L_{Ax}$, $L_{\B{1}y}$, and $L_{\B{2}z}$ are necessarily non-zero. 
Alternative choices are available; see \cref{app:E7E9} for a derivation and a more complete discussion.

For compactness of expressions, we will often drop the $B$ in the subscript of $L_{\B{i}w}$ so that
\begin{equation}\label{eqn:Liw-def}
    L_{iw} \equiv L_{\B{i}w}, \quad \mbox{for } i\in \{1,2\}, \, w\in\{x,y,z\}.
\end{equation}

\noindent\textit{Associated \E{1}:}
In addition to $\vec{T}^{\E{7}}_2$ and $\vec{T}^{\E{7}}_3$ defined above, another independent translation can be defined from
\begin{equation}
    \label{eqn:E7assocE1}
    (g^{\E{7}}_A)^2: \vec{x} \to \vec{x} + \vec{T}^{\E{7}}_1 ,
\end{equation}
so that
\begin{equation}
    \vec{T}^{\E{7}}_1 \equiv \begin{pmatrix} 2 L_{Ax} \\ 0 \\ 0 \end{pmatrix} , \quad
    \vec{T}^{\E{7}}_2 \equiv \begin{pmatrix} 0 \\ L_{1y} \\ 0 \end{pmatrix}, \quad
    \vec{T}^{\E{7}}_3 \equiv \begin{pmatrix} L_{2x} \\ 0 \\ L_{2z} \end{pmatrix} .
\end{equation}
The three vectors $\vec{T}^{\E{7}}_1$, $\vec{T}^{\E{7}}_2$, and $\vec{T}^{\E{7}}_3$ define the associated \E{1}.
Note that the associated \E{1} is generically tilted, a possibility that has been largely ignored in the previous literature. The only times it is untilted is when $L_{2x}=0$, which is a zero-measured set of possible parameter choices.
\\
\noindent \textit{Volume}:
\begin{equation}
    \label{eqn:VE7}
    V_{\E{7}} = \frac{1}{2}\vert(\vec{T}^{\E{7}}_{1}\times\vec{T}^{\E{7}}_{2})\cdot\vec{T}^{\E{7}}_{3}\vert 
      = \vert L_{Ax} L_{1y} L_{2z} \vert\,.
\end{equation}

\noindent \textit{Tilts versus origin position}:
When shifting the origin, ${x}^{\E{7}}_{0y}$ changes $L_{Ay}$. The values of ${x}^{\E{7}}_{0x}$ and ${x}^{\E{7}}_{0z}$ are irrelevant as they define the same reflection plane.
\\

\noindent\textit{Real parameters (5 independent):}
There are 5 independent parameters required to fully define \E{7}.
As noted above, some are redundant with shifting the origin.
Thus we have:
\begin{itemize}
    \item $L_{Ax}$, $L_{1y}$, $L_{2x}$, and $L_{2z}$ are intrinsic parameters of the manifold;
    \item $L_{Ay}$ can be adjusted using ${x}^{\E{7}}_{0y}$;
    \item the standard (special origin) form is $L_{Ay} = 0$;
    \item The choice $L_{2x} = 0$ has frequently been made; however, this is not generic as it removes one intrinsic parameter.
\end{itemize}

\noindent \textit{Parameter ranges}: We want to ensure that we do not double-count parameter choices that appear different but actually generate the same lattice of clones.
To do so we impose the following conditions:
\begin{enumerate}
    \item $0< L_{Ax}$, $0 < L_{\B{1}y} \equiv L_{1y}$, and $0 < L_{\B{2}z} \equiv L_{2z}$, i.e., choice of orientation;
    \item $\vert\vec{T}^{\E{7}}_A\cdot\vec{T}^{\E{7}}_2\vert \leq \frac{1}{2}\vert\vec{T}^{\E{7}}_2\vert^2$, i.e., $\vec{T}^{\E{7}}_{A}$ cannot be shortened by adding or subtracting $\vec{T}^{\E{7}}_2$;
    \item $\vert\vec{T}^{\E{7}}_{\B{2}}\cdot\vec{T}^{\E{7}}_1\vert \leq \frac{1}{2}\vert\vec{T}^{\E{7}}_1\vert^2$, i.e., $\vec{T}^{\E{7}}_{\B{2}}$ cannot be shortened by adding or subtracting $\vec{T}^{\E{7}}_1$.
\end{enumerate}
In terms of the parameters, the necessary conditions become:\footnote{Note that strictly the $\leq$ signs in conditions 2 and 3 allow for a double counting, e.g., $L_{Ay}=\pm L_{1y}/2$ are actually indistinguishable.  However, this is a set of measure zero in the parameter space.}
\begin{enumerate}
    \item $0 < L_{Ax}$, $0 < L_{1y}$, and $0 < L_{2z}$;
    \item $\vert L_{Ay} \vert \leq L_{1y}/2$;
    \item $\vert L_{2x} \vert \leq L_{Ax}$.
\end{enumerate}

\subsection{\E{8}: Klein space with horizontal flip}
\label{secn:topologyE8}

\noindent \textit{Properties}: As listed in \cref{tab:properties}, manifolds of this topology are compact, non-orientable, inhomogeneous, and anisotropic.
\\

\noindent \textit{Generators}: The conventional choice for the generators of \E{8} can be written as
\begin{align} 
    \label{eqn:E8generalT}
    & \mat{M}^{\E{8}}_A = \diag(1, -1, 1), \quad \mat{M}^{\E{8}}_B = \diag(-1,1,1), \quad \mat{M}^{\E{8}}_C = \identity, \quad \mbox{with} \nonumber \\
    & \vec{T}^{\E{8}}_A = \begin{pmatrix} L_{Ax} \\ L_{Ay} \\ 0 \end{pmatrix}, \quad    
    \vec{T}^{\E{8}}_B = \begin{pmatrix} L_{Bx} \\ 0 \\ L_{Bz} \end{pmatrix}, \quad    
    \vec{T}^{\E{8}}_C = \begin{pmatrix} 0 \\ L_{Cy} \\ 0 \end{pmatrix} \equiv \vec{T}^{\E{8}}_2, \,
\end{align}
where $L_{Ax}$, $L_{Bz}$, and $L_{Cy}$ are necessarily non-zero.
The rationale for the name of the space is clear with this choice of generators.
Alternative choices are available; see \cref{app:E8E10} for a derivation and a more complete discussion. \\

\noindent\textit{Associated \E{1}:}
In addition to $\vec{T}^{\E{8}}_2$ defined above, two other independent translations can be defined from
\begin{align}
    \label{eqn:E8assocE1}
    (g^{\E{8}}_A)^2: \vec{x} & \to \vec{x} + \vec{T}^{\E{8}}_1 , \nonumber \\
    (g^{\E{8}}_B)^2: \vec{x} & \to \vec{x} + \vec{T}^{\E{8}}_3 ,
\end{align}
so that
\begin{equation}
    \vec{T}^{\E{8}}_1 \equiv \begin{pmatrix} 2 L_{Ax} \\ 0 \\ 0 \end{pmatrix} , \quad
    \vec{T}^{\E{8}}_2 \equiv \begin{pmatrix} 0 \\ L_{Cy} \\ 0 \end{pmatrix}, \quad
    \vec{T}^{\E{8}}_3 \equiv \begin{pmatrix} 0 \\ 0 \\ 2 L_{Bz} \end{pmatrix} .
\end{equation}
The three vectors $\vec{T}^{\E{8}}_1$, $\vec{T}^{\E{8}}_2$, and $\vec{T}^{\E{8}}_3$ define the associated \E{1}. Note that this is always an \emph{untilted} associated \E{1}, unlike for \E{7}, which can be either.
\\

\noindent \textit{Volume}:
\begin{equation}
    \label{eqn:VE8}
    V_{\E{8}} = \frac{1}{4}\vert(\vec{T}^{\E{8}}_{1}\times\vec{T}^{\E{8}}_{2})\cdot\vec{T}^{\E{8}}_{3}\vert 
      = \vert L_{Ax} L_{Cy} L_{Bz} \vert\,.
\end{equation}

\noindent \textit{Tilts versus origin position}:
When shifting the origin $x^{\E{8}}_{0x}$ shifts $L_{Bx}$ and ${x}^{\E{8}}_{0y}$ shifts $L_{Ay}$. The value of ${x}^{\E{8}}_{0z}$ is irrelevant as it defines the same reflection planes.
\\

\noindent\textit{Real parameters (5 independent):}
There are 5 independent parameters required to fully define \E{8}.
As noted above, some are redundant with shifting the origin.
Thus we have:
\begin{itemize}
    \item $L_{Ax}$, $L_{Cy}$, and $L_{Bz}$ are intrinsic parameters of the manifold;
    \item $L_{Ay}$ can be adjusted using ${x}^{\E{8}}_{0y}$ and $L_{Bx}$ can be adjusted using ${x}^{\E{8}}_{0x}$;
    \item the standard (special origin, i.e., ``untilted'') form is $0 = L_{Ay} = L_{Bx}$.
\end{itemize}

\noindent \textit{Parameter ranges}: We want to ensure that we do not double-count parameter choices that appear different but actually generate the same lattice of clones.
Similar to \E{7} we have
\begin{enumerate}
    \item $0 < L_{Ax}$, $0 < L_{Cy}$, and $0 < L_{Bz}$, i.e., choice of orientation;
    \item $\vert\vec{T}^{\E{8}}_A\cdot\vec{T}^{\E{8}}_2\vert \leq \frac{1}{2}\vert\vec{T}^{\E{8}}_2\vert^2$, i.e., $\vec{T}^{\E{8}}_{A}$ cannot be shortened by adding or subtracting $\vec{T}^{\E{8}}_2$;
    \item $\vert\vec{T}^{\E{8}}_B\cdot\vec{T}^{\E{8}}_1\vert \leq \frac{1}{2}\vert\vec{T}^{\E{8}}_1\vert^2$, i.e., $\vec{T}^{\E{8}}_B$ cannot be shortened by adding or subtracting $\vec{T}^{\E{8}}_1$.
\end{enumerate}
In terms of the parameters, the necessary conditions become:
\begin{enumerate}
    \item $0 < L_{Ax}$, $0 < L_{Cy}$, and $0 < L_{Bz}$;
    \item $\vert L_{Ay} \vert \leq L_{Cy}/2$;
    \item $\vert L_{Bx} \vert \leq L_{Ax}$.
\end{enumerate}

\subsection{\E{9}: Klein space with vertical flip}
\label{secn:topologyE9}

\noindent \textit{Properties}: As listed in \cref{tab:properties}, manifolds of this topology are compact, non-orientable, inhomogeneous, and anisotropic.
\\

\noindent \textit{Generators}: The conventional choice for the generators of \E{9} can be written as
\begin{align} 
    \label{eqn:E9generalT}
    & \mat{M}^{\E{9}}_A = \diag(1, -1, 1), \quad \mat{M}^{\E{9}}_B = \identity, \quad \mbox{with} \nonumber \\
    & \vec{T}^{\E{9}}_A = \begin{pmatrix} L_{Ax} \\ L_{Ay} \\ 0 \end{pmatrix}, \quad    
    \vec{T}^{\E{9}}_{\B{1}} = \begin{pmatrix} 0 \\ L_{\B{1}y} \\ 0 \end{pmatrix} \equiv \vec{T}^{\E{9}}_2, \quad    
    \vec{T}^{\E{9}}_{\B{2}} = \begin{pmatrix} L_{\B{2}x} \\ L_{\B{1}y}/2 \\ L_{\B{2}z}\end{pmatrix} \equiv \vec{T}^{\E{9}}_3,
\end{align}
where $L_{Ax}$, $L_{\B{1}y}$, and $L_{\B{2}z}$ are necessarily non-zero. The rationale for the name of the space is not clear with this choice of generators.
Alternative choices are available; see \cref{app:E7E9} for a derivation and a more complete discussion.
\\

For compactness of expressions we will often drop the $B$ in the subscript of $L_{\B{i}w}$ so that 
\begin{equation}
    L_{iw} \equiv L_{\B{i}w}, \quad \mbox{for } i\in \{1,2\}, \, w\in\{x,y,z\}.
\end{equation}

\noindent\textit{Associated \E{1}:}
In addition to $\vec{T}^{\E{9}}_2$ and $\vec{T}^{\E{9}}_3$ defined above, another independent translation can be defined from
\begin{equation}
    \label{eqn:E9assocE1}
    (g^{\E{9}}_{\A{}})^2: \vec{x} \to \vec{x} + \vec{T}^{\E{9}}_1 ,
\end{equation}
so that
\begin{equation}
    \vec{T}^{\E{9}}_1 \equiv \begin{pmatrix} 2 L_{Ax} \\ 0 \\ 0 \end{pmatrix} , \quad
    \vec{T}^{\E{9}}_2 \equiv \begin{pmatrix} 0 \\ L_{1y} \\ 0 \end{pmatrix}, \quad
    \vec{T}^{\E{9}}_3 \equiv \begin{pmatrix} L_{2x} \\ L_{1y}/2 \\ L_{2z} \end{pmatrix} .
\end{equation}
The three vectors $\vec{T}^{\E{9}}_1$, $\vec{T}^{\E{9}}_2$, and $\vec{T}^{\E{9}}_3$ define the associated \E{1}.
Note that  this is \emph{always} a tilted associated \E{1}, i.e., $L_{1y}\neq0$.
We also note that two translations in the associated \E{1} contain a component in the $y$-direction, i.e., the direction normal to the plane of the flip. 
This will be relevant when determining distances between clones of any given point in the manifold.
\\

\noindent \textit{Volume}:
\begin{equation}
    \label{eqn:VE9}
    V_{\E{9}} = \frac{1}{2}\vert(\vec{T}^{\E{9}}_{1}\times\vec{T}^{\E{9}}_{2})\cdot\vec{T}^{\E{9}}_{3}\vert 
      = \vert L_{Ax} L_{1y} L_{2z} \vert.
\end{equation}

\noindent \textit{Tilts versus origin position}:
When shifting the origin, ${x}^{\E{9}}_{0y}$ changes $L_{Ay}$. The values of ${x}^{\E{9}}_{0x}$ and ${x}^{\E{9}}_{0z}$ are irrelevant as they define the same reflection plane.
\\

\noindent\textit{Real parameters (5 independent):}
There are 5 independent parameters required to fully define \E{9}.
As noted above, some are redundant with shifting the origin.
Thus we have:
\begin{itemize}
    \item $L_{Ax}$, $L_{1y}$, $L_{2x}$, and $L_{2z}$ are intrinsic parameters of the manifold;
    \item $L_{Ay}$ can be adjusted using ${x}^{\E{9}}_{0y}$;
    \item the standard (special origin) form is $L_{Ay} = 0$;
    \item the choice $L_{2x} = 0$ has also frequently been made, though this is not generic as it removes one intrinsic parameter.
\end{itemize}

\noindent \textit{Parameter ranges}: We want to ensure that we do not double-count parameter choices that appear different but actually generate the same lattice of clones.
Similar to \E{7}, we therefore require:
\begin{enumerate}
    \item $0< L_{Ax}$, $0 < L_{\B{1}y} \equiv L_{1y}$, and $0 < L_{\B{2}z} \equiv L_{2z}$, i.e., choice of orientation;
    \item $\vert\vec{T}^{\E{9}}_A\cdot\vec{T}^{\E{9}}_2\vert \leq \frac{1}{2}\vert\vec{T}^{\E{9}}_2\vert^2$, i.e., $\vec{T}^{\E{9}}_{A}$ cannot be shortened by adding or subtracting $\vec{T}^{\E{9}}_2$;
    \item $\vert\vec{T}^{\E{9}}_{\B{2}}\cdot\vec{T}^{\E{9}}_1\vert \leq \frac{1}{2}\vert\vec{T}^{\E{9}}_1\vert^2$, i.e., $\vec{T}^{\E{9}}_{\B{2}}$ cannot be shortened by adding or subtracting $\vec{T}^{\E{9}}_1$.
\end{enumerate}
In terms of the parameters, the necessary conditions become:
\begin{enumerate}
    \item $0 < L_{Ax}$, $0 < L_{1y}$, and $0 < L_{2z}$;
    \item $\vert L_{Ay} \vert \leq L_{1y}/2$;
    \item $\vert L_{2x} \vert \leq L_{Ax}$.
\end{enumerate}

\subsection{\E{10}: Klein space with half-turn}
\label{secn:topologyE10}

\noindent \textit{Properties}: As listed in \cref{tab:properties}, manifolds of this topology are compact, non-orientable, inhomogeneous, and anisotropic.
\\

\noindent
\textit{Generators}: The conventional choice for the generators of \E{10} can be written as
\begin{align} 
    \label{eqn:E10generalT}
    & \mat{M}^{\E{10}}_A = \diag(1, -1, 1), \quad \mat{M}^{\E{10}}_B = \diag(-1,1,1), \quad \mat{M}^{\E{10}}_C = \identity, \quad \mbox{with} \nonumber \\
    & \vec{T}^{\E{10}}_A = \begin{pmatrix} L_{Ax} \\ L_{Ay} \\ 0 \end{pmatrix}, \quad    
    \vec{T}^{\E{10}}_B = \begin{pmatrix} L_{Bx} \\ L_{Cy}/2 \\ L_{Bz} \end{pmatrix}, \quad    
    \vec{T}^{\E{10}}_C = \begin{pmatrix} 0 \\ L_{Cy} \\ 0 \end{pmatrix} \equiv \vec{T}^{\E{10}}_2,
\end{align}
where $L_{Ax}$, $L_{Bz}$, and $L_{Cy}$ are necessarily non-zero. The rationale for the name of the space is that $\mat{M}^{\E{10}}_A\mat{M}^{\E{10}}_B$ is a rotation about the $z$-axis by $\pi$.
Alternative choices are available; see \cref{app:E8E10} for a derivation and a more complete discussion.
\\

\noindent\textit{Associated \E{1}:}
In addition to $\vec{T}^{\E{10}}_2$ defined above, two other independent translations can be defined from
\begin{align}
    \label{eqn:E10assocE1}
    (g^{\E{10}}_A)^2: \vec{x} & \to \vec{x} + \vec{T}^{\E{10}}_1 , \nonumber \\
    \inverse{(g^{\E{10}}_C)}(g^{\E{10}}_B)^2: \vec{x} & \to \vec{x} + \vec{T}^{\E{10}}_3 ,
\end{align}
so that
\begin{equation}
    \vec{T}^{\E{10}}_1 \equiv \begin{pmatrix} 2 L_{Ax} \\ 0 \\ 0 \end{pmatrix} , \quad
    \vec{T}^{\E{10}}_2 \equiv \begin{pmatrix} 0 \\ L_{Cy} \\ 0 \end{pmatrix} , \quad
    \vec{T}^{\E{10}}_3 \equiv \begin{pmatrix} 0 \\ 0 \\ 2 L_{Bz} \end{pmatrix}  .
\end{equation}
The three vectors $\vec{T}^{\E{10}}_1$, $\vec{T}^{\E{10}}_2$, and $\vec{T}^{\E{10}}_3$ define the associated \E{1}. Note that, like for \E{8}, this is always an \emph{untilted} associated \E{1}, unlike for \E{9}, which is always tilted, and \E{7}, which can be either.

\noindent \textit{Volume}:
\begin{equation}
    \label{eqn:VE10}
    V_{\E{10}} = \frac{1}{4}\vert(\vec{T}^{\E{10}}_{1}\times\vec{T}^{\E{10}}_{2})\cdot\vec{T}^{\E{10}}_{3}\vert 
      = \vert L_{Ax} L_{Cy} L_{Bz} \vert.
\end{equation}

\noindent \textit{Tilts versus origin position}:
When shifting the origin $x^{\E{10}}_{0x}$ shifts $L_{Bx}$ and ${x}^{\E{10}}_{0y}$ shifts $L_{Ay}$. The value of ${x}^{\E{10}}_{0z}$ is irrelevant as it defines the same reflection planes.
\\

\noindent\textit{Real parameters (5 independent):}
Just like in \E{8}, there are 5 independent parameters required to fully define \E{10}.
As noted above, some are redundant with shifting the origin.
Thus we have:
\begin{itemize}
    \item $L_{Ax}$, $L_{Cy}$, and $L_{Bz}$ are intrinsic parameters of the manifold;
    \item $L_{Ay}$ can be adjusted using ${x}^{\E{10}}_{0y}$ and $L_{Bx}$ can be adjusted using ${x}^{\E{10}}_{0x}$;
    \item the standard (special origin, i.e., ``untilted'') form is $L_{Ay} = L_{Bx} = 0$.
\end{itemize}

\noindent \textit{Parameter ranges}: We want to ensure that we do not double-count parameter choices that appear different but actually generate the same lattice of clones.
Similar to \E{8}, we therefore require:
\begin{enumerate}
    \item $0 < L_{Ax}$, $0 < L_{Cy}$, and $0 < L_{Bz}$, i.e., choice of orientation;
    \item $\vert\vec{T}^{\E{10}}_A\cdot\vec{T}^{\E{10}}_2\vert \leq \frac{1}{2}\vert\vec{T}^{\E{10}}_2\vert^2$, i.e., $\vec{T}^{\E{10}}_{A}$ cannot be shortened by adding or subtracting $\vec{T}^{\E{10}}_2$;
    \item $\vert\vec{T}^{\E{10}}_B\cdot\vec{T}^{\E{10}}_1\vert \leq \frac{1}{2}\vert\vec{T}^{\E{10}}_1\vert^2$, i.e., $\vec{T}^{\E{10}}_B$ cannot be shortened by adding or subtracting $\vec{T}^{\E{10}}_1$.
\end{enumerate}
In terms of the parameters, the necessary conditions become:
\begin{enumerate}
    \item $0 < L_{Ax}$, $0 < L_{Cy}$, and $0 < L_{Bz}$;
    \item $\vert L_{Ay} \vert \leq L_{Cy}/2$;
    \item $\vert L_{Bx} \vert \leq L_{Ax}$.
\end{enumerate}

\subsection{\E{13}: Chimney space with vertical flip}
\label{secn:topologyE13}

\E{13} is the chimney space with a glide reflection and translation perpendicular to the glide axis.
It can be thought of as \E{7} with with one non-compact dimension. \\

\noindent \textit{Properties}:
As listed in \cref{tab:properties}, this manifold has compact cross-sections and is non-orientable, inhomogeneous, and anisotropic. \\

\noindent \textit{Generators}: Since \E{13} only has two compact dimensions, it is described by two generators.
As with the other Klein spaces, there are multiple choices that can be made; see \cref{app:E13E14} for a derivation and a more complete discussion.
For \E{13} we choose the set of generators based on \E{7} with $|\vec{T}^{\E{7}}_{\B{1}}| \to \infty$ written as
\begin{align} 
    \label{eqn:E13generalT}
    & \mat{M}^{\E{13}}_A = \diag(1, -1, 1), \quad \mat{M}^{\E{13}}_B = \identity, \quad\mbox{with} \nonumber \\
    & \vec{T}^{\E{13}}_A = \begin{pmatrix} L_{Ax} \\ L_{Ay} \\ 0 \end{pmatrix} , \quad
    \vec{T}^{\E{13}}_{B} = \begin{pmatrix} L_{Bx} \\ 0 \\ L_{Bz} \end{pmatrix} \equiv \vec{T}^{\E{13}}_{2},  \nonumber
\end{align}
where $L_{Ax}$ and $L_{Bz}$ are necessarily non-zero. 
The rationale for the name of the space is not clear with this choice of generators since we have chosen to orient the one flip across the $xz$-plane instead of the $xy$-plane.

\noindent \textit{Associated \E{11}}: In addition to $\vec{T}^{\E{13}}_B$ defined above, a second independent translation is
\begin{equation}
    ( g^{\E{13}}_A )^2: \vec{x} \to \vec{x} + \vec{T}^{\E{13}}_1,
\end{equation}
so that
\begin{equation}
    \vec{T}^{\E{13}}_1 \equiv \begin{pmatrix} 2 L_{Ax} \\ 0 \\ 0 \end{pmatrix} , \quad
    \vec{T}^{\E{13}}_2 \equiv \begin{pmatrix} L_{Bx} \\ 0 \\ L_{Bz} \end{pmatrix}  .
\end{equation}
The two vectors $\vec{T}^{\E{13}}_1$ and $\vec{T}^{\E{13}}_2$ define the associated \E{11}.
\\

\noindent\textit{Cross-sectional area}: 
Since the chimney spaces have two compact dimensions their volumes are infinite, but their cross-sections perpendicular to the non-compact direction are finite:
\begin{equation}
    \label{eqn:AE13}
    A_{\E{13}} = \frac{1}{2} \vert \vec{T}^{\E{13}}_1 \times \vec{T}^{\E{13}}_2 \vert =  \vert L_{Ax} L_{By} \vert.
\end{equation}

\noindent\textit{Tilts versus origin position}: 
When shifting the origin, ${x}^{\E{13}}_{0y}$ changes $L_{Ay}$. The values of ${x}^{\E{13}}_{0x}$ and ${x}^{\E{13}}_{0z}$ are irrelevant as they define the same reflection plane. \\

\noindent \textit{Real parameters (4 independent)}: There are 4 independent parameters required to fully define \E{13} with one being redundant with shifting the origin.
Thus we have:
\begin{itemize}
    \item $L_{Ax}$, $L_{Bx}$ and $L_{Bz}$ are intrinsic parameters of the manifold;
    \item $L_{Ay}$ can be traded for $x^{\E{13}}_{0y}$;
    \item the standard (special origin) form is $L_{Ay} = 0$;
    \item the choice $L_{Bx} = 0$ has also frequently been made, though this is not generic as it removes one intrinsic parameter.
\end{itemize}

\noindent\textit{Parameter ranges}:  We want to ensure that we do not double-count parameter choices that appear different but actually generate the same lattice of clones.
Since there are only two compact directions there are fewer constraints than in \E{7}.
For \E{13} we require
\begin{enumerate}
    \item $0< L_{Ax}$ and $0 < L_{Bz}$, i.e., choice of orientation;
    \item $\vert\vec{T}^{\E{13}}_B\cdot\vec{T}^{\E{13}}_1\vert \leq \frac{1}{2}\vert\vec{T}^{\E{13}}_1\vert^2$, i.e., $\vec{T}^{\E{13}}_B$ cannot be shortened by adding or subtracting $\vec{T}^{\E{13}}_1$.
\end{enumerate}

In terms of the parameters, the necessary conditions become:
\begin{enumerate}
    \item $0 < L_{Ax}$ and $0 < L_{Bz}$;
    \item $\vert L_{Bx} \vert \leq L_{Ax}$.
\end{enumerate}

\subsection{\E{14}: Chimney space with horizontal flip}
\label{secn:topologyE14}

\E{14} is the chimney space with a glide reflection and translation parallel to the glide axis.
It can be thought of as \E{7} with one non-compact dimension. \\

\noindent \textit{Properties}:
As listed in \cref{tab:properties}, this manifold has compact cross-sections and is non-orientable, inhomogeneous, and anisotropic. \\

\noindent \textit{Generators}: \E{14} is similar to \E{13} and can be described by the same $O(3)$ structure; see \cref{app:E13E14} for a derivation and a more complete discussion.
For \E{14} we choose the set of generators based on \E{7} with $|\vec{T}^{\E{7}}_{\B{2}}| \to \infty$
\begin{align} 
    \label{eqn:E14generalT}
    & \mat{M}^{\E{14}}_A = \diag(1, -1, 1), \quad \mat{M}^{\E{14}}_B = \identity, \quad\mbox{with} \nonumber \\
    & \vec{T}^{\E{14}}_A = \begin{pmatrix} L_{Ax} \\ L_{Ay} \\ 0 \end{pmatrix} , \quad
    \vec{T}^{\E{14}}_{B} = \begin{pmatrix} 0 \\ L_B \\ 0 \end{pmatrix} \equiv \vec{T}^{\E{14}}_{2}, \\ \nonumber
\end{align}
where $L_{Ax}$ and $L_{B}$ are necessarily non-zero. The name of the space seems logical with this choice of generators since we have chosen to orient the one flip across $xz$-plane.

\noindent \textit{Associated \E{11}}: In addition to $\vec{T}^{\E{14}}_B$ defined above, a second independent translation is
\begin{equation}
    ( g^{\E{14}}_A )^2: \vec{x} \to \vec{x} + \vec{T}^{\E{14}}_1,
\end{equation}
so that
\begin{equation}
    \vec{T}^{\E{14}}_1 \equiv \begin{pmatrix} 2 L_{Ax} \\ 0 \\ 0 \end{pmatrix} , \quad
    \vec{T}^{\E{14}}_2 \equiv \begin{pmatrix} 0 \\ L_{B} \\ 0 \end{pmatrix}  .
\end{equation}
The two vectors $\vec{T}^{\E{14}}_1$ and $\vec{T}^{\E{14}}_2$ define the associated \E{11}.
\\

\noindent\textit{Cross-sectional area}: 
Since the chimney spaces have two compact dimensions their volumes are infinite, but their cross-sections perpendicular to the non-compact direction are finite:
\begin{equation}
    \label{eqn:AE14}
    A_{\E{14}} = \frac{1}{2} \vert \vec{T}^{\E{14}}_1 \times \vec{T}^{\E{14}}_2 \vert =  \vert L_{Ax} L_B \vert.
\end{equation}

\noindent\textit{Tilts versus origin position}: 
When shifting the origin, ${x}^{\E{14}}_{0y}$ changes $L_{Ay}$. The values of ${x}^{\E{14}}_{0x}$ and ${x}^{\E{14}}_{0z}$ are irrelevant as they define the same reflection plane.\\

\noindent \textit{Real parameters (3 independent)}: There are 3 independent parameters required to fully define \E{14} with one being redundant with shifting the origin.
Thus we have:
\begin{itemize}
    \item $L_{Ax}$ and $L_B$ are intrinsic parameters of the manifold;
    \item $L_{Ay}$ can be traded for $x^{\E{14}}_{0y}$;
    \item the standard (special origin) form is $L_{Ay} = 0$.
\end{itemize}

\noindent \textit{Parameter ranges}: We want to ensure that we do not double-count parameter choices that appear different but actually generate the same lattice of clones.
Since there are only two compact directions there are fewer constraints than in \E{7}.
For \E{14} we require
\begin{enumerate}
    \item $0< L_{Ax}$ and $0 < L_B$, i.e., choice of orientation;
    \item $\vert\vec{T}^{\E{14}}_A\cdot\vec{T}^{\E{14}}_2\vert \leq \frac{1}{2}\vert\vec{T}^{\E{14}}_2\vert^2$, i.e., $\vec{T}^{\E{14}}_{A}$ cannot be shortened by adding or subtracting $\vec{T}^{\E{14}}_2$.
\end{enumerate}
In terms of the parameters, the necessary conditions become:
\begin{enumerate}
    \item $0 < L_{Ax}$ and $0 < L_B$;
    \item $\vert L_{Ay} \vert \leq L_B/2$.
\end{enumerate}

\subsection{\E{15}: Chimney Klein with half-turn and flip}
\label{secn:topologyE15}

The chimney space \E{15} is the chimney space with glide reflections along two orthogonal axes.
It can be thought of as \E{8} with one non-compact dimension. \\

\noindent \textit{Properties}:
As listed in \cref{tab:properties}, this manifold has compact cross-sections and is non-orientable, inhomogeneous, and anisotropic. \\

\noindent \textit{Generators}: \E{15} is similar to \E{13} in that it has two generators and can be described in multiple ways; see \cref{app:E15} for a derivation and a more complete discussion.
For \E{15} we choose the set of generators based on \E{8} with $|\vec{T}^{\E{8}}_{C}| \to \infty$
\begin{align} 
    \label{eqn:E15generalT}
    & \mat{M}^{\E{15}}_A = \diag(1, -1, 1), \quad \mat{M}^{\E{15}}_B = \diag(-1, 1, 1), \quad\mbox{with} \nonumber \\
    & \vec{T}^{\E{15}}_A = \begin{pmatrix} L_{Ax} \\ L_{Ay} \\ 0 \end{pmatrix} , \quad
    \vec{T}^{\E{15}}_{B} = \begin{pmatrix} L_{Bx} \\ 0 \\ L_{Bz} \end{pmatrix}, \\ \nonumber
\end{align}
where $L_{Ax}$ and $L_{Bz}$ are necessarily non-zero. The rationale for the name of the space is not clear with this choice of generators since we have chosen to have two glide reflections (flips) instead of one glide reflection and one corkscrew motion. 
However, since $\mat{M}^{\E{15}}_A \mat{M}^{\E{15}}_B$ is a rotation about the $z$-axis, we could have replaced one of the glide reflection with a  half-turn corkscrew motion.

\noindent \textit{Associated \E{11}}: Two translations can be constructed from
\begin{equation}
    ( g^{\E{15}}_A )^2: \vec{x} \to \vec{x} + \vec{T}^{\E{15}}_1,
    ( g^{\E{15}}_B )^2: \vec{x} \to \vec{x} + \vec{T}^{\E{15}}_2,
\end{equation}
for
\begin{equation}
    \label{eqn:E15assocE11}
    \vec{T}^{\E{15}}_1 \equiv \begin{pmatrix} 2 L_{Ax} \\ 0 \\ 0 \end{pmatrix}
    \quad\mbox{and}\quad
    \vec{T}^{\E{15}}_2 \equiv \begin{pmatrix} 0 \\ 0 \\ 2 L_{Bz} \end{pmatrix}
    .
\end{equation}
The two vectors $\vec{T}^{\E{15}}_1$ and $\vec{T}^{\E{15}}_2$ define the associated \E{11}.
\\

\noindent\textit{Cross-sectional area}: 
Since the chimney spaces have two compact dimensions their volumes are infinite, but their cross-sections perpendicular to the non-compact direction are finite:
\begin{equation}
    \label{eqn:AE15}
    A_{\E{15}} = \frac{1}{4} \vert \vec{T}^{\E{15}}_1 \times \vec{T}^{\E{15}}_2 \vert =  \vert L_{Ax} L_{Bz} \vert.
\end{equation}

\noindent\textit{Tilts versus origin position}: 
When shifting the origin $x^{\E{15}}_{0x}$ shifts $L_{Bx}$ and ${x}^{\E{15}}_{0y}$ shifts $L_{Ay}$. The value of ${x}^{\E{15}}_{0z}$ is irrelevant as it defines the same reflection planes. \\

\noindent \textit{Real parameters (4 independent)}: There are 4 independent parameters required to fully define \E{15} with two being redundant with shifting the origin.
Thus we have:
\begin{itemize}
    \item $L_{Ax}$ and $L_{Bz}$ are intrinsic parameters of the manifold;
    \item $L_{Ay}$ can be traded for $x^{\E{15}}_{0y}$ and $L_{Bx}$ can be traded for $x^{\E{15}}_{0x}$;
    \item the standard (special origin) form is $L_{Ay} = L_{Bx} = 0$.
\end{itemize}

\noindent\textit{Parameter ranges}:  We want to ensure that we do not double-count parameter choices that appear different but actually generate the same lattice of clones.
Since there are only two compact directions there are fewer constraints than in \E{8}.
For \E{15} we require
\begin{enumerate}
    \item $0 < L_{Ax}$ and $0 < L_{Bz}$, i.e., choice of orientation;
    \item $\vert\vec{T}^{\E{15}}_B\cdot\vec{T}^{\E{15}}_1\vert \leq \frac{1}{2}\vert\vec{T}^{\E{15}}_1\vert^2$, i.e., $\vec{T}^{\E{15}}_B$ cannot be shortened by adding or subtracting $\vec{T}^{\E{15}}_1$.
\end{enumerate}
In terms of the parameters, the necessary conditions become:
\begin{enumerate}
    \item $0 < L_{Ax}$ and $0 < L_{Bz}$;
    \item $\vert L_{Bx} \vert \leq L_{Ax}$.
\end{enumerate}

\subsection{\E{17}: Slab space with flip}
\label{secn:topologyE17}

The slab space \E{17} is the slab space with one glide reflection.
It can be thought of as \E{7} with two non-compact dimensions. \\

\noindent \textit{Properties:} As listed in \cref{tab:properties}, this manifold has a compact length and is non-orientable, inhomogeneous, and anisotropic. \\

\noindent \textit{Generators}: In general, since \E{17} has one compact dimension it is described by one generator, which we may take to be a glide reflection in the $xz$-plane,
\begin{equation} 
    \label{eqn:E17generalT}
    \mat{M}^{\E{17}}_A = \diag(1,-1,1) \quad \mbox{with} \quad
    \vec{T}^{\E{17}}_A = \begin{pmatrix} 0 \\ L_y \\ L_z \end{pmatrix},
\end{equation}
where $L_{z}$ is necessarily non-zero.

\noindent \textit{Associated \slabh}: A pure translation (the associated \slabh) can be defined for \E{17} as
\begin{equation}
    \label{eqn:E16iassocE16h}
    g^{\E{17}}_1 \equiv (g^{\E{17}}_A)^2: \vec{x} \to \vec{x} + \vec{T}^{\E{17}}_1, \quad \mbox{for }
    \vec{T}^{\E{17}}_1 \equiv \transpose{(0, 0, 2 L_{z})} .
\end{equation}
\\
\noindent \textit{Length}: Since the slab spaces have only one compact dimension their volumes and cross-sectional areas are infinite.
Given \eqref{eqn:E16iassocE16h}, the length of \E{17} is 
\begin{equation}
    \label{eqn:LE17}
    L_{\E{17}} =  \vert L_z \vert.
\end{equation}

\noindent \textit{Tilts versus origin position}: 
When shifting the origin, ${x}^{\E{17}}_{0y}$ changes $L_{Ay}$. The values of ${x}^{\E{17}}_{0x}$ and ${x}^{\E{17}}_{0z}$ are irrelevant as they define the same reflection plane.
 \\

\noindent \textit{Real parameters (2 independent)}: There are 2 independent parameters required to fully define \E{17} with 1 parameter interchangeable with a shift of origin.
Thus we have:
\begin{itemize}
    \item $L_z$ is an intrinsic parameter of the manifold;
    \item $L_y$ can be traded for $x^{\E{17}}_{0y}$;
    \item the standard (special origin, i.e., ``untilted'') form is $L_y = 0$.
\end{itemize}

\noindent \textit{Parameter ranges}: We want to ensure that we do not double-count parameter choices that appear different but actually generate the same lattice of clones.
Similar to \slabh, we can always require $0 < L_y$ and $0 < L_z$ through orientation of the coordinate system.

\section{Eigenmodes of the scalar Laplacian and correlation matrices}
\label{secn:eigenmodes}

Cosmological perturbation theory is usually developed in a basis of the  scalar, vector, and tensor eigenmodes of the Laplacian, for which theories typically give statistical predictions of the amplitudes \cite{Harrison:1967zza,Lyth:1995cw,Hu:1997mn}.

In this section, we present the scalar eigenmodes for the non-orientable Euclidean manifolds in their full generality, following closely the structure of the analogous section of \cite{COMPACT:2023rkp}.
Such eigenmodes have been presented before \cite{Inoue1999:cqg,Lehoucq2002:cqg, Lachieze-rey2005, Riazuelo2004:prd,Weeks2006:cqg}, but not including the full parameter space associated with each topology.
We are not faithful to the notational conventions of those prior works, so any comparisons should be made carefully.

As in \cite{COMPACT:2023rkp}, we first, but even more briefly, review the situation in the covering space \E{18}, where a conventional set of eigenmodes of the scalar Laplacian with a convenient normalization are Fourier modes
\begin{equation}
    \label{eqn:EuclideanFourierBasis}
    \Upsilon^{\E{18}}_{\vec{k}}(\vec{x}) = \eexp^{i\vec{k}\cdot(\vec{x}-\vec{x}_0)}.
\end{equation}
Here $\vec{k}=\transpose{(k_x,k_y,k_z)}$, referred to as the wavevector, is any triplet of real values (with units inverse to those of $\vec{x}$),
while $\vec{x}_0$ is arbitrary since \E{18} is homogeneous.
The explicit inclusion of $\vec{x}_0$ is significant in the inhomogeneous spaces that we explore here.
The Laplacian eigenvalue $-\vert\vec{k}\vert^2\equiv -k^2$ of $\Upsilon^{\E{18}}_{\vec{k}}$  can assume any non-positive real value.

In standard inflationary cosmological theory, the adiabatic curvature perturbation field $\delta^{\mathcal{R}}(\vec{x})$ is the superposition of the eigenmodes $\Upsilon^{\E{18}}_{\vec{k}}(\vec{x})$ with amplitudes $\delta^{\mathcal{R}}(\vec{k})$  described by (approximately) Gaussian random variables of zero mean and dimensionless power spectrum ${\mathcal{P}}^{\mathcal{R}}(k)$.
The resulting three-dimensional scalar field can be written as
\begin{equation}
    \label{eqn:scalarfieldinE18Cartesian}
    \delta^{\mathcal{R}}(\vec{x}) = \int \frac{\dderiv^3 k}{(2\pi)^3}\; \delta^{\mathcal{R}}(\vec{k}) \Upsilon^{\E{18}}_{\vec{k}}(\vec{x})\,.
\end{equation}
Observables are often tied to other scalar fields $\delta^X$ that are linearly related to $\delta^{\mathcal{R}}$ by a transfer function, which, as in \cite{COMPACT:2023rkp}, we write as $\Delta^{X}(\vec{k})$. 
The expectation value of any pair of $\delta^X(\vec{k})$ is
\begin{equation}
    \label{eqn:CE18XY}
    C^{\E{18};XY}_{\vec{k} \vec{k}'} 
        \equiv \langle \delta^X(\vec{k}) \delta^{Y*}(\vec{k}')\rangle 
        = (2\pi)^3\frac{2\pi^2}{k^3} {\mathcal{P}}^{\mathcal{R}}(k) \Delta^{X}(\vec{k}) \DeltaYstar(\vec{k}')
        \Ddelta(\vec{k}-\vec{k}')\; ,
\end{equation}
where $\Ddelta (\vec{k}-\vec{k}')$ is the three-dimensional Dirac delta function.
We will write the primordial power spectrum of the adiabatic curvature $\delta^{\mathcal{R}}$ as\footnote{
    The normalization by ${2\pi^2}/{k^3}$ in \eqref{eqn:CE18XY} is a common, but not universal, convention.
    ${\mathcal{P}}^X(k)$ is then the contribution to the variance per logarithmic interval of wavenumber: the total variance of $\delta^X$ is $\int \dderiv(\ln k)\; {\mathcal{P}}^X(k)$.
    In the large-scale-structure literature, the matter power spectrum is usually denoted by the quantity $P(k)=2\pi^2 {\mathcal{P}}(k)\Delta^{m}(k)\Delta^{m}(k)/k^3$, where $\Delta^{m}(k)$ is the matter transfer function.} 
\begin{equation}
    {\mathcal{P}}^{\mathcal{R}}(k) = A_\mathrm{s} \left(\frac{k}{k_*}\right)^{n_\mathrm{s}-1}\;,
\end{equation}
with the scalar amplitude $A_\mathrm{s}$ defined at the fiducial wavenumber $k_*$, and the scalar spectral tilt $n_\mathrm{s}$.
As in \cite{COMPACT:2023rkp}, we do not add a topology label to ${\mathcal{P}}^{\mathcal{R}}$.

It is often useful to expand the plane waves of \eqref{eqn:EuclideanFourierBasis} in terms of spherical Bessel functions $j_\ell$ and spherical harmonics $Y_{\ell m}$:
\begin{equation}
    \label{eqn:planewaveinsphericalcoords}
    \eexp^{i\vec{k}\cdot(\vec{x}-\vec{x}_0)} 
        = 4\pi \eexp^{-i\vec{k}\cdot\vec{x}_0}\sum_{\ell m} i^\ell j_\ell(kr) Y_{\ell m}^*(\unitvec{k}) Y_{\ell m}(\theta, \phi)\,.
\end{equation}
This reflects the practicalities of making cosmological observations from our single observational location.
Thus we can also write a Fourier mode as
\begin{equation}
    \Upsilon^{\E{i}}_{\vec{k}}(\vec{x}) =  4\pi \sum_{\ell m} j_\ell(kr) \xi^{\E{i};\unitvec{k}}_{k \ell m} Y_{\ell m}(\theta, \phi)\,,
\end{equation}
where, for \E{18}
\begin{equation}
    \xi^{\E{18};\unitvec{k}}_{k \ell m} = \eexp^{-i\vec{k}\cdot\vec{x}_0} i^\ell  Y_{\ell m}^*(\unitvec{k}).
\end{equation}
For observations that project $\delta^X$ onto the sphere of the sky, 
we often write
\begin{equation}
    \label{eqn:observable}
    \delta^X (\theta, \phi) = \sum_{\ell m} a^{\E{18};X}_{\ell m} Y_{\ell m} (\theta, \phi).
\end{equation}
Here $\delta^X$ has been integrated along the line of sight with an appropriate weighting function, the transfer function; in Fourier representation $\Delta^X(\vec{k})$.
In harmonic representation,
\begin{align}
    \label{eqn:almE18}
    a^{\E{18};X}_{\ell m} 
    &= \frac{4\pi}{(2\pi)^3}
    \int \dderiv^3k ~\delta^{\mathcal{R}}({\vec{k}})
    \xi^{\E{18};\unitvec{k}}_{k \ell m}
     \Delta^X_\ell(k).
\end{align}
$\Delta^X_\ell(k)$ is the spherical-harmonic transfer function from $\mathcal{R}$ to $X$,
and, relative to $\Delta^X(\vec{k})$, absorbs the $j_\ell(kr)$ that contributed to the integrand of the radial integral. 
When the background metric and any relevant microphysics is isotropic, then $\Delta^X_\ell(\vec{k})$ depends only on $\vert\vec{k}\vert$ and $\unitvec{k}\cdot\unitvec{n}$ (where $\unitvec{n}$ is the unit vector to the line-of-sight),
not on the full vector $\vec{k}$.
In this case it conventional to absorb into  $\Delta^X_\ell$ factors of $\unitvec{k}\cdot\unitvec{n}$. 
This allows us to choose to make $\Delta^X_\ell$ a function only of the magnitude of $\vec{k}$.
We will assume throughout this paper that this simplification is possible, and so write $\Delta^X_\ell(k)$.

The isotropy of \E{18} means that if $\delta^X(\vec{k})$ are independent Gaussian random variables of zero mean with variance a function only of $k$, then $a^X_{\ell m}$ are independent Gaussian random variables with variance  dependent only on $\ell$.
(However, for $X$ a real scalar quantity, for example CMB temperature fluctuations, $a_{\ell m}^{X *} = (-1)^m a^X_{\ell~ -m}$, and only  $a^X_{\ell m}$ with $m\geq 0$ encode unique physical information.)
Statistical isotropy is consequently 
\begin{equation}
    \label{eqn:SI-2d}
    C^{\E{18};XY}_{\ell m \ell' m'} \equiv 
    \langle a^{\E{18};X}_{\ell m}a^{\E{18};Y*}_{\ell' m'} \rangle = C^{\E{18};XY}_{\ell} \Kdelta_{\ell\ell'}\Kdelta_{mm'},
\end{equation}
where $\Kdelta_{ij}$ is the Kronecker delta.

As emphasized in \cite{COMPACT:2023rkp}, non-trivial topological boundary conditions have two important effects on the Laplacian eigenmodes:
\begin{enumerate}
    \item Only certain wavevectors $\vec{k}$ are allowed by the boundary conditions.
    For the fully compact non-orientable topologies \E{7}--\E{10}, the allowed wavevectors form a discrete lattice\footnote{
        This means that, for \E{7}--\E{10}, only certain values of $-k^2$ are eigenvalues of the Laplacian, and the multiplicity of the eigen-subspace of each eigenvalue is finite.}
    so we write $\delta^X_{\vec{k}}$, not $\delta^X(\vec{k})$.
    Thus $\langle\delta^X_{\vec{k}}\delta^{X*}_{\vec{k}'}\rangle$ contains terms involving ${\mathcal{P}}^X(k) \Kdelta_{\vec{k} \vec{k}'}$ (i.e., a Kronecker, rather than Dirac, delta, although there can be a mix of the two for  spaces with a mix of finite and infinite directions, here \E{13}--\E{15} and \E{17}).
    \item The eigenmodes are not the usual single covering-space eigenmodes (Fourier modes) but linear combinations thereof, with different $\vec{k}$ of the same magnitude. This induces extra terms in the correlator, coupling, with Kronecker or Dirac deltas, each $\vec{k}$ to its images under the rotations and reflections, $M_a^{\E{i}}\in O(3)$, that appear in that topology's group actions \eqref{eqn:actionofgenerator}.
\end{enumerate}
These violations of statistical isotropy mean that
 $C^{\E{i};XY}_{\vec{k} \vec{k}'}\equiv \langle \delta^X(\vec{k}) \delta^{Y*}(\vec{k}')\rangle $ is no longer proportional to a Dirac delta function of $\vec{k}$ and $\vec{k}'$.
 Instead it vanishes except for certain allowed $\vec{k}$ and generically connects all pairs of allowed $\vec{k}$ of equal magnitude with correlations of equal magnitude and location-dependent phase.
Consequently,
\begin{equation}
    \label{eqn:noSI-2d}
    C^{\E{i};XY}_{\ell m \ell' m'} \equiv \langle a^{\E{i};X}_{\ell m} a^{\E{i};Y*}_{\ell' m'}\rangle 
\end{equation}
is no longer diagonal.
Even taking into account the reality condition on the spherical-harmonic coefficients themselves,  the correlation components $C^{XY}_{\ell~ -m \ell'm'} \equiv \langle a^X_{\ell~ -m} a^{Y*}_ {\ell' m'} \rangle =  (-1)^m \langle a^{X*}_ {\ell m} a^{Y*}_{\ell' m'} \rangle$ does contain independent information.
In general the $C^{XY}_{\ell m\ell'm'}$ matrix is Hermitian in the $(X, \ell m),(Y, \ell'm')$ index sets.

In the subsections below, we present the eigenmodes and eigenspectra of the non-orientable Euclidean manifolds as functions of their manifold parameters in their full generality.
Assuming\footnote{
    In the standard, i.e., \E{18}, case, the justification for this assumption is  the inflationary origin of the $\delta^{\cal R}(\vec{k})$.
    In the case of non-trivial topology, inflation may not be responsible for the generation of the primordial fluctuations.
} 
that it is the amplitudes of these eigenmodes that are Gaussian random variables of zero mean and dimensionless power spectrum ${\mathcal{P}}^{\mathcal{R}}(k)$, we present the correlation matrices for Fourier-mode amplitudes $C^{\E{i};XY}_{\vec{k} \vec{k}'}$ and spherical-harmonic amplitudes $C^{\E{i};XY}_{\ell m \ell' m'}$.
The important results for each topology are boxed.
The generality of the results employs the orientation and other choices described in \cref{secn:topologiesmanifolds}, but also includes both an arbitrary origin for the definition of the manifold parameters and an arbitrary location for the observer.
As discussed in \cref{secn:topologiesmanifolds}, there are redundancies in these choices.
Any comprehensive search over parameters must take care to avoid overweighting some parts of parameter space.
In practice, it is often convenient to use the freedom discussed above of shifting the origin of the coordinate system  to move parameters between $\vec{x}_0$ and the components of the $\vec{T}^{\E{i}}_{aj}$.
Though not necessary, it allows one to 
\begin{enumerate}
    \item choose the plane(s) of reflection or axes of rotation to pass through the origin, $\vec{x}_0 = \vec{0}$, and use the ``tilted'' parameters of the manifold, or
    \item choose the simplified set of parameters, but the plane(s) of reflection or axes of rotation may not pass through the origin and some of the components of their location, $\vec{x}_0$, become significant.
\end{enumerate}

\subsection{General considerations for eigenmodes}

The eigenmodes of the scalar Laplacian $\Upsilon^{\E{i}}_{\vec{k}}(\vec{x})$ must be invariant under every possible  group transformation $G_\alpha \in \Gamma^{\E{i}}$
\begin{equation}
    \label{eqn:generaleigenmodeinvariance}
    \Upsilon^{\E{i}}_{\vec{k}}(G_\alpha\vec{x}) = \Upsilon^{\E{i}}_{\vec{k}}(\vec{x}) \,.
\end{equation}
As described in detail in \cite{COMPACT:2023rkp}, in the Euclidean case, this means that each eigenmode is a linear combination of a small number of Fourier modes $\Upsilon^{\E{18}}_{\vec{k}}(\vec{x})$
that are related to one another by group elements $G_\alpha \in \mathcal{G}^{\E{i}}$,
\begin{equation}
    \label{eqn:generaleigenmodeformula}
    \Upsilon^{\E{i}}_{\vec{k}}(\vec{x}) = \frac{\eexp^{i\Phi^{\E{i}}_{\vec{k}}}}{\sqrt{N(\mathcal{G}^{\E{i}})}}\sum_{G_\alpha \in \mathcal{G}^{\E{i}}} \eexp^{i\vec{k}\cdot G_\alpha\vec{x}} \,,
\end{equation}
where $N(\mathcal{G}^{\E{i}})$ is the number of elements in $\mathcal{G}^{\E{i}}$ and $\Phi^{\E{i}}_{\vec{k}}$ represents a phase that will be chosen to simplify the reality condition of the field, which varies with topology.\footnote{
    This random phase was not included for the scalar modes of the orientable topologies in \cite{COMPACT:2023rkp} but was introduced for the tensor modes in \cite{Samandar:2025kuf}.
}
The subset $\mathcal{G}^{\E{i}}$ of $\Gamma^{\E{i}}$ includes one group element for each of the $O(3)$ matrices $\mat{M}^{(G_\alpha)}$ that appears when we explicitly write the action of the group elements,
\begin{equation}
    \label{eqn:actionofGalpha}
    G_\alpha:(\vec{x}-\vec{x}_0)\to \mat{M}^{(G_\alpha)} (\vec{x}-\vec{x}_0) + \vec{v}^{(G_\alpha)}\,.
\end{equation}
These $\mat{M}^{(G_\alpha)}$ are then just the matrices $\mat{M}^{\E{i}}_a$ that appear in the generators, as described in \cref{secn:topologiesmanifolds-general}, plus all non-identical $O(3)$ matrices that can be built from arbitrary products of those $\mat{M}^{\E{i}}_a$.
Below, we will present explicit versions of \eqref{eqn:generaleigenmodeformula} for each \E{i}.

Equation \eqref{eqn:generaleigenmodeinvariance} must still be satisfied for every group element $G_\alpha \in \Gamma^{\E{i}}$.
In particular, $\Gamma^{\E{i}}$ has a subgroup of pure translations, which are all the integer linear combinations of the $\vec{T}^{\E{i}}_j$, i.e., the translations of the associated homogeneous space (\E{1}, \E{11}, or \slabh) of that manifold.
Thus (as explained in detail in \cite{COMPACT:2023rkp}),
the allowed wavevectors $\vec{k}$ must satisfy
\begin{equation}
    \label{eqn:associatedhomogeneousdiscretization}
    \vec{k}\cdot\vec{T}^{\E{i}}_j = 2\pi n_j\,, \quad \mbox{for } n_j\in\integers.
\end{equation}
In other words, the eigenmodes of the scalar Laplacian on an \E{i} manifold are linear combinations of the Fourier modes that are eigenmodes of the associated homogeneous space (\E{1}, \E{11}, or \slabh) of \E{i}.
For each \E{i} below, we present those discretization conditions.

Equation \eqref{eqn:generaleigenmodeformula} satisfies the invariance condition \eqref{eqn:generaleigenmodeinvariance} for all $\vec{k}$ allowed by 
\eqref{eqn:associatedhomogeneousdiscretization}, however in some cases the sum over $G_\alpha \in \mathcal{G}^{\E{i}}$ yields more than one identical term.
This occurs when $\mat{M}^{(G_\alpha)}\vec{k}=\vec{k}$ for certain $\vec{k}$ allowed by \eqref{eqn:associatedhomogeneousdiscretization}.
We consider those cases explicitly for each \E{i}.

\subsection{\E{1}: 3-torus}
\label{secn:eigenmodesE1}

Although we presented its properties in detail in \cite{COMPACT:2023rkp}, the 3-torus is the simplest of the compact Euclidean topologies and will serve as a model for determining the eigenspectrum and eigenmodes of the non-orientable Euclidean three-manifolds considered below.
Therefore, in this subsection we recall the eigenvalues and eigenmodes of the scalar Laplacian in \E{1}, and then the Fourier-space and spherical-harmonic-space correlation matrices of any fluctuations that are linearly related to independent Gaussian random fluctuations of the amplitudes of those eigenmodes.
 
The \E{1} eigenmodes are the subset of the \E{18} eigenmodes \eqref{eqn:EuclideanFourierBasis} that respect the \E{1} symmetries, 
\begin{equation}
    \label{eqn:E1eigenmodeinvariance}
    \Upsilon^{\E{1}}_{\vec{k}}(g^{\E{1}}_{\A{j}}\vec{x}) = \Upsilon^{\E{1}}_{\vec{k}}(\vec{x}).
\end{equation}
This follows since all the group elements of $\Gamma^{\E{1}}$ are pure translations, i.e.,
$\mat{M}^{(G_\alpha)}=\identity$ for all $G_\alpha\in \mathcal{G}^{\E{1}}$, 
so $\mat{M}^{(G_\alpha)}\vec{k}=\vec{k}$ trivially.

As discussed above (cf.\ \eqref{eqn:associatedhomogeneousdiscretization}), the symmetry condition \eqref{eqn:E1eigenmodeinvariance} leads to the discretization of the allowed $\vec{k}$ in \E{1}: 
\begin{align}
    2\pi n_1 &= (\vec{k}_{\vec{n}})_{x} L_{1x} , \nonumber \\
    2\pi n_2 &= (\vec{k}_{\vec{n}})_{x} L_{2x}+ (\vec{k}_{\vec{n}})_{y} L_{2y},  \\
    2\pi n_3 &= (\vec{k}_{\vec{n}})_{x} L_{3x}
        + (\vec{k}_{\vec{n}})_{y} L_{3y}
        + (\vec{k}_{\vec{n}})_{z} L_{3z} .  \nonumber
\end{align}
Since the wavenumbers are now discretized they are labeled by integers $n_i \in \integers$ and we denote this explicitly by writing the wavevector as $\vec{k}_{\vec{n}}$ for $\vec{n} = (n_1, n_2, n_3)$.
Here and below we will use either the $n_j$ or $(\vec{k}_{\vec{n}})_j$ labels as convenient for the situation.
Inverting these requirements, the components of the wavevectors are 
\begin{align}
    \label{eqn:E1_ki}
    (\vec{k}_{\vec{n}})_{x} &= \frac{2\pi n_1}{L_{1x}} , \nonumber \\
    (\vec{k}_{\vec{n}})_{y} &= \frac{2\pi n_2}{L_{2y}} - \frac{2\pi n_1}{L_1}\frac{L_{2x}}{L_{2y}},\\
    (\vec{k}_{\vec{n}})_{z} &= \frac{2\pi n_3}{L_{3z}}
        - 
        \frac{2\pi n_2}{L_{2y}}
          \frac{L_{3y}}{L_{3z}} 
        - 
        \frac{2\pi n_1}{L_{1x}}
            \frac{L_{2y}L_{3x}-L_{2x}L_{3y}}{L_{2y}L_{3z}}.
          \nonumber    
\end{align}
Clearly the eigenvalues $-k_{\vec{n}}^2 = -\vert\vec{k}_{\vec{n}}\vert^2$ are a quadratic form in the $n_j$.

Summarizing, this gives
\begin{empheq}[box=\fbox]{equation}
    \Upsilon^{\E{1}}_{\vec{k}_{\vec{n}}}(\vec{x}) = \eexp^{i\vec{k}_{\vec{n}}\cdot(\vec{x}-\vec{x}_0)}, \quad \mbox{for } \vec{n}\in \setN^{\E{1}},
\end{empheq}
where
\begin{empheq}[box=\fbox]{equation}
    \setN^{\E{1}} \equiv \{(n_1,n_2,n_3) \vert n_i\in\integers\}\setminus (0,0,0),
\end{empheq}
and
\begin{empheq}[box=\fbox]{equation}
    \xi^{\E{1};\unitvec{k}_{\vec{n}}}_{k_{\vec{n}} \ell m} \equiv \eexp^{-i\vec{k}_{\vec{n}}\cdot\vec{x}_0} i^\ell  Y_{\ell m}^*(\unitvec{k}_{\vec{n}}).
\end{empheq}
Throughout this paper, we exclude the $(0,0,0)$ mode, as it contributes only a constant to the monopole term and is irrelevant for cosmological perturbations, the focus of this paper.
Following \eqref{eqn:CE18XY} the Fourier-mode correlation matrix for \E{1} is
\begin{empheq}[box=\fbox]{align}
    \label{eqn:CE1XY}
    C^{\E{1};XY}_{\vec{k}_{\vec{n}} \vec{k}_{\vecnp}} 
    &= V_{\E{1}}\frac{2\pi^2}{k_{\vec{n}}^3}{\mathcal{P}}^{\mathcal{R}}(k_{\vec{n}}) 
    \Delta^X(\vec{k}_{\vec{n}}) \DeltaYstar(\vec{k}_{\vec{n}'}) 
    \Kdelta_{\vec{k}_{\vec{n}}\vec{k}_{\vecnp}}\;.
\end{empheq}
In transitioning from the covering space \E{18} we have replaced $(2\pi)^3\Ddelta(\vec{k}-\vec{k}')$ with $V_{\E{1}}\Kdelta_{\vec{k}_{\vec{n}}\vec{k}_{\vecnp}}$, where the volume factor 
\begin{equation}
    \label{eqn:E1volume}
    V_{\E{1}} 
        = \vert(\vec{T}^{\E{1}}_{{1}}\times\vec{T}^{\E{1}}_{{2}})\cdot\vec{T}^{\E{1}}_{{3}}\vert = \vert L_{{1}x} L_{{2}y} L_{{3}z}\vert. 
\end{equation}

As for \E{18} above, we can project the field $\delta^X$ onto the sky by performing a radial integral with suitable weight function and transfer function, giving
\begin{align}
    \label{eqn:almE1}
    a^{\E{1};X}_{\ell m} 
    &=  \frac{4\pi}{V_{\E{1}}}
    \sum_{\vec{n}\in\setN^{\E{1}} } \delta^{\mathcal{R}}_{\vec{k}_{\vec{n}}} 
     \xi^{\E{1};\unitvec{k}_{\vec{n}}}_{k_{\vec{n}} \ell m}
     \Delta^X_\ell(k_{\vec{n}}).
\end{align}
Because $\setN^{\E{1}}$ labels only a discrete set of $\vec{k}_{\vec{n}}$, the integral over $\dderiv^3 k$ in \cref{eqn:almE18} is replaced by a sum over $\vec{n}\in\setN^{\E{1}}$.

For the compact topologies \E{i} with $i \in \{1, \ldots, 10\}$, the harmonic space covariance matrix has the general form\footnote{
    Note that, while $\Delta^Y(\vec{k})$ is complex, $\Delta^Y_{\ell}$ is real for the usual cases of CMB temperature and polarization; nevertheless we retain the complex conjugate for generic $Y$.}
\begin{empheq}[box=\fbox]{equation}
    \label{eqn:HarmonicCovariance}
    C^{\E{i};XY}_{\ell m\ell'm'}  =
     \frac{(4\pi)^2}{{V_{\E{i}}}}
    \sum_{\vec{n}\in \setN^{\E{i}}} 
    \Delta^X_{\ell}(k_{\vec{n}})
    \DeltaYstarforell_{\ell'}(k_{\vec{n}})
    \frac{2\pi^2 \mathcal{P}^{\mathcal{R}}(k_{\vec{n}})}{k_{\vec{n}}^3}
    \xi^{\E{i};\unitvec{k}_{\vec{n}}}_{k_{\vec{n}} \ell m}
    \xi^{\E{i};\unitvec{k}_{\vec{n}}*}_{k_{\vec{n}} \ell' m'} .
\end{empheq}

\subsection{\E{7}: Klein space}
\label{secn:eigenmodesE7}

The eigenspectrum and eigenmodes of the Klein space can be determined in a manner analogous to that of the orientable manifolds \cite{COMPACT:2023rkp}.
We could begin from the covering space, but it is more expedient to recognize that \E{7} is \E{1} with extra symmetries imposed.
With this, the eigenspectrum of \E{7} will be discretized with wavevectors $\vec{k}_{\vec{n}}$ and the eigenfunctions $\Upsilon^{\E{7}}_{\vec{k}_{\vec{n}}}(\vec{x})$ will be linear combinations of $\Upsilon^{\E{1}}_{\vec{k_{\vec{n}}}}(\vec{x})$.
For \E{7} the discretization condition \eqref{eqn:associatedhomogeneousdiscretization} from the translation vectors $\vec{T}^{\E{7}}_i$ leads to the components of the allowed wavevectors,
\begin{align}
    \label{eqn:E7_kni}
    (\vec{k}_{\vec{n}})_{x} &= \frac{2\pi n_1}{2 L_{Ax}}, \nonumber \\
    (\vec{k}_{\vec{n}})_{y} &= \frac{2\pi n_2}{L_{1y}}, \\
    (\vec{k}_{\vec{n}})_{z} &= \frac{2\pi}{L_{2z}} \left( n_3 - \frac{L_{2x}}{2L_{Ax}} n_1 \right). \nonumber
\end{align}

Unlike in \E{1}, the eigenmodes of \E{7} can include a linear combination of two \E{1} eigenmodes.
This follows in the application of \cref{eqn:generaleigenmodeformula} since the condition $(\mat{M}^{\E{7}}_A)^N\vec{k}_{\vec{n}} = \vec{k}_{\vec{n}}$ has more than one solution for the minimum positive $N$, depending on $\vec{k}_{\vec{n}}$,
namely, $N=1$ for $(\vec{k}_{\vec{n}})_{y} = 0$ and $N=2$ otherwise.
Written explicitly,
\begin{description}
    \item[\textbf{$N=1$ eigenmodes: }]  $\vec{k}_{\vec{n}}=\transpose{((\vec{k}_{\vec{n}})_x, 0, (\vec{k}_{\vec{n}})_z)}$, 
        i.e., $\vec{n}=(n_1,0,n_3)$, 
        $n_1 \in 2\integers$, $n_3 \in \integers$ with at least one of $n_1 \neq 0$ or $n_3 \neq 0$ and
    \begin{empheq}[box=\fbox]{equation}
        \Upsilon^{\E{7}}_{(n_1,0,n_3)}(\vec{x}) = \eexp^{i\vec{k}_{\vec{n}} \cdot (\vec{x}-\vec{x}_0)} = \eexp^{i(\vec{k}_{\vec{n}})_{x} (x-x_0) + i(\vec{k}_{\vec{n}})_{z} (z-z_0)}, 
    \end{empheq}
    \item[\textbf{$N=2$ eigenmodes: }] $(\vec{k}_{\vec{n}})_{y} \neq 0$, 
        i.e., $n_2 \neq 0$, and per \cref{eqn:generaleigenmodeformula},
    \begin{empheq}[box=\fbox]{equation}
        \Upsilon^{\E{7}}_{\vec{n}}(\vec{x}) = \frac{1}{\sqrt{2}} \eexp^{-i(\vec{k}_{\vec{n}}\cdot \vec{T}^{\E{7}}_A/2 + |\vec{k}_{\vec{n}}\cdot \vec{T}^{\E{7}}_1|/4)} \left[ \eexp^{i\vec{k}_{\vec{n}}\cdot(\vec{x}-\vec{x}_0)} + \eexp^{i\vec{k}_{\vec{n}}\cdot (\mat{M}^{\E{7}}_A(\vec{x}-\vec{x}_0)+\vec{T}^{\E{7}}_A)} \right].
    \end{empheq}
\end{description}
For the $N=2$ modes, we have chosen the phase $\Phi^{\E{7}}_{\vec{k}} = -\vec{k}_{\vec{n}}\cdot \vec{T}^{\E{7}}_A/2 - |\vec{k}_{\vec{n}}\cdot \vec{T}^{\E{7}}_1|/4$, defined in \eqref{eqn:generaleigenmodeformula}  to simplify the reality condition of the field.
 For this  choice, the relationship between the coefficients $c_{n_1, n_2, n_3}$ of eigenmodes $\Upsilon^{\E{7}}_{{n_1, n_2, n_3}}(\vec{x})$ that guarantees the reality of the resulting field is just $c_{-n_1,n_2,-n_3} = c_{n_1, n_2, n_3}^*$, without any extra factors.
 
 In the following topologies, phases will be chosen to bring about similar simplifications.
 
The linear combination in the $N=2$ modes requires some care.
Notice that $\transpose{\vec{k}}_{\vec{n}}\mat{M}^{\E{7}}_A=((\vec{k}_{\vec{n}})_{x},-(\vec{k}_{\vec{n}})_{y},(\vec{k}_{\vec{n}})_{z})$, i.e., $\mat{M}^{\E{7}}_A$ maps $(n_1,n_2,n_3)\to(n_1,-n_2,n_3)$.
One implication of this is that summing over $(n_1,n_2,n_3)$ would double-count eigenmodes if all $n_2\in\integers$ were included.
Hence, we define two sets of allowed modes, one for $N=1$ and another for $N=2$,
\begin{empheq}[box=\fbox]{align}
    \setN^{\E{7}}_1 &= \{(n_1,0,n_3)|n_1\in 2\integers, n_3\in \integers \} \setminus (0, 0, 0), \nonumber\\
    \setN^{\E{7}}_2 &= \{(n_1,n_2,n_3)| n_1 \in \integers, n_2 \in \integers^{>0}, n_3 \in \integers\} , \\
    \setN^{\E{7}} &= \setN^{\E{7}}_1 \cup \setN^{\E{7}}_2 . \nonumber
\end{empheq}
With these the Fourier-mode correlation matrix can now be expressed as
\begin{empheq}[box=\fbox]{align}
\label{eqn:E7FourierCovarianceStandardConvention}
    C^{\E{7};XY}_{\vec{k}_{\vec{n}}\vec{k}_{\vecnp}} 
     = {} & V_{\E{7}} \frac{2\pi^2}{k^3_{\vec{n}}}
        {\mathcal{P}}^{\mathcal{R}}(k_{\vec n})\Delta^X(\vec{k}_{\vec{n}})\DeltaYstar(\vec{k}_{\vec{n}'})
        \eexp^{i(\vec{k}_{\vecnp}-\vec{k}_{\vec{n}})\cdot\vec{x}_0}
        \left[
            \sum_{\vec{\tilde{n}} \in \setN^{\E{7}}_1}
                 \Kdelta_{\vec{k}_{\vec{n}}\vec{k}_{\vec{\tilde{n}}}}
                 \Kdelta_{\vec{k}_{\vecnp}\vec{k}_{\vec{\tilde{n}}}} + {}
        \right.
            \\
            & \left. \quad
                {} + \frac{1}{2}
                \sum_{\vec{\tilde{n}}\in \setN^{\E{7}}_2}
                \sum_{a=0}^1\sum_{b=0}^1 
                \eexp^{i\vec{k}_{\vec{\tilde{n}}}\cdot(\vec{T}^{(a)}-\vec{T}^{(b)})}
                 \Kdelta_{\vec{k}_{\vec{n}}([(\mat{M}^{\E{7}}_A){}^T]^a\vec{k}_{\vec{\tilde{n}}})}
                 \Kdelta_{\vec{k}_{\vecnp}([(\mat{M}^{\E{7}}_A){}^T]^b\vec{k}_{\vec{\tilde{n}}})} 
         \right], 
         \nonumber
\end{empheq}
where $V_{\E{7}}$ is given in \eqref{eqn:VE7}, $\vec{T}^{(0)}\equiv\vec{0}$, and $\vec{T}^{(1)}\equiv\vec{T}^{\E{7}}_A$.\footnote{
    In  \cref{eqn:E7FourierCovarianceStandardConvention}, $\vec{k}_{\vec{n}}$ and $\vec{k}_{\vecnp}$ are wavevectors of the associated \E{1}, as specified \eqref{eqn:E7_kni}.
    $C^{\E{7};XY}_{\vec{k}_{\vec{n}}\vec{k}_{\vecnp}}$ describes correlations between amplitudes of the plane waves that comprise the eigenmodes of a specific manifold -- i.e., of a specific topology, with specific values of its parameters.
    It is this object that would be used, for example, in creating realizations of initial conditions for large-scale structure simulations.
    If one was, instead, constructing a likelihood function to compare data with expectations from \E{7} manifolds, one would need to convolve $C^{\E{7};XY}_{\vec{k}_{\vec{n}}\vec{k}_{\vecnp}}$ with a kernel characterizing the Fourier structure of the survey of interest.
    }
Note that $C^{\E{7};XY}_{\vec{k}_{\vec{n}}\vec{k}_{\vecnp}}=0$ for $\vert k_{\vecnp}\vert\neq\vert k_{\vec{n}}\vert$.

Another implication of the reflection in $\mat{M}^{\E{7}}_A$ comes when representing the eigenmodes in the harmonic basis.
Here we will combine modes with the same eigenvalue $-k_{\vec{n}}^2$ and with orientations $\unitvec{k}_{\vec{n}}$ and $(\mat{M}^{\E{7}}_A){}^T \unitvec{k}_{\vec{n}}$.
Since the flip is a reflection across the $xz$-plane we can use the properties of the spherical harmonics to explicitly write our expressions.
In particular
\begin{equation}
    Y^*_{\ell m} ((\mat{M}^{\E{7}}_A){}^T \unitvec{k}_{\vec{n}}) = Y^*_{\ell m} (\theta, -\phi) = Y_{\ell m} (\unitvec{k}_{\vec{n}}).
\end{equation}
This gives for the eigenmodes in the harmonic basis
\begin{empheq}[box=\fbox]{align}
    \xi^{\E{7};\unitvec{k}_{\vec{n}}}_{k_{\vec{n}}\ell m} 
    &= i^\ell  Y^*_{\ell m} (\unitvec{k}_{\vec{n}}) \eexp^{-i \vec{k}_{\vec{n}}\cdot\vec{x}_0}, \quad \mbox{for } \vec{n} \in \setN^{\E{7}}_1 , \nonumber\\
    \xi^{\E{7};\unitvec{k}_{\vec{n}}}_{k_{\vec{n}}\ell m} 
     &= \frac{1}{\sqrt{2}} i^\ell \eexp^{-i(\vec{k}_{\vec{n}}\cdot \vec{T}^{\E{7}}_A/2 + |\vec{k}_{\vec{n}}\cdot \vec{T}^{\E{7}}_1|/4)} \left[ Y^*_{\ell m}(\unitvec{k}_{\vec{n}}) \eexp^{-i\vec{k}_{\vec{n}}\cdot \vec{x}_0}
        + {} \right. \\
     & \qquad\qquad \left. {} + Y_{\ell m} (\unitvec{k}_{\vec{n}}) \eexp^{-i\vec{k}_{\vec{n}}\cdot (\mat{M}^{\E{7}}_A \vec{x}_0 - \vec{T}^{\E{7}}_A)} \right], \quad \mbox{for } \vec{n}\in \setN^{\E{7}}_2 \nonumber
\end{empheq}
and the harmonic space covariance matrix has the form \eqref{eqn:HarmonicCovariance}.

\subsection{\E{8}: Klein space with horizontal flip}
\label{secn:eigenmodesE8}

The eigenspectrum and eigenmodes of the Klein space with horizontal flip can be determined in a manner analogous to that for \E{7} though the derivation more closely follows that of \E{6}, the Hantzsche-Wendt space, (c.f., section 4.7 of \cite{COMPACT:2023rkp}).
For \E{8} the discretization condition \eqref{eqn:associatedhomogeneousdiscretization} from the translation vectors $\vec{T}^{\E{8}}_i$ leads to the components of the allowed wavevectors,
\begin{equation}
    (\vec{k}_{\vec{n}})_{x} = \frac{2\pi n_1}{2 L_{Ax} }, \quad
    (\vec{k}_{\vec{n}})_{y} = \frac{2\pi n_2}{L_{Cy}},  \quad
     (\vec{k}_{\vec{n}})_{z} = \frac{2\pi n_3}{2 L_{Bz}}.
\end{equation}

As in \E{7}, the eigenmodes of \E{8} can include linear combinations of \E{1} eigenmodes.
Here we have that $(\mat{M}^{\E{8}}_A)^{N_A} \vec{k}_{\vec{n}} = \vec{k}_{\vec{n}}$, $(\mat{M}^{\E{8}}_B)^{N_B} \vec{k}_{\vec{n}} = \vec{k}_{\vec{n}}$, and $(\mat{M}^{\E{8}}_A \mat{M}^{\E{8}}_B)^{N_{AB}} \vec{k}_{\vec{n}} = \vec{k}_{\vec{n}}$ can have more than one solution for the minimum positive integers $N_a$, for $a \in \{A, B, AB\}$, depending on $\vec{k}_{\vec{n}}$.
Working through the cases this leads to
\begin{description}
    \item[\textbf{$N_A=1$ eigenmodes: }]  $\vec{k}_{\vec{n}}=\transpose{((\vec{k}_{\vec{n}})_x,0,(\vec{k}_{\vec{n}})_z)}$, 
        i.e., $\vec{n}=(n_1,0,n_3)$, 
        $n_1 \in 2\integers^{\neq 0}$, $n_3 \in \integers$, with
    \begin{empheq}[box=\fbox]{align}
        \Upsilon^{\E{8}}_{(n_1,0,n_3)}(\vec{x}) &= \frac{1}{\sqrt{2}} \eexp^{-i(\vec{k}_{\vec{n}} \cdot \vec{T}^{\E{8}}_B/2 + |\vec{k}_{\vec{n}} \cdot \vec{T}^{\E{8}}_3|/4)}
        \left[ \eexp^{i\vec{k}_{\vec{n}} \cdot (\vec{x}-\vec{x}_0)} 
        + {} \right. \\
        & \left. \qquad\qquad {} + \eexp^{i\vec{k}_{\vec{n}} \cdot (\mat{M}^{\E{8}}_B (\vec{x}-\vec{x}_0) + \vec{T}^{\E{8}}_B)} \right], \nonumber
    \end{empheq}
    \item[\textbf{$N_B=1$ eigenmodes: }]  $\vec{k}_{\vec{n}}=\transpose{(0,(\vec{k}_{\vec{n}})_y,(\vec{k}_{\vec{n}})_z)}$, 
        i.e., $\vec{n}=(0,n_2,n_3)$, 
        $n_2 \in \integers^{\neq 0}$, $n_3 \in 2\integers$, with
    \begin{empheq}[box=\fbox]{equation}
        \Upsilon^{\E{8}}_{(0,n_2,n_3)}(\vec{x}) = \frac{1}{\sqrt{2}} \eexp^{-i\vec{k}_{\vec{n}} \cdot \vec{T}^{\E{8}}_A/2}
        \left[ \eexp^{i\vec{k}_{\vec{n}} \cdot (\vec{x}-\vec{x}_0)} + \eexp^{i\vec{k}_{\vec{n}} \cdot (\mat{M}^{\E{8}}_A (\vec{x}-\vec{x}_0) + \vec{T}^{\E{8}}_A)} \right], 
    \end{empheq}
    \item[\textbf{$N_{AB}=1$ eigenmodes: }]  $\vec{k}_{\vec{n}}=\transpose{(0,0,(\vec{k}_{\vec{n}})_z)}$, 
        i.e., $\vec{n}=(0,0,n_3)$, 
        $n_3 \in 2\integers^{\neq 0}$, with
    \begin{empheq}[box=\fbox]{equation}
        \Upsilon^{\E{8}}_{(0,0,n_3)}(\vec{x}) = \eexp^{i\vec{k}_{\vec{n}} \cdot (\vec{x}-\vec{x}_0)}, 
    \end{empheq}
    \item[\textbf{$N=2$ eigenmodes: }] $((\vec{k}_{\vec{n}})_{x}, (\vec{k}_{\vec{n}})_{y}) \neq (0,0)$, 
        i.e., $(n_1,n_2) \neq (0,0)$, $n_3 \in \integers$, with
    \begin{empheq}[box=\fbox]{align}
        \Upsilon^{\E{8}}_{\vec{n}}(\vec{x}) &= \frac{1}{\sqrt{4}} \eexp^{-i\vec{k}_{\vec{n}} \cdot [(\mat{M}^{\E{8}}_A \vec{T}^{\E{8}}_B +\vec{T}^{\E{8}}_A)/2 + |\vec{k}_{\vec{n}} \cdot \vec{T}^{\E{8}}_3|/4]} \left[
            \eexp^{i\vec{k}_{\vec{n}}\cdot (\vec{x}-\vec{x}_0)} 
            + {} \right. \nonumber \\
            & \qquad\quad {} 
            + \eexp^{i\vec{k}_{\vec{n}}\cdot (\mat{M}^{\E{8}}_A(\vec{x}-\vec{x}_0)+\vec{T}^{\E{8}}_A)}
            + \eexp^{i\vec{k}_{\vec{n}}\cdot (\mat{M}^{\E{8}}_B(\vec{x}-\vec{x}_0)+\vec{T}^{\E{8}}_B)}
            + {} \\
            & \left. \qquad\quad {} 
            + \eexp^{i\vec{k}_{\vec{n}}\cdot (\mat{M}^{\E{8}}_A \mat{M}^{\E{8}}_B(\vec{x}-\vec{x}_0)+ \mat{M}^{\E{8}}_A \vec{T}^{\E{8}}_B +\vec{T}^{\E{8}}_A)}
            \right] \nonumber .
    \end{empheq}
\end{description}
For the $N_A=1$, $N_B=1$, and $N=2$ modes, we have chosen the separate phases $\Phi^{\E{8}}_{\vec{k}}$ in \eqref{eqn:generaleigenmodeformula} to simplify the reality condition of the fields.
The eigenmodes for $N=2$ are written using $g^{\E{8}}_A g^{\E{8}}_B$ for the last term.
If $g^{\E{8}}_B g^{\E{8}}_A$ were used instead this term would not change since 
\begin{equation}
    \exp\left[ i \vec{k}_{\vec{n}}\cdot g^{\E{8}}_A g^{\E{8}}_B(\vec{x}) \right]
    = \exp\left[ i \vec{k}_{\vec{n}}\cdot g^{\E{8}}_B g^{\E{8}}_A(\vec{x}) \right] .
\end{equation}
This follows from the facts that $M^{\E{8}}_A M^{\E{8}}_B = M^{\E{8}}_B M^{\E{8}}_A$, $M^{\E{8}}_A \vec{T}^{\E{8}}_B + \vec{T}^{\E{8}}_A = M^{\E{8}}_B \vec{T}^{\E{8}}_A + \vec{T}^{\E{8}}_B + \vec{T}^{\E{8}}_1$, and $\exp(i \vec{k}_{\vec{n}}\cdot \vec{T}^{\E{8}}_1) = 1$.

As in \E{7} the cyclic properties of $\mat{M}^{\E{8}}_a$ would lead to repeated counting of eigenmodes if all $n_1\in\integers$ and $n_2\in\integers$ were included.
In this case, under the action of $\mat{M}^{\E{8}}_A$ we have the mapping $n_2\to -n_2$, under the action of $\mat{M}^{\E{8}}_B$ we have the mapping $n_1\to -n_1$, and under the action of $\mat{M}^{\E{8}}_A \mat{M}^{\E{8}}_B$ we have the mapping $(n_1, n_2)\to (-n_1, -n_2)$.
To avoid this, we define sets of allowed modes as
\begin{empheq}[box=\fbox]{align}
    \setN^{\E{8}}_{1A} &= \{(n_1,0,n_3)|n_1\in 2\integers^{>0}, n_3\in \integers \} , \nonumber \\
    \setN^{\E{8}}_{1B} &= \{(0,n_2,n_3)|n_2\in \integers^{>0}, n_3\in 2\integers \} , \nonumber \\
    \setN^{\E{8}}_{1AB} &= \{(0,0,n_3)|n_3\in 2\integers^{\neq 0} \} , \\
    \setN^{\E{8}}_2 &= \{(n_1,n_2,n_3)| n_1 \in \integers^{> 0}, n_2 \in \integers^{>0}, n_3 \in \integers\} , \nonumber \\
    \setN^{\E{8}} &= \setN^{\E{8}}_{1A} \cup \setN^{\E{8}}_{1B} \cup \setN^{\E{8}}_{1AB} \cup \setN^{\E{8}}_2 . \nonumber
\end{empheq}
With these the Fourier-mode correlation matrix can now be expressed as
\begin{empheq}[box=\fbox]{align}
\label{eqn:E8FourierCovarianceStandardConvention}
    C^{\E{8};XY}_{\vec{k}_{\vec{n}}\vec{k}_{\vecnp}} 
    = {} & V_{\E{8}} \frac{2\pi^2}{k^3_{\vec{n}}}
        {\mathcal{P}}^{\mathcal{R}}(k_{\vec n})\Delta^X(\vec{k}_{\vec{n}})\DeltaYstar(\vec{k}_{\vec{n}'})
        \eexp^{i(\vec{k}_{\vecnp}-\vec{k}_{\vec{n}})\cdot\vec{x}_0} \times {} \\
        & {} \quad \times \left[
            \frac{1}{2} \sum_{\vec{\tilde{n}} \in \setN^{\E{8}}_{1A}}
                \smashoperator[r]{\sum_{a,b\in\{0, B\}}} \eexp^{i\vec{k}_{\vec{n}} \cdot (\vec{T}^{(a)} - \vec{T}^{(b)})}
                 \Kdelta_{\vec{k}_{\vec{n}}((\mat{M}^{\E{8}}_a){}^T \vec{k}_{\tilde{\vec{n}}})}
                 \Kdelta_{\vec{k}_{\vecnp}((\mat{M}^{\E{8}}_b){}^T \vec{k}_{\tilde{\vec{n}}})} + {}
            \right. \nonumber \\
        & {} \qquad + \frac{1}{2} \sum_{\vec{\tilde{n}} \in \setN^{\E{8}}_{1B}}
                \smashoperator[r]{\sum_{a,b\in\{0, A\}}} \eexp^{i\vec{k}_{\vec{n}} \cdot (\vec{T}^{(a)} - \vec{T}^{(b)})}
                 \Kdelta_{\vec{k}_{\vec{n}}((\mat{M}^{\E{8}}_a){}^T \vec{k}_{\tilde{\vec{n}}})}
                 \Kdelta_{\vec{k}_{\vecnp}((\mat{M}^{\E{8}}_b){}^T \vec{k}_{\tilde{\vec{n}}})} + {} \nonumber \\
        & {} \qquad + \frac{1}{2} \sum_{\vec{\tilde{n}} \in \setN^{\E{8}}_{1AB}}
                \smashoperator[r]{\sum_{a,b\in\{0, AB\}}} \eexp^{i\vec{k}_{\vec{n}} \cdot (\vec{T}^{(a)} - \vec{T}^{(b)})}
                 \Kdelta_{\vec{k}_{\vec{n}}((\mat{M}^{\E{8}}_a){}^T \vec{k}_{\tilde{\vec{n}}})}
                 \Kdelta_{\vec{k}_{\vecnp}((\mat{M}^{\E{8}}_b){}^T \vec{k}_{\tilde{\vec{n}}})} + {} \nonumber \\
        & \left. \qquad
                {} + \frac{1}{4}
                \sum_{\vec{\tilde{n}}\in \setN^{\E{8}}_2}
                \smashoperator[r]{\sum_{a,b\in\{0, A, B, AB\}}} \eexp^{i\vec{k}_{\vec{n}} \cdot (\vec{T}^{(a)} - \vec{T}^{(b)})}
                 \Kdelta_{\vec{k}_{\vec{n}}((\mat{M}^{\E{8}}_a){}^T \vec{k}_{\tilde{\vec{n}}})}
                 \Kdelta_{\vec{k}_{\vecnp}((\mat{M}^{\E{8}}_b){}^T \vec{k}_{\tilde{\vec{n}}})} 
         \right] ,
         \nonumber
\end{empheq}
where $V_{\E{8}}$ is given in \eqref{eqn:VE8}, $\vec{T}^{(0)} \equiv \vec{0}$, $\vec{T}^{(A)} \equiv \vec{T}^{\E{8}}_A$, $\vec{T}^{(B)} \equiv \vec{T}^{\E{8}}_B$, $\vec{T}^{(AB)} \equiv \mat{M}^{\E{8}}_A \vec{T}^{\E{8}}_B + \vec{T}^{\E{8}}_A$, and $\mat{M}^{\E{8}}_{AB} \equiv \mat{M}^{\E{8}}_A \mat{M}^{\E{8}}_B$.

Also as in \E{7}, we can use the transformation properties of the spherical harmonics along with the fact that the relevant transformations are reflections or rotations to note that
\begin{align}
    Y^*_{\ell m}((\mat{M}^{\E{8}}_A){}^T\unitvec{k}_{\vec{n}}) &= Y_{\ell m} (\unitvec{k}_{\vec{n}}), \nonumber \\
    Y^*_{\ell m}((\mat{M}^{\E{8}}_B){}^T\unitvec{k}_{\vec{n}}) &= Y^*_{\ell m}(\theta, \pi-\phi) = (-1)^m Y_{\ell m} (\unitvec{k}_{\vec{n}}), \\
    Y^*_{\ell m}((\mat{M}^{\E{8}}_A \mat{M}^{\E{8}}_B){}^T\unitvec{k}_{\vec{n}}) &= Y^*_{\ell m}(\theta, \pi+\phi) = (-1)^m Y^*_{\ell m} (\unitvec{k}_{\vec{n}}) \nonumber .
\end{align}
This gives for the eigenmodes in the harmonic basis
\begin{empheq}[box=\fbox]{align}
    \xi^{\E{8};\unitvec{k}_{\vec{n}}}_{k_{\vec{n}}\ell m} 
    &= \frac{1}{\sqrt{2}} i^\ell \eexp^{-i(\vec{k}_{\vec{n}} \cdot \vec{T}^{\E{8}}_B/2 + |\vec{k}_{\vec{n}} \cdot \vec{T}^{\E{8}}_3|/4)} \left[ Y^*_{\ell m}(\unitvec{k}_{\vec{n}}) \eexp^{-i \vec{k}_{\vec{n}}\cdot\vec{x}_0} 
    + {} \right. \nonumber \\
    & \left. \qquad\qquad\qquad {} + (-1)^m Y_{\ell m}(\unitvec{k}_{\vec{n}}) \eexp^{-i \vec{k}_{\vec{n}} \cdot (\mat{M}^{\E{8}}_B \vec{x}_0 - \vec{T}^{\E{8}}_B)} \right] , \; \mbox{for } \vec{n} \in \setN^{\E{8}}_{1A} , \nonumber \\
    \xi^{\E{8};\unitvec{k}_{\vec{n}}}_{k_{\vec{n}}\ell m} 
    &= \frac{1}{\sqrt{2}} i^\ell \eexp^{-i\vec{k}_{\vec{n}} \cdot \vec{T}^{\E{8}}_A/2} \left[ Y^*_{\ell m}(\unitvec{k}_{\vec{n}}) \eexp^{-i \vec{k}_{\vec{n}}\cdot\vec{x}_0} 
    + {} \right. \nonumber \\
    & \left. \qquad\qquad\qquad {} + Y_{\ell m}(\unitvec{k}_{\vec{n}}) \eexp^{-i \vec{k}_{\vec{n}} \cdot (\mat{M}^{\E{8}}_A \vec{x}_0 - \vec{T}^{\E{8}}_A)} \right] , \; \mbox{for } \vec{n} \in \setN^{\E{8}}_{1B} , \nonumber \\
    \xi^{\E{8};\unitvec{k}_{\vec{n}}}_{k_{\vec{n}}\ell m} 
    &= i^\ell Y^*_{\ell m}(\unitvec{k}_{\vec{n}}) \eexp^{-i \vec{k}_{\vec{n}}\cdot\vec{x}_0} , \quad \mbox{for } \vec{n} \in \setN^{\E{8}}_{1AB} , \\
    \xi^{\E{8};\unitvec{k}_{\vec{n}}}_{k_{\vec{n}}\ell m} 
     &= \frac{1}{\sqrt{4}} i^\ell \eexp^{-i\vec{k}_{\vec{n}} \cdot [(\mat{M}^{\E{8}}_A \vec{T}^{\E{8}}_B +\vec{T}^{\E{8}}_A)/2 + |\vec{k}_{\vec{n}} \cdot \vec{T}^{\E{8}}_3|/4]} 
     \left[ Y^*_{\ell m}(\unitvec{k}_{\vec{n}}) \left( \eexp^{-i \vec{k}_{\vec{n}}\cdot\vec{x}_0} 
     + {} \right. \right. \nonumber \\
     & \left. \qquad\qquad\qquad\qquad {} + (-1)^m \eexp^{-i \vec{k}_{\vec{n}} \cdot (\mat{M}^{\E{8}}_A \mat{M}^{\E{8}}_B \vec{x_0} - \mat{M}^{\E{8}}_A \vec{T}^{\E{8}}_B - \vec{T}^{\E{8}}_A)} \right)  
     + {} \nonumber \\
     & \qquad\qquad {} + Y_{\ell m}(\unitvec{k}_{\vec{n}}) \left(
         \eexp^{-i \vec{k}_{\vec{n}} \cdot (\mat{M}^{\E{8}}_A \vec{x_0} - \vec{T}^{\E{8}}_A)}
         + {} \right. \nonumber \\
     & \left.\left. \qquad\qquad\qquad\qquad {} + (-1)^m \eexp^{-i \vec{k}_{\vec{n}} \cdot (\mat{M}^{\E{8}}_B \vec{x_0} - \vec{T}^{\E{8}}_B)} \right) \right], \; \mbox{for } \vec{n}\in \setN^{\E{8}}_2 \nonumber
\end{empheq}
and the harmonic space covariance matrix has the form \eqref{eqn:HarmonicCovariance}.

\subsection{\E{9}: Klein space with vertical flip}
\label{secn:eigenmodesE9}

The eigenspectrum and eigenmodes of \E{9} follow directly from those of $\E{7}$ in \cref{secn:eigenmodesE7}.
The only difference is that $(\vec{T}^{\E{9}}_{\B{2}})_y \neq 0$.
This changes $(\vec{k}_{\vec{n}})_{z}$.
For \E{9} the discretization condition \eqref{eqn:associatedhomogeneousdiscretization} from the translation vectors $\vec{T}^{\E{9}}_i$ leads to the components of the allowed wavevectors,
\begin{equation}
    \label{eqn:E9_kni}
    (\vec{k}_{\vec{n}})_{x} = \frac{2\pi n_1}{2 L_{Ax}}, \quad
    (\vec{k}_{\vec{n}})_{y} = \frac{2\pi n_2}{L_{1y}}, \quad
    (\vec{k}_{\vec{n}})_{z} = \frac{2\pi}{L_{2z}} \left( n_3 - \frac{L_{2x}}{2L_{Ax}} n_1 -\frac{1}{2} n_2 \right).
\end{equation}

The eigenmodes of \E{9} are identical to those of \E{7}.
The results are copied here.
\begin{description}
    \item[\textbf{$N=1$ eigenmodes: }]  $\vec{k}_{\vec{n}}=\transpose{((\vec{k}_{\vec{n}})_x, 0, (\vec{k}_{\vec{n}})_z)}$, 
        i.e., $\vec{n}=(n_1,0,n_3)$, 
        $n_1 \in 2\integers$, $n_3 \in \integers$ with at least one of $n_1 \neq 0$ or $n_3 \neq 0$ and
    \begin{empheq}[box=\fbox]{equation}
        \Upsilon^{\E{9}}_{(n_1,0,n_3)}(\vec{x}) = \eexp^{i\vec{k}_{\vec{n}} \cdot (\vec{x}-\vec{x}_0)} = \eexp^{i(\vec{k}_{\vec{n}})_{x} (x-x_0) + i(\vec{k}_{\vec{n}})_{z} (z-z_0)}, 
    \end{empheq}
    \item[\textbf{$N=2$ eigenmodes: }] $(\vec{k}_{\vec{n}})_{y} \neq 0$, 
        i.e., $n_2 \neq 0$, and per \cref{eqn:generaleigenmodeformula},
    \begin{empheq}[box=\fbox]{align}
        \label{eqn:E9_eigenmode_N2}
        \Upsilon^{\E{9}}_{\vec{n}}(\vec{x}) &= \frac{1}{\sqrt{2}} \eexp^{-i(\vec{k}_{\vec{n}}\cdot \vec{T}^{\E{9}}_A/2 + |\vec{k}_{\vec{n}}\cdot \vec{T}^{\E{9}}_1|/4)} \left[ \eexp^{i\vec{k}_{\vec{n}}\cdot(\vec{x}-\vec{x}_0)} 
        + {} \right. \\
        & \left. \qquad\quad {} + \eexp^{i\vec{k}_{\vec{n}}\cdot (\mat{M}^{\E{9}}_A(\vec{x}-\vec{x}_0)+\vec{T}^{\E{9}}_A)} \right]. \nonumber
    \end{empheq}
\end{description}
For the $N=2$ modes, we have chosen the phase $\Phi^{\E{9}}_{\vec{k}} = -\vec{k}_{\vec{n}}\cdot \vec{T}^{\E{9}}_A/2 - |\vec{k}_{\vec{n}}\cdot \vec{T}^{\E{9}}_1|/4$, defined in \eqref{eqn:generaleigenmodeformula} to simplify the reality condition of the field.
With this choice $\Upsilon^{\E{9}}_{-n_1,n_2,n_3} = \Upsilon^{\E{9}*}_{n_1,n_2,-n_3+n_2}$.
Again as in \E{7}, summing over $(n_1,n_2,n_3)$ would double-count eigenmodes if all $n_2\in\integers$ were included.
Hence, we define two sets of allowed modes, one for $N=1$ and another for $N=2$,
\begin{empheq}[box=\fbox]{align}
    \setN^{\E{9}}_1 &= \{(n_1,0,n_3)|n_1\in 2\integers, n_3\in \integers \} \setminus (0, 0, 0), \nonumber\\
    \setN^{\E{9}}_2 &= \{(n_1,n_2,n_3)| n_1 \in \integers, n_2 \in \integers^{>0}, n_3 \in \integers\} , \\
    \setN^{\E{9}} &= \setN^{\E{9}}_1 \cup \setN^{\E{9}}_2 . \nonumber
\end{empheq}
With these the Fourier-mode correlation matrix can now be expressed as
\begin{empheq}[box=\fbox]{align}
\label{eqn:E9FourierCovarianceStandardConvention}
    C^{\E{9};XY}_{\vec{k}_{\vec{n}}\vec{k}_{\vecnp}} 
     = {} & V_{\E{9}} \frac{2\pi^2}{k^3_{\vec{n}}}
        {\mathcal{P}}^{\mathcal{R}}(k_{\vec n})\Delta^X(\vec{k}_{\vec{n}})\DeltaYstar(\vec{k}_{\vec{n}'})
        \eexp^{i(\vec{k}_{\vecnp}-\vec{k}_{\vec{n}})\cdot\vec{x}_0}
        \left[
            \sum_{\vec{\tilde{n}} \in \setN^{\E{9}}_1}
                 \Kdelta_{\vec{k}_{\vec{n}}\vec{k}_{\vec{\tilde{n}}}}
                 \Kdelta_{\vec{k}_{\vecnp}\vec{k}_{\vec{\tilde{n}}}} + {}
        \right.
            \\
            & \left. \quad
                {} + \frac{1}{2}
                \sum_{\vec{\tilde{n}}\in \setN^{\E{9}}_2}
                \sum_{a=0}^1\sum_{b=0}^1 
                \eexp^{i\vec{k}_{\vec{\tilde{n}}}\cdot(\vec{T}^{(a)}-\vec{T}^{(b)})}
                 \Kdelta_{\vec{k}_{\vec{n}}([(\mat{M}^{\E{9}}_A){}^T]^a\vec{k}_{\vec{\tilde{n}}})}
                 \Kdelta_{\vec{k}_{\vecnp}([(\mat{M}^{\E{9}}_A){}^T]^b\vec{k}_{\vec{\tilde{n}}})} 
         \right], 
         \nonumber
\end{empheq}
where $V_{\E{9}}$ is given in \eqref{eqn:VE9}, $\vec{T}^{(0)}\equiv\vec{0}$, and $\vec{T}^{(1)}\equiv\vec{T}^{\E{9}}_A$.

The eigenmodes in the harmonic basis are again identical to those from \E{7}
\begin{empheq}[box=\fbox]{align}
    \xi^{\E{9};\unitvec{k}_{\vec{n}}}_{k_{\vec{n}}\ell m} 
    &= i^\ell  Y^*_{\ell m} (\unitvec{k}_{\vec{n}}) \eexp^{-i \vec{k}_{\vec{n}}\cdot\vec{x}_0}, \quad \mbox{for } \vec{n} \in \setN^{\E{9}}_1 ,\\
    \xi^{\E{9};\unitvec{k}_{\vec{n}}}_{k_{\vec{n}}\ell m} 
     &= \frac{1}{\sqrt{2}} i^\ell \eexp^{-i(\vec{k}_{\vec{n}}\cdot \vec{T}^{\E{9}}_A/2 + |\vec{k}_{\vec{n}}\cdot \vec{T}^{\E{9}}_1|/4)} \left[ Y^*_{\ell m}(\unitvec{k}_{\vec{n}}) \eexp^{-i\vec{k}_{\vec{n}}\cdot \vec{x}_0}
        + {} \right. \nonumber \\
    & \left. \qquad\qquad\qquad\qquad {} + Y_{\ell m} (\unitvec{k}_{\vec{n}}) \eexp^{-i\vec{k}_{\vec{n}}\cdot (\mat{M}^{\E{9}}_A \vec{x}_0 - \vec{T}^{\E{9}}_A)} \right],\quad \mbox{for } \vec{n}\in \setN^{\E{9}}_2 . \nonumber
\end{empheq}
Finally, the harmonic space covariance matrix has the form \eqref{eqn:HarmonicCovariance}.

\subsection{\E{10}: Klein space with half-turn and flip}
\label{secn:eigenmodesE10}

The eigenspectrum and eigenmodes of \E{10} follow directly from those of \E{8} in \cref{secn:eigenmodesE8}.
For \E{10} the discretization condition \eqref{eqn:associatedhomogeneousdiscretization} from the translation vectors $\vec{T}^{\E{10}}_i$ again leads to the components of the allowed wavevectors,
\begin{equation}
    (\vec{k}_{\vec{n}})_{x} = \frac{2\pi n_1}{2 L_{Ax} }, \quad
    (\vec{k}_{\vec{n}})_{y} = \frac{2\pi n_2}{L_{Cy}},  \quad
     (\vec{k}_{\vec{n}})_{z} = \frac{2\pi n_3}{2 L_{Bz}}.
\end{equation}

The form of the eigenmodes of \E{10} are identical to those of \E{8} though the set of allowed integers for $N_B=1$ has been modified.
\begin{description}
    \item[\textbf{$N_A=1$ eigenmodes: }]  $\vec{k}_{\vec{n}} = \transpose{((\vec{k}_{\vec{n}})_x,0,(\vec{k}_{\vec{n}})_z)}$, 
        i.e., $\vec{n}=(n_1,0,n_3)$, 
        $n_1 \in 2\integers^{\neq 0}$, $n_3 \in \integers$, with
    \begin{empheq}[box=\fbox]{align}
        \Upsilon^{\E{10}}_{(n_1,0,n_3)}(\vec{x}) &= \frac{1}{\sqrt{2}} \eexp^{-i(\vec{k}_{\vec{n}} \cdot \vec{T}^{\E{10}}_B/2 + |\vec{k}_{\vec{n}} \cdot \vec{T}^{\E{10}}_3|/4)}
        \left[ \eexp^{i\vec{k}_{\vec{n}} \cdot (\vec{x}-\vec{x}_0)} 
        + {} \right. \\
        & \left. \qquad\qquad {} + \eexp^{i\vec{k}_{\vec{n}} \cdot (\mat{M}^{\E{10}}_B (\vec{x}-\vec{x}_0) + \vec{T}^{\E{10}}_B)} \right], \nonumber
    \end{empheq}
    \item[\textbf{$N_B=1$ eigenmodes: }]  $\vec{k}_{\vec{n}}=\transpose{(0,(\vec{k}_{\vec{n}})_y,(\vec{k}_{\vec{n}})_z)}$, 
        i.e., $\vec{n}=(0,n_2,n_3)$, 
        $n_2 \in \integers^{\neq 0}$, $n_3 \in \integers$, $n_2 + n_3 \in 2\integers$, 
        with
    \begin{empheq}[box=\fbox]{equation}
        \Upsilon^{\E{10}}_{(0,n_2,n_3)}(\vec{x}) = \frac{1}{\sqrt{2}} \eexp^{-i\vec{k}_{\vec{n}} \cdot \vec{T}^{\E{10}}_A/2}
        \left[ \eexp^{i\vec{k}_{\vec{n}} \cdot (\vec{x}-\vec{x}_0)} + \eexp^{i\vec{k}_{\vec{n}} \cdot (\mat{M}^{\E{10}}_A (\vec{x}-\vec{x}_0) + \vec{T}^{\E{10}}_A)} \right], 
    \end{empheq}
    \item[\textbf{$N_{AB}=1$ eigenmodes: }]  $\vec{k}_{\vec{n}}=\transpose{(0,0,(\vec{k}_{\vec{n}})_z)}$, 
        i.e., $\vec{n}=(0,0,n_3)$, 
        $n_3 \in 2\integers^{\neq 0}$, with
    \begin{empheq}[box=\fbox]{equation}
        \Upsilon^{\E{10}}_{(0,0,n_3)}(\vec{x}) = \eexp^{i\vec{k}_{\vec{n}} \cdot (\vec{x}-\vec{x}_0)}, 
    \end{empheq}
    \item[\textbf{$N=2$ eigenmodes: }] $((\vec{k}_{\vec{n}})_{x}, (\vec{k}_{\vec{n}})_{y}) \neq (0,0)$, 
        i.e., $(n_1,n_2) \neq (0,0)$, $n_3 \in \integers$, with
    \begin{empheq}[box=\fbox]{align}
        \Upsilon^{\E{10}}_{\vec{n}}(\vec{x}) &= \frac{1}{\sqrt{4}} \eexp^{-i\vec{k}_{\vec{n}} \cdot [(\mat{M}^{\E{10}}_A \vec{T}^{\E{10}}_B +\vec{T}^{\E{10}}_A)/2 + |\vec{k}_{\vec{n}} \cdot \vec{T}^{\E{10}}_3|/4]} \left[
            \eexp^{i\vec{k}_{\vec{n}}\cdot (\vec{x}-\vec{x}_0)} 
            + {} \right. \nonumber \\
            & \qquad\quad {} 
            + \eexp^{i\vec{k}_{\vec{n}}\cdot (\mat{M}^{\E{10}}_A(\vec{x}-\vec{x}_0)+\vec{T}^{\E{10}}_A)}
            + \eexp^{i\vec{k}_{\vec{n}}\cdot (\mat{M}^{\E{10}}_B(\vec{x}-\vec{x}_0)+\vec{T}^{\E{10}}_B)}
            + {} \\
            & \left. \qquad\quad {} 
            + \eexp^{i\vec{k}_{\vec{n}}\cdot (\mat{M}^{\E{10}}_A \mat{M}^{\E{10}}_B(\vec{x}-\vec{x}_0)+ \mat{M}^{\E{10}}_A \vec{T}^{\E{10}}_B +\vec{T}^{\E{10}}_A)}
            \right] \nonumber .
    \end{empheq}
\end{description}
For the $N_A=1$, $N_B=1$, and $N=2$ modes, we have chosen the separate phases $\Phi^{\E{10}}_{\vec{k}}$ in \eqref{eqn:generaleigenmodeformula} to simplify the reality condition of the fields.
Again as in \E{8}, summing over $(n_1, n_2, n_3)$ would double-count eigenmodes.
To avoid this, we define sets of allowed modes as
\begin{empheq}[box=\fbox]{align}
    \setN^{\E{10}}_{1A} &= \{(n_1,0,n_3)|n_1\in 2\integers^{>0}, n_3\in \integers \} , \nonumber \\
    \setN^{\E{10}}_{1B} &= \{(0,n_2,n_3)|n_2\in \integers^{>0}, n_3\in \integers, n_2 + n_3 \in 2\integers \} , \nonumber \\
    \setN^{\E{10}}_{1AB} &= \{(0,0,n_3)|n_3\in 2\integers^{\neq 0} \} , \\
    \setN^{\E{10}}_2 &= \{(n_1,n_2,n_3)| n_1 \in \integers^{> 0}, n_2 \in \integers^{>0}, n_3 \in \integers\} , \nonumber \\
    \setN^{\E{10}} &= \setN^{\E{10}}_{1A} \cup \setN^{\E{10}}_{1B} \cup \setN^{\E{10}}_{1AB} \cup \setN^{\E{10}}_2 . \nonumber
\end{empheq}
With these the Fourier-mode correlation matrix can now be expressed as
\begin{empheq}[box=\fbox]{align}
\label{eqn:E10FourierCovarianceStandardConvention}
    C^{\E{10};XY}_{\vec{k}_{\vec{n}}\vec{k}_{\vecnp}} 
    = {} & V_{\E{10}} \frac{2\pi^2}{k^3_{\vec{n}}}
        {\mathcal{P}}^{\mathcal{R}}(k_{\vec n})\Delta^X(\vec{k}_{\vec{n}})\DeltaYstar(\vec{k}_{\vec{n}'})
        \eexp^{i(\vec{k}_{\vecnp}-\vec{k}_{\vec{n}})\cdot\vec{x}_0} \times {} \\
        & {} \ \times \left[
            \frac{1}{2} \sum_{\vec{\tilde{n}} \in \setN^{\E{10}}_{1A}}
                \smashoperator[r]{\sum_{a,b\in\{0, B\}}} \eexp^{i\vec{k}_{\vec{n}} \cdot (\vec{T}^{(a)} - \vec{T}^{(b)})}
                 \Kdelta_{\vec{k}_{\vec{n}}((\mat{M}^{\E{10}}_a){}^T \vec{k}_{\tilde{\vec{n}}})}
                 \Kdelta_{\vec{k}_{\vecnp}((\mat{M}^{\E{10}}_b){}^T \vec{k}_{\tilde{\vec{n}}})} + {}
            \right. \nonumber \\
        & {} \quad + \frac{1}{2} \sum_{\vec{\tilde{n}} \in \setN^{\E{10}}_{1B}}
                \smashoperator[r]{\sum_{a,b\in\{0, A\}}} \eexp^{i\vec{k}_{\vec{n}} \cdot (\vec{T}^{(a)} - \vec{T}^{(b)})}
                 \Kdelta_{\vec{k}_{\vec{n}}((\mat{M}^{\E{10}}_a){}^T \vec{k}_{\tilde{\vec{n}}})}
                 \Kdelta_{\vec{k}_{\vecnp}((\mat{M}^{\E{10}}_b){}^T \vec{k}_{\tilde{\vec{n}}})} + {} \nonumber \\
        & {} \quad + \frac{1}{2} \sum_{\vec{\tilde{n}} \in \setN^{\E{10}}_{1AB}}
                \smashoperator[r]{\sum_{a,b\in\{0, AB\}}} \eexp^{i\vec{k}_{\vec{n}} \cdot (\vec{T}^{(a)} - \vec{T}^{(b)})}
                 \Kdelta_{\vec{k}_{\vec{n}}((\mat{M}^{\E{10}}_a){}^T \vec{k}_{\tilde{\vec{n}}})}
                 \Kdelta_{\vec{k}_{\vecnp}((\mat{M}^{\E{10}}_b){}^T \vec{k}_{\tilde{\vec{n}}})} + {} \nonumber \\
        & \left. \quad
                {} + \frac{1}{4}
                \sum_{\vec{\tilde{n}}\in \setN^{\E{10}}_2}
                \smashoperator[r]{\sum_{a,b\in\{0, A, B, AB\}}} \eexp^{i\vec{k}_{\vec{n}} \cdot (\vec{T}^{(a)} - \vec{T}^{(b)})}
                 \Kdelta_{\vec{k}_{\vec{n}}((\mat{M}^{\E{10}}_a){}^T \vec{k}_{\tilde{\vec{n}}})}
                 \Kdelta_{\vec{k}_{\vecnp}((\mat{M}^{\E{10}}_b){}^T \vec{k}_{\tilde{\vec{n}}})} 
         \right] ,
         \nonumber
\end{empheq}
where $V_{\E{10}}$ is given in \eqref{eqn:VE10}, $\vec{T}^{(0)} \equiv \vec{0}$, $\vec{T}^{(A)} \equiv \vec{T}^{\E{10}}_A$, $\vec{T}^{(B)} \equiv \vec{T}^{\E{10}}_B$, $\vec{T}^{(AB)} \equiv \mat{M}^{\E{10}}_A \vec{T}^{\E{10}}_B + \vec{T}^{\E{10}}_A$, and $\mat{M}^{\E{10}}_{AB} \equiv \mat{M}^{\E{10}}_A \mat{M}^{\E{10}}_B$.

The eigenmodes in the harmonic basis are again identical to those from \E{8}
\begin{empheq}[box=\fbox]{align}
    \xi^{\E{10};\unitvec{k}_{\vec{n}}}_{k_{\vec{n}}\ell m} 
    &= \frac{1}{\sqrt{2}} i^\ell \eexp^{-i(\vec{k}_{\vec{n}} \cdot \vec{T}^{\E{10}}_B/2 + |\vec{k}_{\vec{n}} \cdot \vec{T}^{\E{10}}_3|/4)} \left[ Y^*_{\ell m}(\unitvec{k}_{\vec{n}}) \eexp^{-i \vec{k}_{\vec{n}}\cdot\vec{x}_0} 
    {} + \right. \nonumber \\
    & \left. \qquad\qquad\qquad {} + (-1)^m Y_{\ell m}(\unitvec{k}_{\vec{n}}) \eexp^{-i \vec{k}_{\vec{n}} \cdot (\mat{M}^{\E{10}}_B \vec{x}_0 - \vec{T}^{\E{10}}_B)} \right] , \quad \mbox{for } \vec{n} \in \setN^{\E{10}}_{1A} , \nonumber \\
    \xi^{\E{10};\unitvec{k}_{\vec{n}}}_{k_{\vec{n}}\ell m} 
    &= \frac{1}{\sqrt{2}} i^\ell \eexp^{-i\vec{k}_{\vec{n}} \cdot \vec{T}^{\E{10}}_A/2} \left[ Y^*_{\ell m}(\unitvec{k}_{\vec{n}}) \eexp^{-i \vec{k}_{\vec{n}}\cdot\vec{x}_0} 
    {} + \right. \nonumber \\
    & \left. \qquad\qquad\qquad {} + Y_{\ell m}(\unitvec{k}_{\vec{n}}) \eexp^{-i \vec{k}_{\vec{n}} \cdot (\mat{M}^{\E{10}}_A \vec{x}_0 - \vec{T}^{\E{10}}_A)} \right] , \quad \mbox{for } \vec{n} \in \setN^{\E{10}}_{1B} , \nonumber \\
    \xi^{\E{10};\unitvec{k}_{\vec{n}}}_{k_{\vec{n}}\ell m} 
    &= i^\ell Y^*_{\ell m}(\unitvec{k}_{\vec{n}}) \eexp^{-i \vec{k}_{\vec{n}}\cdot\vec{x}_0} , \quad \mbox{for } \vec{n} \in \setN^{\E{10}}_{1AB} , \\
    \xi^{\E{10};\unitvec{k}_{\vec{n}}}_{k_{\vec{n}}\ell m} 
     &= \frac{1}{\sqrt{4}} i^\ell \eexp^{-i\vec{k}_{\vec{n}} \cdot [(\mat{M}^{\E{10}}_A \vec{T}^{\E{10}}_B +\vec{T}^{\E{10}}_A)/2 + |\vec{k}_{\vec{n}} \cdot \vec{T}^{\E{10}}_3|/4]} \left[ Y^*_{\ell m}(\unitvec{k}_{\vec{n}})
        \left( \eexp^{-i \vec{k}_{\vec{n}} \cdot \vec{x}_0} \vphantom{\eexp^{-i \vec{k}_{\vec{n}} \cdot \mat{M}^{\E{10}}_A}}
        {} + \right. \right. \nonumber \\
    & \left. \qquad\qquad\qquad\qquad {} + (-1)^m \eexp^{-i \vec{k}_{\vec{n}} \cdot (\mat{M}^{\E{10}}_A \mat{M}^{\E{10}}_B \vec{x_0} - \mat{M}^{\E{10}}_A \vec{T}^{\E{10}}_B - \vec{T}^{\E{10}}_A)} \right) + {} \nonumber \\
      & {} \qquad\qquad + Y_{\ell m}(\unitvec{k}_{\vec{n}}) \left(
        \eexp^{-i \vec{k}_{\vec{n}} \cdot (\mat{M}^{\E{10}}_A \vec{x_0} - \vec{T}^{\E{10}}_A)}
        \right. + {} \nonumber \\
      & {} \left. \left. \qquad \qquad \qquad \qquad {} 
        + (-1)^m \eexp^{-i \vec{k}_{\vec{n}} \cdot (\mat{M}^{\E{10}}_B \vec{x_0} - \vec{T}^{\E{10}}_B)} \right) \right], \quad \mbox{for } \vec{n}\in \setN^{\E{10}}_2 , \nonumber
\end{empheq}
and the harmonic space covariance matrix has the form \eqref{eqn:HarmonicCovariance}.

\subsection{\E{13}: Chimney space with vertical flip}
\label{secn:eigenmodesE13}

The eigenspectrum and eigenmodes of the chimney space with vertical flip can be determined in a manner analogous to that of the chimney space \E{11} from \cite{COMPACT:2023rkp} or from the fact that \E{13} is a limit of \E{7} with $|\vec{T}^{\E{7}}_{\B{1}}| \to \infty$.
We will follow the latter path taking $L_{1y} \to \infty$ so that $(\vec{k}_{\vec{n}})_y \to k_y$ becomes continuous with the other two components of $\vec{k}_{\vec{n}}$ discrete.
To remain consistent with \E{7} we will continue to use $n_1$ and $n_3$ as the two integers.
With this the discretization condition \eqref{eqn:associatedhomogeneousdiscretization} leads to the components of the allowed wavevectors,
\begin{equation}
    (\vec{k}_{\vec{n}})_{x} = \frac{2\pi n_1}{L_{Ax}} , \quad (\vec{k}_{\vec{n}})_{z} = \frac{2\pi}{L_{Bz}} \left( n_3 - \frac{L_{Bx}}{2 L_{Ax}} n_1 \right),
\end{equation}
and $k_y$ unconstrained.
We will write $\vec{k}_{\vec{n}}$ as a shorthand for the wavevector parametrized by the integer array $\vec{n}=(n_1,n_3)$ and the real variable $k_y$.

The eigenmodes of \E{13} now follow directly from those of \E{7}.
There are still two solutions to $(\mat{M}^{\E{13}}_A)^N \vec{k}_{\vec{n}} = \vec{k}_{\vec{n}}$ for $N=1$ and $N=2$ written explicitly as
\begin{description}
    \item[\textbf{$N=1$ eigenmodes: }]  $\vec{k}_{\vec{n}}=\transpose{((\vec{k}_{\vec{n}})_x, 0, (\vec{k}_{\vec{n}})_z)}$, 
        i.e., $\vec{n}=(n_1,n_3)$, 
        $n_1 \in 2\integers$, $n_3 \in \integers$ with at least one of $n_1 \neq 0$ or $n_3 \neq 0$, with
    \begin{empheq}[box=\fbox]{equation}
        \Upsilon^{\E{13}}_{(n_1;0;n_3)}(\vec{x}) = \eexp^{i\vec{k}_{\vec{n}} \cdot (\vec{x}-\vec{x}_0)} = \eexp^{i(\vec{k}_{\vec{n}})_{x} (x-x_0) + i(\vec{k}_{\vec{n}})_{z} (z-z_0)}, 
    \end{empheq}
    \item[\textbf{$N=2$ eigenmodes: }] $\vec{k}_{\vec{n}}=\transpose{((\vec{k}_{\vec{n}})_x, k_y, (\vec{k}_{\vec{n}})_z)}$, 
        i.e., $\vec{n}=(n_1,n_3)$, $k_y \neq 0$, with
    \begin{empheq}[box=\fbox]{align}
        \Upsilon^{\E{13}}_{\vec{n}}(\vec{x}) &= \frac{1}{\sqrt{2}} \eexp^{-i(\vec{k}_{\vec{n}}\cdot \vec{T}^{\E{13}}_A/2 +  |\vec{k}_{\vec{n}}\cdot \vec{T}^{\E{13}}_1|/4)} \left[ \eexp^{i\vec{k}_{\vec{n}}\cdot(\vec{x}-\vec{x}_0)} 
        + {} \right. \\
        & \left. \qquad\qquad {} + \eexp^{i\vec{k}_{\vec{n}}\cdot (\mat{M}^{\E{13}}_A(\vec{x}-\vec{x}_0)+\vec{T}^{\E{13}}_A)} \right]. \nonumber
    \end{empheq}
\end{description}
For the $N=2$ modes we have chosen the phase $\Phi^{\E{13}}_{\vec{k}} = -\vec{k}_{\vec{n}}\cdot \vec{T}^{\E{13}}_A/2 - |\vec{k}_{\vec{n}}\cdot \vec{T}^{\E{13}}_1|/4$ in \eqref{eqn:generaleigenmodeformula} to simplify the reality condition of the field.
As in \E{7} the two sets of allowed modes are defined by
\begin{empheq}[box=\fbox]{align}
    \setN^{\E{13}}_1 &= \{(n_1,n_3)|n_1\in 2\integers, n_3\in \integers \} \setminus (0, 0), \nonumber\\
    \setN^{\E{13}}_2 &= \{(n_1,n_3)| n_1 \in \integers, n_3 \in \integers\} , \\
    \setN^{\E{13}} &= \setN^{\E{13}}_1 \cup \setN^{\E{13}}_2 . \nonumber
\end{empheq}
Note that for $\setN^{\E{13}}_1$ we require $k_y = 0$ and for $\setN^{\E{13}}_2$ we require $k_y > 0$.
With these the Fourier-mode correlation matrix can now be expressed as
\begin{empheq}[box=\fbox]{align}
\label{eqn:E13FourierCovarianceStandardConvention}
    C^{\E{13};XY}_{\vec{k}_{\vec{n}}\vec{k}_{\vecnp}} 
     = {} & 2\pi A_{\E{13}} \frac{2\pi^2}{k^3_{\vec{n}}}
        {\mathcal{P}}^{\mathcal{R}}(k_{\vec n})\Delta^X(\vec{k}_{\vec{n}})\DeltaYstar(\vec{k}_{\vec{n}'})
        \eexp^{i(\vec{k}_{\vecnp}-\vec{k}_{\vec{n}})\cdot\vec{x}_0} \times {} \nonumber \\
        & {} \; \times
            \frac{1}{2} 
                \smashoperator[l]{\sum_{(\tilde{n}_1, \tilde{n}_3)\in \setN^{\E{13}}_2}}
                \int_{0}^{\infty} \dderiv\tilde{k}_y \, \sum_{a=0}^1\sum_{b=0}^1
                \eexp^{i\vec{k}_{\vec{\tilde{n}}}\cdot(\vec{T}^{(a)}-\vec{T}^{(b)})} 
                \Kdelta_{(\vec{k}_{\vec{n}})_x (\vec{k}^{(a)}_{\vec{\tilde{n}}})_x}
                \Kdelta_{(\vec{k}_{\vecnp})_x (\vec{k}^{(b)}_{\vec{\tilde{n}}})_x}
                \\
            & \qquad\quad {} \times
                 \Kdelta_{(\vec{k}_{\vec{n}})_z (\vec{k}^{(a)}_{\vec{\tilde{n}}})_z}
                 \Kdelta_{(\vec{k}_{\vecnp})_z (\vec{k}^{(b)}_{\vec{\tilde{n}}})_z}
                 \Ddelta(k_y - \tilde{k}^{(a)}_y)
                 \Ddelta(k'_y - \tilde{k}^{(b)}_y)
         , 
         \nonumber
\end{empheq}
where the terms with $\vec{n} \in \setN^{\E{13}}_1$ are of measure zero and therefore do not contribute to the integral.
$A_{\E{13}}$ is given in \eqref{eqn:AE13}, $\vec{T}^{(0)}\equiv\vec{0}$, $\vec{T}^{(1)}\equiv\vec{T}^{\E{13}}_A$, and $\vec{k}^{(a)}_{\tilde{\vec{n}}} \equiv [(\mat{M}^{\E{13}}_A){}^T]^a \vec{k}_{\vec{\tilde{n}}}$, so that $(\vec{k}^{(a)}_{\tilde{\vec{n}}})_x = (\vec{k}_{\tilde{\vec{n}}})_x$, $(\vec{k}^{(a)}_{\tilde{\vec{n}}})_z = (\vec{k}_{\tilde{\vec{n}}})_z$, and 
$\tilde{k}^{(a)}_y \equiv (\vec{k}^{(a)}_{\tilde{\vec{n}}})_y = (-1)^a \tilde{k}_y$.

Again following \E{7} the eigenmodes in the harmonic basis are given by
\begin{empheq}[box=\fbox]{align}
    \xi^{\E{13};\unitvec{k}_{\vec{n}}}_{k_{\vec{n}}\ell m} 
    &= i^\ell  Y^*_{\ell m} (\unitvec{k}_{\vec{n}}) \eexp^{-i \vec{k}_{\vec{n}}\cdot\vec{x}_0}, \quad \mbox{for } \vec{n} \in \setN^{\E{13}}_1, k_y = 0 ,\\
    \xi^{\E{13};\unitvec{k}_{\vec{n}}}_{k_{\vec{n}}\ell m} 
    &= \frac{1}{\sqrt{2}} i^\ell \eexp^{-i(\vec{k}_{\vec{n}}\cdot \vec{T}^{\E{13}}_A/2 +  |\vec{k}_{\vec{n}}\cdot \vec{T}^{\E{13}}_1|/4)} \left[ Y^*_{\ell m}(\unitvec{k}_{\vec{n}}) \eexp^{-i\vec{k}_{\vec{n}}\cdot \vec{x}_0}
       + {} \right. \nonumber \\
    & \left. \qquad\qquad\qquad {} +  Y_{\ell m} (\unitvec{k}_{\vec{n}})\eexp^{-i\vec{k}_{\vec{n}}\cdot (\mat{M}^{\E{13}}_A \vec{x}_0 - \vec{T}^{\E{13}}_A)} \right], \mbox{for } \vec{n}\in \setN^{\E{13}}_2, k_y > 0 . \nonumber
\end{empheq}
The harmonic space covariance matrix is calculated from these eigenmodes is
\begin{empheq}[box=\fbox]{equation}
    \label{eqn:HarmonicCovarianceE13}
    C^{\E{13};XY}_{\ell m\ell'm'}  =
     \frac{(4\pi)^2}{{2\pi A_{\E{13}}}}
    \sum_{(n_1,n_3)\in \setN^{\E{13}}} \int_{0}^{\infty} \dderiv k_y \,
    \Delta^X_{\ell}(k_{\vec{n}})
    \DeltaYstarforell_{\ell'}(k_{\vec{n}})
    \frac{2\pi^2 \mathcal{P}^{\mathcal{R}}(k_{\vec{n}})}{k_{\vec{n}}^3}
    \xi^{\E{13};\unitvec{k}_{\vec{n}}}_{k_{\vec{n}} \ell m}
    \xi^{\E{13};\unitvec{k}_{\vec{n}}*}_{k_{\vec{n}} \ell' m'} .
\end{empheq}

\subsection{\E{14}: Chimney space with horizontal flip}
\label{secn:eigenmodesE14}

The eigenspectrum and eigenmodes of the chimney space with horizontal flip can be determined in a manner analogous to that of \E{13} since \E{14} is a limit of \E{7} with $|\vec{T}^{\E{7}}_{\B{2}}| \to \infty$.
This means that $(\vec{k}_{\vec{n}})_z \to k_z$ becomes continuous with the other two components of $\vec{k}_{\vec{n}}$ discrete.
To remain consistent with \E{7} we will continue to use $n_1$ and $n_2$ as the two integers.
With this the discretization condition \eqref{eqn:associatedhomogeneousdiscretization} leads to the components of the allowed wavevectors,
\begin{equation}
    (\vec{k}_{\vec{n}})_{x} = \frac{2\pi n_1}{2 L_{Ax}} , \quad (\vec{k}_{\vec{n}})_{y} = \frac{2\pi n_2}{L_B},
\end{equation}
and $k_z$ unconstrained.
We will write $\vec{k}_{\vec{n}}$ as a shorthand for the wavevector parametrized by the integer array $\vec{n}=(n_1,n_2)$ and the real variable $k_z$.

The eigenmodes of \E{14} now follow directly from those of \E{7}.
There are still two solutions to $(\mat{M}^{\E{14}}_A)^N \vec{k}_{\vec{n}} = \vec{k}_{\vec{n}}$ for $N=1$ and $N=2$ written explicitly as
\begin{description}
    \item[\textbf{$N=1$ eigenmodes: }]  $\vec{k}_{\vec{n}}=\transpose{((\vec{k}_{\vec{n}})_x, 0, k_z)}$, 
        i.e., $\vec{n}=(n_1,0)$, 
        $n_1 \in 2\integers$ with at least one of $n_1 \neq 0$ or $k_z \neq 0$ and
    \begin{empheq}[box=\fbox]{equation}
        \Upsilon^{\E{14}}_{\vec{n}}(\vec{x}) = \eexp^{i\vec{k}_{\vec{n}} \cdot (\vec{x}-\vec{x}_0)} = \eexp^{i(\vec{k}_{\vec{n}})_{x} (x-x_0) + i k_{z} (z-z_0)}, 
    \end{empheq}
    \item[\textbf{$N=2$ eigenmodes: }] $\vec{k}_{\vec{n}}=\transpose{((\vec{k}_{\vec{n}})_x, (\vec{k}_{\vec{n}})_y, k_z)}$, 
        i.e., $\vec{n}=(n_1,n_2)$, $n_2 \neq 0$, with
    \begin{empheq}[box=\fbox]{align}
        \Upsilon^{\E{14}}_{\vec{n}}(\vec{x}) &= \frac{1}{\sqrt{2}} \eexp^{-i(\vec{k}_{\vec{n}}\cdot \vec{T}^{\E{14}}_A/2 + |\vec{k}_{\vec{n}}\cdot \vec{T}^{\E{14}}_1|/4)} \left[ \eexp^{i\vec{k}_{\vec{n}}\cdot(\vec{x}-\vec{x}_0)} 
        + {} \right. \\
        & \left. \qquad\qquad {} + \eexp^{i\vec{k}_{\vec{n}}\cdot(\mat{M}^{\E{14}}_A(\vec{x}-\vec{x}_0)+\vec{T}^{\E{14}}_A)} \right]. \nonumber
    \end{empheq}
\end{description}
For the $N=2$ modes we have chosen the phase $\Phi^{\E{14}}_{\vec{k}} = -\vec{k}_{\vec{n}}\cdot \vec{T}^{\E{14}}_A/2 - |\vec{k}_{\vec{n}}\cdot \vec{T}^{\E{14}}_1|/4$ in \eqref{eqn:generaleigenmodeformula} to simplify the reality condition of the field.
As in \E{7} the two sets of allowed modes are defined by
\begin{empheq}[box=\fbox]{align}
    \setN^{\E{14}}_1 &= \{(n_1,0)|n_1\in 2\integers \} , \nonumber\\
    \setN^{\E{14}}_2 &= \{(n_1,n_2)| n_1 \in \integers, n_2 \in \integers^{>0}\} , \\
    \setN^{\E{14}} &= \setN^{\E{14}}_1 \cup \setN^{\E{14}}_2 . \nonumber
\end{empheq}
Note that for $(n_1, n_2)\in \setN^{\E{14}}_1$ we require either $n_1 \neq 0$ or $k_z \neq 0$.
With these the Fourier-mode correlation matrix can now be expressed as
\begin{empheq}[box=\fbox]{align}
\label{eqn:E14FourierCovarianceStandardConvention}
    C^{\E{14};XY}_{\vec{k}_{\vec{n}}\vec{k}_{\vecnp}} 
     = {} & 2\pi A_{\E{14}} \frac{2\pi^2}{k^3_{\vec{n}}}
        {\mathcal{P}}^{\mathcal{R}}(k_{\vec n})\Delta^X(\vec{k}_{\vec{n}})\DeltaYstar(\vec{k}_{\vec{n}'})
        \eexp^{i(\vec{k}_{\vecnp}-\vec{k}_{\vec{n}})\cdot\vec{x}_0}
        \times {} \nonumber \\
        & {} \; \times
            \left[
                \smashoperator[r]{\sum_{(\tilde{n}_1, \tilde{n}_2)\in \setN^{\E{14}}_1}}
                \Kdelta_{(\vec{k}_{\vec{n}})_x (\vec{k}_{\vec{\tilde{n}}})_x}
                \Kdelta_{(\vec{k}_{\vecnp})_x (\vec{k}_{\vec{\tilde{n}}})_x}
                \Kdelta_{(\vec{k}_{\vec{n}})_y (\vec{k}_{\vec{\tilde{n}}})_y}
                \Kdelta_{(\vec{k}_{\vecnp})_y (\vec{k}_{\vec{\tilde{n}}})_y}
                 \Ddelta(k_z - k'_z)
            \right. + {} \\
            \nonumber
        & {} \quad +
                \frac{1}{2} \smashoperator[l]{\sum_{(\tilde{n}_1, \tilde{n}_2)\in \setN^{\E{14}}_2}}
                \int_{-\infty}^{\infty} \dderiv\tilde{k}_z \, \sum_{a=0}^1\sum_{b=0}^1
                \eexp^{i\vec{k}_{\vec{\tilde{n}}}\cdot(\vec{T}^{(a)}-\vec{T}^{(b)})}
                \Kdelta_{(\vec{k}_{\vec{n}})_x (\vec{k}^{(a)}_{\vec{\tilde{n}}})_x}
                \Kdelta_{(\vec{k}_{\vecnp})_x (\vec{k}^{(b)}_{\vec{\tilde{n}}})_x}
                \times {} \\
            & \qquad\quad {} \times
            \left.
                 \Kdelta_{(\vec{k}_{\vec{n}})_y (\vec{k}^{(a)}_{\vec{\tilde{n}}})_y}
                 \Kdelta_{(\vec{k}_{\vecnp})_y (\vec{k}^{(b)}_{\vec{\tilde{n}}})_y}
                 \Ddelta(k_z - \tilde{k}^{(a)}_z)
                 \Ddelta(k'_z - \tilde{k}^{(b)}_z)
            \right]
         , 
         \nonumber
\end{empheq}
where $A_{\E{14}}$ is given in \eqref{eqn:AE14}, $\vec{T}^{(0)}\equiv\vec{0}$, $\vec{T}^{(1)}\equiv\vec{T}^{\E{14}}_A$, and $\vec{k}^{(a)}_{\tilde{\vec{n}}} \equiv [(\mat{M}^{\E{14}}_A){}^T]^a \vec{k}_{\vec{\tilde{n}}}$, so that $(\vec{k}^{(a)}_{\tilde{\vec{n}}})_x = (\vec{k}_{\tilde{\vec{n}}})_x$, $(\vec{k}^{(a)}_{\tilde{\vec{n}}})_y = (-1)^a (\vec{k}_{\tilde{\vec{n}}})_y$, and $\tilde{k}^{(a)}_z \equiv (\vec{k}^{(a)}_{\tilde{\vec{n}}})_z = \tilde{k}_z$.
Unlike in \E{13} the terms with $\vec{n} \in \setN^{\E{14}}_1$ are not of measure zero and have thus been retained.

Again following \E{7} the eigenmodes in the harmonic basis are given by
\begin{empheq}[box=\fbox]{align}
    \xi^{\E{14};\unitvec{k}_{\vec{n}}}_{k_{\vec{n}}\ell m} 
    &= i^\ell  Y^*_{\ell m} (\unitvec{k}_{\vec{n}}) \eexp^{-i \vec{k}_{\vec{n}}\cdot\vec{x}_0}, \quad \mbox{for } \vec{n} \in \setN^{\E{14}}_1 ,\\
    \xi^{\E{14};\unitvec{k}_{\vec{n}}}_{k_{\vec{n}}\ell m} 
    &= \frac{1}{\sqrt{2}} i^\ell \eexp^{-i(\vec{k}_{\vec{n}}\cdot \vec{T}^{\E{14}}_A/2 + |\vec{k}_{\vec{n}}\cdot \vec{T}^{\E{14}}_1|/4)} \left[ Y^*_{\ell m}(\unitvec{k}_{\vec{n}}) \eexp^{-i\vec{k}_{\vec{n}}\cdot \vec{x}_0}
     + {} \right. \nonumber \\
    & \left. \qquad\qquad\qquad {} + Y_{\ell m} (\unitvec{k}_{\vec{n}}) \eexp^{-i\vec{k}_{\vec{n}}\cdot (\mat{M}^{\E{14}}_A \vec{x}_0 - \vec{T}^{\E{14}}_A)} \right],\quad \mbox{for } \vec{n}\in \setN^{\E{14}}_2 . \nonumber
\end{empheq}
The harmonic space covariance matrix is calculated from these eigenmodes is
\begin{empheq}[box=\fbox]{equation}
    \label{eqn:HarmonicCovarianceE14}
    C^{\E{14};XY}_{\ell m\ell'm'}  =
     \frac{(4\pi)^2}{{2\pi A_{\E{14}}}}
    \sum_{(n_1,n_2)\in \setN^{\E{14}}} \int_{-\infty}^{\infty} \dderiv k_z \,
    \Delta^X_{\ell}(k_{\vec{n}})
    \DeltaYstarforell_{\ell'}(k_{\vec{n}})
    \frac{2\pi^2 \mathcal{P}^{\mathcal{R}}(k_{\vec{n}})}{k_{\vec{n}}^3}
    \xi^{\E{14};\unitvec{k}_{\vec{n}}}_{k_{\vec{n}} \ell m}
    \xi^{\E{14};\unitvec{k}_{\vec{n}}*}_{k_{\vec{n}} \ell' m'} .
\end{empheq}

\subsection{\E{15}: Chimney space with half-turn and flip}
\label{secn:eigenmodesE15}

The eigenspectrum and eigenmodes of the chimney space with half-turn and flip can be determined in a manner analogous to that of \E{13} and \E{14}, now recognizing that \E{15} is a limit of \E{8} with $|\vec{T}^{\E{8}}_C| \to \infty$.
This means taking $L_{Cy} \to \infty$ so that $(\vec{k}_{\vec{n}})_y \to k_y$ becomes continuous with the other two components of $\vec{k}_{\vec{n}}$ discrete.
To remain consistent with \E{8} we will continue to use $n_1$ and $n_3$ as the two integers.
With this the discretization condition \eqref{eqn:associatedhomogeneousdiscretization} leads to the components of the allowed wavevectors,
\begin{equation}
    (\vec{k}_{\vec{n}})_{x} = \frac{2\pi n_1}{L_{Ax}} , \quad (\vec{k}_{\vec{n}})_{z} = \frac{2\pi n_3}{2 L_{Bz}} ,
\end{equation}
and $k_y$ unconstrained.
We will write $\vec{k}_{\vec{n}}$ as a shorthand for the wavevector parametrized by the integer array $\vec{n}=(n_1,n_3)$ and the real variable $k_y$.

The eigenmodes of \E{15} now follow directly from those of \E{8}.
There are still multiple solutions to $(\mat{M}^{\E{15}}_a)^{N_a} \vec{k}_{\vec{n}} = \vec{k}_{\vec{n}}$ for $a\in\{A, B, AB\}$.
Written explicitly we have
\begin{description}
    \item[\textbf{$N_A=1$ eigenmodes: }]  $\vec{k}_{\vec{n}}=\transpose{((\vec{k}_{\vec{n}})_x,0,(\vec{k}_{\vec{n}})_z)}$, 
        i.e., $\vec{n}=(n_1,n_3)$, $n_1 \in 2\integers^{\neq 0}$, $n_3 \in \integers$, with
    \begin{empheq}[box=\fbox]{align}
        \Upsilon^{\E{15}}_{(n_1;0;n_3)}(\vec{x}) &= \frac{1}{\sqrt{2}} \eexp^{-i(\vec{k}_{\vec{n}} \cdot \vec{T}^{\E{15}}_B/2 + |\vec{k}_{\vec{n}} \cdot \vec{T}^{\E{15}}_2|/4)}
        \left[ \eexp^{i\vec{k}_{\vec{n}} \cdot (\vec{x}-\vec{x}_0)} 
        + {} \right. \\
        & \left. \qquad\qquad {} + \eexp^{i\vec{k}_{\vec{n}} \cdot (\mat{M}^{\E{15}}_B (\vec{x}-\vec{x}_0) + \vec{T}^{\E{15}}_B)} \right], \nonumber
    \end{empheq}
    \item[\textbf{$N_B=1$ eigenmodes: }]  $\vec{k}_{\vec{n}}=\transpose{(0,k_y,(\vec{k}_{\vec{n}})_z)}$, 
        i.e., $\vec{n}=(0,n_3)$, $n_3 \in 2\integers$, $k_y \neq 0$, with
    \begin{empheq}[box=\fbox]{equation}
        \Upsilon^{\E{15}}_{\vec{n}}(\vec{x}) = \frac{1}{\sqrt{2}} \eexp^{-i\vec{k}_{\vec{n}} \cdot \vec{T}^{\E{15}}_A/2}
        \left[ \eexp^{i\vec{k}_{\vec{n}} \cdot (\vec{x}-\vec{x}_0)} + \eexp^{i\vec{k}_{\vec{n}} \cdot (\mat{M}^{\E{15}}_A (\vec{x}-\vec{x}_0) + \vec{T}^{\E{15}}_A)} \right], 
    \end{empheq}
    \item[\textbf{$N_{AB}=1$ eigenmodes: }]  $\vec{k}_{\vec{n}}=\transpose{(0,0,(\vec{k}_{\vec{n}})_z)}$, 
        i.e., $\vec{n}=(0,n_3)$, $n_3 \in 2\integers^{\neq 0}$, with
    \begin{empheq}[box=\fbox]{equation}
        \Upsilon^{\E{15}}_{(0,0,n_3)}(\vec{x}) = \eexp^{i\vec{k}_{\vec{n}} \cdot (\vec{x}-\vec{x}_0)}, 
    \end{empheq}
    \item[\textbf{$N=2$ eigenmodes: }] $((\vec{k}_{\vec{n}})_{x}, k_y) \neq (0,0)$, i.e., $n_1 \in \integers^{\neq 0}$, $n_3 \in \integers$, $k_y \neq 0$, with
    \begin{empheq}[box=\fbox]{align}
        \Upsilon^{\E{15}}_{\vec{n}}(\vec{x}) &= \frac{1}{\sqrt{4}} \eexp^{-i\vec{k}_{\vec{n}} \cdot [(\mat{M}^{\E{15}}_A \vec{T}^{\E{15}}_B +\vec{T}^{\E{15}}_A)/2 + |\vec{k}_{\vec{n}} \cdot \vec{T}^{\E{15}}_2|/4]} 
        \left[ \eexp^{i\vec{k}_{\vec{n}}\cdot (\vec{x}-\vec{x}_0)} 
        + {} \right. \nonumber \\
        & \qquad\quad {}
            + \eexp^{i\vec{k}_{\vec{n}}\cdot (\mat{M}^{\E{15}}_A(\vec{x}-\vec{x}_0)+\vec{T}^{\E{15}}_A)}
             + \eexp^{i\vec{k}_{\vec{n}}\cdot (\mat{M}^{\E{15}}_B(\vec{x}-\vec{x}_0)+\vec{T}^{\E{15}}_B)}
            + {} \\
            & \left. \qquad\quad {} 
            + \eexp^{i\vec{k}_{\vec{n}}\cdot (\mat{M}^{\E{15}}_A \mat{M}^{\E{15}}_B(\vec{x}-\vec{x}_0)+ \mat{M}^{\E{15}}_A \vec{T}^{\E{15}}_B +\vec{T}^{\E{15}}_A)}
            \right] . \nonumber
    \end{empheq}
\end{description}
For the $N_A=1$, $N_B=1$, and $N=2$ modes, we have chosen the separate phases $\Phi^{\E{15}}_{\vec{k}}$ in \eqref{eqn:generaleigenmodeformula} to simplify the reality condition of the fields.
As in \E{8} the sets of allowed modes are defined by
\begin{empheq}[box=\fbox]{align}
    \setN^{\E{15}}_{1A} &= \{(n_1,n_3)|n_1\in 2\integers^{>0}, n_3\in \integers \} , \nonumber \\
    \setN^{\E{15}}_{1B} &= \{(0,n_3)| n_3\in 2\integers \} , \nonumber \\
    \setN^{\E{15}}_{1AB} &= \{(0,n_3)|n_3\in 2\integers^{\neq 0} \} , \\
    \setN^{\E{15}}_2 &= \{(n_1,n_3)| n_1 \in \integers^{> 0}, n_3 \in \integers\} , \nonumber \\
    \setN^{\E{15}} &= \setN^{\E{15}}_{1A} \cup \setN^{\E{15}}_{1B} \cup \setN^{\E{15}}_{1AB} \cup \setN^{\E{15}}_2 . \nonumber
\end{empheq}
Note that for $\setN^{\E{15}}_{1A}$ and $\setN^{\E{15}}_{1AB}$ we require $k_y=0$ and for $\setN^{\E{15}}_{1B}$ and $\setN^{\E{15}}_{2}$ we require $k_y > 0$.
With these the Fourier-mode correlation matrix can now be expressed as
\begin{empheq}[box=\fbox]{align}
\label{eqn:E15FourierCovarianceStandardConvention}
    C^{\E{15};XY}_{\vec{k}_{\vec{n}}\vec{k}_{\vecnp}} 
    = {} & 2\pi A_{\E{15}} \frac{2\pi^2}{k^3_{\vec{n}}}
        {\mathcal{P}}^{\mathcal{R}}(k_{\vec n})\Delta^X(\vec{k}_{\vec{n}})\DeltaYstar(\vec{k}_{\vec{n}'})
        \eexp^{i(\vec{k}_{\vecnp}-\vec{k}_{\vec{n}})\cdot\vec{x}_0} \times {} \\
        & {} \times \left[
            \frac{1}{2} \sum_{(\tilde{n}_1, \tilde{n}_3)\in \setN^{\E{15}}_{1B}} \int_{0}^{\infty} \dderiv\tilde{k}_y \,
            \smashoperator{\sum_{a,b\in\{0, A\}}} \eexp^{i\vec{k}_{\vec{n}} \cdot (\vec{T}^{(a)} - \vec{T}^{(b)})}
                \Kdelta_{(\vec{k}_{\vec{n}})_x (\vec{k}^{(a)}_{\vec{\tilde{n}}})_x}
                \Kdelta_{(\vec{k}_{\vecnp})_x (\vec{k}^{(b)}_{\vec{\tilde{n}}})_x}
            \right.
                \times {} \nonumber \\
            & \quad\quad {} \times
                 \Kdelta_{(\vec{k}_{\vec{n}})_z (\vec{k}^{(a)}_{\vec{\tilde{n}}})_z}
                 \Kdelta_{(\vec{k}_{\vecnp})_z (\vec{k}^{(b)}_{\vec{\tilde{n}}})_z}
                 \Ddelta(k_y - \tilde{k}^{(a)}_y)
                 \Ddelta(k'_y - \tilde{k}^{(b)}_y)
                 + {} \nonumber \\
        & {} \quad
            + \frac{1}{4}
            \sum_{(\tilde{n}_1, \tilde{n}_3)\in \setN^{\E{15}}_2} \int_{0}^{\infty} \dderiv\tilde{k}_y \,
            \smashoperator{\sum_{a,b\in\{0, A, B, AB\}}} \eexp^{i\vec{k}_{\vec{n}} \cdot (\vec{T}^{(a)} - \vec{T}^{(b)})}
                \Kdelta_{(\vec{k}_{\vec{n}})_x (\vec{k}^{(a)}_{\vec{\tilde{n}}})_x}
                \Kdelta_{(\vec{k}_{\vecnp})_x (\vec{k}^{(b)}_{\vec{\tilde{n}}})_x}
                \times {} \nonumber \\
            & \quad\quad {} \times
            \left.
                 \Kdelta_{(\vec{k}_{\vec{n}})_z (\vec{k}^{(a)}_{\vec{\tilde{n}}})_z}
                 \Kdelta_{(\vec{k}_{\vecnp})_z (\vec{k}^{(b)}_{\vec{\tilde{n}}})_z}
                 \Ddelta(k_y - \tilde{k}^{(a)}_y)
                 \Ddelta(k'_y - \tilde{k}^{(b)}_y)
            \right]
         , \nonumber
\end{empheq}
where $A_{\E{15}}$ is given in \eqref{eqn:AE15}, $\vec{T}^{(0)} \equiv \vec{0}$, $\vec{T}^{(A)} \equiv \vec{T}^{\E{15}}_A$, $\vec{T}^{(B)} \equiv \vec{T}^{\E{15}}_B$, $\vec{T}^{(AB)} \equiv \mat{M}^{\E{15}}_A \vec{T}^{\E{15}}_B + \vec{T}^{\E{15}}_A$, $\vec{k}^{(a)}_{\vec{\tilde{n}}} \equiv (\mat{M}^{\E{15}}_a)^{T} \vec{k}_{\vec{\tilde{n}}}$, and $\mat{M}^{\E{15}}_{AB} \equiv \mat{M}^{\E{15}}_A \mat{M}^{\E{15}}_B$.
Similar to \E{13} the terms with $\vec{n} \in \setN^{\E{15}}_{1A}$ and $\vec{n} \in \setN^{\E{15}}_{1AB}$ are of measure zero and have been dropped and similar to \E{14} the term with $\vec{n} \in \setN^{\E{15}}_{1B}$ is not of measure zero and has been retained.
The eigenmodes in the harmonic basis follow from those in \E{8}
\begin{empheq}[box=\fbox]{align}
    \xi^{\E{15};\unitvec{k}_{\vec{n}}}_{k_{\vec{n}}\ell m} 
    &= \frac{1}{\sqrt{2}} i^\ell \eexp^{-i(\vec{k}_{\vec{n}} \cdot \vec{T}^{\E{15}}_B/2 + |\vec{k}_{\vec{n}} \cdot \vec{T}^{\E{15}}_2|/4)} Y^*_{\ell m}(\unitvec{k}_{\vec{n}}) \left[ \eexp^{-i \vec{k}_{\vec{n}}\cdot\vec{x}_0} 
     + {} \right. \nonumber \\
    & \left. \qquad\qquad\qquad {} + (-1)^m \eexp^{-i \vec{k}_{\vec{n}} \cdot (\mat{M}^{\E{15}}_B \vec{x}_0 - \vec{T}^{\E{15}}_B)} \right] , \mbox{ for } \vec{n} \in \setN^{\E{15}}_{1A}, k_y = 0 , \nonumber \\
    \xi^{\E{15};\unitvec{k}_{\vec{n}}}_{k_{\vec{n}}\ell m} 
    &= \frac{1}{\sqrt{2}} i^\ell \eexp^{-i\vec{k}_{\vec{n}} \cdot \vec{T}^{\E{15}}_A/2} Y^*_{\ell m}(\unitvec{k}_{\vec{n}}) \left[ \eexp^{-i \vec{k}_{\vec{n}}\cdot\vec{x}_0} 
     + {} \right. \nonumber \\
    & \left. \qquad\qquad\qquad {} + (-1)^m \eexp^{-i \vec{k}_{\vec{n}} \cdot (\mat{M}^{\E{15}}_A \vec{x}_0 - \vec{T}^{\E{15}}_A)} \right] , \mbox{ for } \vec{n} \in \setN^{\E{15}}_{1B}, k_y > 0 , \nonumber \\
    \xi^{\E{15};\unitvec{k}_{\vec{n}}}_{k_{\vec{n}}\ell m} 
    &= i^\ell Y^*_{\ell m}(\unitvec{k}_{\vec{n}}) \eexp^{-i \vec{k}_{\vec{n}}\cdot\vec{x}_0} , \quad \mbox{for } \vec{n} \in \setN^{\E{15}}_{1AB}, k_y = 0 , 
\end{empheq}
\begin{empheq}[box=\fbox]{align}
    \xi^{\E{15};\unitvec{k}_{\vec{n}}}_{k_{\vec{n}}\ell m} 
     &= \frac{1}{\sqrt{4}} i^\ell \eexp^{-i\vec{k}_{\vec{n}} \cdot [(\mat{M}^{\E{15}}_A \vec{T}^{\E{15}}_B +\vec{T}^{\E{15}}_A)/2 + |\vec{k}_{\vec{n}} \cdot \vec{T}^{\E{15}}_2|/4]} \left[ Y^*_{\ell m}(\unitvec{k}_{\vec{n}})
        \left( \eexp^{-i \vec{k}_{\vec{n}}\cdot\vec{x}_0} \vphantom{\eexp^{-i \vec{k}_{\vec{n}} \cdot (\mat{M}^{\E{15}}_A}}
        + {} \right.\right. \nonumber \\
    & \left. \qquad\qquad\qquad\qquad {} + (-1)^m \eexp^{-i \vec{k}_{\vec{n}} \cdot (\mat{M}^{\E{15}}_A \mat{M}^{\E{15}}_B \vec{x_0} - \mat{M}^{\E{15}}_A \vec{T}^{\E{15}}_B - \vec{T}^{\E{15}}_A)} \right) + {} \nonumber \\
      & {} \qquad + Y_{\ell m}(\unitvec{k}_{\vec{n}}) \left(
        \eexp^{-i \vec{k}_{\vec{n}} \cdot (\mat{M}^{\E{15}}_A \vec{x_0} - \vec{T}^{\E{15}}_A)}
        \right. + {} \nonumber \\
      & {} \left. \left. \qquad \qquad \qquad \quad {} 
        + (-1)^m \eexp^{-i \vec{k}_{\vec{n}} \cdot (\mat{M}^{\E{15}}_B \vec{x_0} - \vec{T}^{\E{15}}_B)} \right) \right], \mbox{ for } \vec{n}\in \setN^{\E{15}}_2, k_y > 0 , \nonumber
\end{empheq}
and the harmonic space covariance matrix has the form \eqref{eqn:HarmonicCovarianceE13}.

\subsection{\E{17}: Slab space with a flip}

The eigenspectrum and eigenmodes of \E{17} can be determined from limits of \E{7}, \E{13}, or \E{14} in a manner similar to the chimney spaces.
Since there is only one compact direction, and remaining consistent with \E{7}, the discretization condition \eqref{eqn:associatedhomogeneousdiscretization} leads to the component of the wavevector
\begin{equation}
    (\vec{k}_{\vec{n}})_z = \frac{2\pi n_3}{2 L_z},
\end{equation}
with $k_x$ and $k_y$ unconstrained.

The eigenmodes of \E{17} now follow directly from those of \E{7}.
There are still two solutions to $(\mat{M}^{\E{17}}_A)^N \vec{k}_{\vec{n}} = \vec{k}_{\vec{n}}$ for $N=1$ and $N=2$ written explicitly as
\begin{description}
    \item[\textbf{$N=1$ eigenmodes: }]  $\vec{k}_{\vec{n}}=\transpose{((k_x, 0, (\vec{k}_{\vec{n}})_z)}$, 
        i.e., $\vec{n}=(n_3)$, $n_3 \in 2\integers$ where at least one of $n_3 \neq 0$ or $k_x \neq 0$, with
    \begin{empheq}[box=\fbox]{equation}
        \Upsilon^{\E{17}}_{\vec{n}}(\vec{x}) = \eexp^{i\vec{k}_{\vec{n}} \cdot (\vec{x}-\vec{x}_0)} = \eexp^{i(k_{x} (x-x_0) + i(\vec{k}_{\vec{n}})_{z} (z-z_0)}, 
    \end{empheq}
    \item[\textbf{$N=2$ eigenmodes: }] $\vec{k}_{\vec{n}}=\transpose{(k_x, k_y, (\vec{k}_{\vec{n}})_z)}$, 
        i.e., $\vec{n}=(n_3)$, $n_3 \in \integers$, $k_y \neq 0$, with
    \begin{empheq}[box=\fbox]{align}
        \Upsilon^{\E{17}}_{\vec{n}}(\vec{x}) &= \frac{1}{\sqrt{2}} \eexp^{-i(\vec{k}_{\vec{n}}\cdot \vec{T}^{\E{17}}_A/2 + |\vec{k}_{\vec{n}}\cdot \vec{T}^{\E{17}}_1|/4)}
        \left[ \eexp^{i\vec{k}_{\vec{n}}\cdot(\vec{x}-\vec{x}_0)} 
        + {} \right. \\
        & \left. \qquad\qquad {} + \eexp^{i\vec{k}_{\vec{n}}\cdot(\mat{M}^{\E{17}}_A(\vec{x}-\vec{x}_0)+\vec{T}^{\E{17}}_A)} \right]. \nonumber
    \end{empheq}
\end{description}
For the $N=2$ modes we have chosen the phase $\Phi^{\E{17}}_{\vec{k}} = -\vec{k}_{\vec{n}}\cdot \vec{T}^{\E{17}}_A/2 - |\vec{k}_{\vec{n}}\cdot \vec{T}^{\E{17}}_1|/4$ in \eqref{eqn:generaleigenmodeformula} to simplify the reality condition of the field.
As in \E{7} the two sets of allowed modes are defined by
\begin{empheq}[box=\fbox]{align}
    \setN^{\E{17}}_1 &= \{n_3\in 2\integers \}, \nonumber\\
    \setN^{\E{17}}_2 &= \{n_3 \in \integers\} , \\
    \setN^{\E{17}} &= \setN^{\E{17}}_1 \cup \setN^{\E{17}}_2 . \nonumber
\end{empheq}
Note that for $n_3\in \setN^{\E{17}}_1$ we require $k_y=0$ and either $n_3 \neq 0$ or $k_x \neq 0$ and for $n_3\in \setN^{\E{17}}_1$ we require $k_y > 0$.
With these the Fourier-mode correlation matrix can now be expressed as
\begin{empheq}[box=\fbox]{align}
\label{eqn:E17FourierCovarianceStandardConvention}
    C^{\E{17};XY}_{\vec{k}_{\vec{n}}\vec{k}_{\vecnp}} 
     = {} & (2\pi)^2 L_{\E{17}} \frac{2\pi^2}{k^3_{\vec{n}}}
        {\mathcal{P}}^{\mathcal{R}}(k_{\vec n})\Delta^X(\vec{k}_{\vec{n}})\DeltaYstar(\vec{k}_{\vec{n}'})
        \eexp^{i(\vec{k}_{\vecnp}-\vec{k}_{\vec{n}})\cdot\vec{x}_0} \times {} \nonumber \\
        & {}  \times
            \frac{1}{2}
                \smashoperator[l]{\sum_{\tilde{n} \in \setN^{\E{17}}_2}}
                \int_{-\infty}^{\infty} \dderiv \tilde{k}_x \, \int_{0}^{\infty} \dderiv\tilde{k}_y \, \sum_{a=0}^1\sum_{b=0}^1
                \eexp^{i\vec{k}_{\vec{\tilde{n}}}\cdot(\vec{T}^{(a)}-\vec{T}^{(b)})} 
                \Kdelta_{(\vec{k}_{\vec{n}})_z (\vec{k}^{(a)}_{\vec{\tilde{n}}})_z}
                \Kdelta_{(\vec{k}_{\vecnp})_z (\vec{k}^{(b)}_{\vec{\tilde{n}}})_z}
                \\
            & \qquad\quad {} \times
                 \Ddelta(k_x - \tilde{k}^{(a)}_x)
                 \Ddelta(k'_x - \tilde{k}^{(b)}_x)
                 \Ddelta(k_y - \tilde{k}^{(a)}_y)
                 \Ddelta(k'_y - \tilde{k}^{(b)}_y)
         , 
         \nonumber
\end{empheq}
where the terms with $n \in \setN^{\E{17}}_1$ are of measure zero and have been dropped, $L_{\E{17}}$ is given in \eqref{eqn:LE17}, $\vec{T}^{(0)}\equiv\vec{0}$, $\vec{T}^{(1)}\equiv\vec{T}^{\E{17}}_A$, and $\vec{k}^{(a)}_{\tilde{\vec{n}}} \equiv [(\mat{M}^{\E{17}}_A){}^T]^a \vec{k}_{\vec{\tilde{n}}}$, so that $\tilde{k}^{(a)}_x \equiv (\vec{k}^{(a)}_{\tilde{\vec{n}}})_x = \tilde{k}_x$, $\tilde{k}^{(a)}_y \equiv (\vec{k}^{(a)}_{\tilde{\vec{n}}})_y = (-1)^a \tilde{k}_y$, and $(\vec{k}^{(a)}_{\tilde{\vec{n}}})_z = (\vec{k}_{\tilde{\vec{n}}})_z$.
Again following \E{7} the eigenmodes in the harmonic basis are given by
\begin{empheq}[box=\fbox]{align}
    \xi^{\E{17};\unitvec{k}_{\vec{n}}}_{k_{\vec{n}}\ell m} 
    &= i^\ell  Y^*_{\ell m} (\unitvec{k}_{\vec{n}}) \eexp^{-i \vec{k}_{\vec{n}}\cdot\vec{x}_0} , \quad \mbox{for } \vec{n} \in \setN^{\E{17}}_1, k_y = 0 ,\\
    \xi^{\E{17};\unitvec{k}_{\vec{n}}}_{k_{\vec{n}}\ell m} 
    &= \frac{1}{\sqrt{2}} i^\ell \eexp^{-i(\vec{k}_{\vec{n}}\cdot \vec{T}^{\E{17}}_A/2 + |\vec{k}_{\vec{n}}\cdot \vec{T}^{\E{17}}_1|/4)} \left[ Y^*_{\ell m}(\unitvec{k}_{\vec{n}}) \eexp^{-i\vec{k}_{\vec{n}}\cdot \vec{x}_0}
     + {} \right. \nonumber \\
    & \left. \qquad\qquad\qquad\quad {} + Y_{\ell m} (\unitvec{k}_{\vec{n}}) \eexp^{-i\vec{k}_{\vec{n}}\cdot (\mat{M}^{\E{17}}_A \vec{x}_0 - \vec{T}^{\E{17}}_A)} \right], \mbox{for } \vec{n}\in \setN^{\E{17}}_2, k_y > 0 . \nonumber
\end{empheq}
The harmonic space covariance matrix is calculated from these eigenmodes is
\begin{empheq}[box=\fbox]{equation}
    \label{eqn:HarmonicCovarianceE17}
    C^{\E{17};XY}_{\ell m\ell'm'}  =
     \frac{(4\pi)^2}{{(2\pi)^2 L_{\E{17}}}}
    \sum_{n \in \setN^{\E{17}}} \int_{-\infty}^{\infty} \dderiv k_x \int_{0}^{\infty} \dderiv k_y \,
    \Delta^X_{\ell}(k_{\vec{n}})
    \DeltaYstarforell_{\ell'}(k_{\vec{n}})
    \frac{2\pi^2 \mathcal{P}^{\mathcal{R}}(k_{\vec{n}})}{k_{\vec{n}}^3}
    \xi^{\E{17};\unitvec{k}_{\vec{n}}}_{k_{\vec{n}} \ell m}
    \xi^{\E{17};\unitvec{k}_{\vec{n}}*}_{k_{\vec{n}} \ell' m'} .
\end{empheq}

\section{Numerical analysis}
\label{secn:numerical_results}

In the preceding section, we presented the covariance matrices for observables arising from scalar fluctuations across the non-orientable, Euclidean topologies of $E^3$, as functions of the parameters of the manifold for each topology.
This was done assuming that any isotropy violation is a result of the non-trivial topology and not microphysics.
As an application of these results, we focus on the CMB scalar temperature anisotropies resulting from Gaussian random scalar fluctuations at the epoch of last scattering and transfer functions
appropriate to  cosmological parameters consistent with the {\it Planck} results \cite{Planck:2019kim}.

In the standard isotropic covering space, the covariance simplifies to
\begin{equation}
 C_{\ell m\ell'm'}^{\E{18};TT} = \langle a_{\ell m}^{\E{18};T} a_{\ell' m'}^{\E{18};T*} \rangle = C_\ell^{\E{18};TT} \delta_{\ell\ell'}\delta_{mm'},
 \end{equation}
 i.e., only diagonal terms are non-zero and they are independent of $m$.
In contrast, non-orientable Euclidean topologies break isotropy and parity symmetries, resulting 
generically in all off-diagonal elements of the covariance matrix, potentially being non-zero, though residual symmetries may still make some elements vanish.

Following the methodology in \rcite{COMPACT:2023rkp}, where we computed correlation matrix elements for the compact, orientable Euclidean topologies \E{1}--\E{6}, we now extend this analysis to the compact, non-orientable topologies \E{7}--\E{10}.
We evaluate the covariance matrices for representative manifolds of these topologies and assess their distinguishability from the isotropic covering space by computing the Kullback-Leibler (KL) divergence.

\subsection{Evaluation of the CMB temperature covariance matrices}
\label{subsecn:numerical_covariances}

For non-trivial, fully compact, non-orientable Euclidean topologies, 
we use \cref{eqn:HarmonicCovariance}, for the auto-correlation of CMB temperature ($T$) fluctuations:
\begin{align}
\label{eqn:general_covariance_matrix}
    C^{\E{i};\,TT}_{\ell m\ell'm'} = 
   \frac{(4\pi)^2}{{V_{\E{i}}}}
    \sum_{\vec{n}\in \setN^{\E{i}}} 
    \Delta^{T}_{\ell}(k_{\vec{n}})
    \Delta^{T}_{\ell'}(k_{\vec{n}})
    \frac{2\pi^2 \mathcal{P}^{\mathcal{R}}(k_{\vec{n}})}{k_{\vec{n}}^3}
    \xi^{\E{i};\unitvec{k}_{\vec{n}}}_{k_{\vec{n}} \ell m}
    \xi^{\E{i};\unitvec{k}_{\vec{n}}*}_{k_{\vec{n}} \ell' m'}.
\end{align}
For brevity, we omit the $T$ and $TT$ labels in this section.
The infinite summation over wavevectors is infeasable numerically.
Extending our previous work on scalar eigenmodes in \rcite{COMPACT:2023rkp} to the non-orientable topologies $\E{7}$--$\E{10}$, we restrict the summation to a multipole-dependent maximum wavevector $|\vec{k}_{\mathrm{max}}(\ell)|$.
This cutoff is chosen to ensure sufficient accuracy for each topological scale and each topology  under consideration -- the larger the topology scale, the higher the necessary value of $|\vec{k}_{\mathrm{max}}(\ell)|$. 
The cutoff is determined from the ratio
\begin{equation}
    \label{eqn:cut_off_definition}
    R_\ell(|\vec{k}|) = \frac{C^{|\vec{k}|}_\ell}{C^{\Lambda\textrm{CDM}}_\ell},
\end{equation}
where
\begin{equation}
    C^{|\vec{k}|}_\ell = 4\pi \int^{|\vec{k}|}_0 \mathrm{d} k'\,\frac{\mathcal{P}^{\mathcal{R}}(k')}{k'}\Delta_\ell(k')^2,
\end{equation}
and $C^{\Lambda\textrm{CDM}}_{\ell}$ is the standard $\Lambda$CDM angular power spectrum generated by \texttt{CAMB} \cite{Lewis:1999bs,2011ascl.soft02026L} for $\E{18}$.
Consistent with prior studies in \rcite{COMPACT:2023rkp,Samandar:2025kuf}, we select $|\vec{k}_{\mathrm{max}}(\ell)|$ as the smallest $|\vec{k}|$ satisfying $R_\ell(|\vec{k}_{\mathrm{max}}(\ell)|) \geq 0.99$, i.e., the finite integral produces a power spectrum within 1\% of the true value as $L \to \infty$.
For off-diagonal elements ($\ell \neq \ell'$), we adopt $\max\!\left(|\vec{k}_{\mathrm{max}}(\ell)|,|\vec{k}_{\mathrm{max}}(\ell')|\right)$ as our cutoff, which turns out to be the same as $|\vec{k}_{\mathrm{max}}(\max(\ell, \ell'))|$.
Further precision beyond 99\% minimally affects KL divergence results.

We use the \textit{Planck} 2018 $\Lambda$CDM parameters \cite{Planck:2018vyg} to determine the primordial power spectrum $\mathcal{P}^{\mathcal{R}}(k)$  and as inputs to \texttt{CAMB} to compute the transfer function $\Delta_\ell(k)$.
It is worth noting that the pattern of non-zero elements in the covariance matrix depends on the orientation of the coordinate system,
though, since rotations mix only $a_{\ell m}$ of different $m$ within a given $\ell$, many aspects of the correlation matrix will stay the same, such as the vanishing of certain blocks of fixed  $\ell$ and $\ell'$.
However, since our focus is on comparing non-trivial topology correlations with the rotationally invariant covering space $\E{18}$, the KL divergence remains independent of coordinate orientation in the idealized case of noise-free observations over the full sky.

To facilitate comparison with the covering space, we plot the \emph{rescaled} covariance matrix:
\begin{equation}
    \Xi^{\E{i}}_{\ell m\ell'm'} = \frac{C^{\E{i}}_{\ell m\ell'm'}}{\sqrt{C^{\Lambda \mathrm{CDM}}_{\ell} C^{\Lambda \mathrm{CDM}}_{\ell'}}},
\end{equation}
which is critical for KL divergence analysis, as the eigenvalues of this matrix enter into its computation (see \cref{eq:DKLqp}).
The modulus of these rescaled $TT$ covariance matrices for topologies $\E{7}$--$\E{10}$ is presented in the first row of \cref{fig:E7_corr_kl,fig:E9_corr_kl,fig:E8_corr_kl,fig:E10_corr_kl}, using $\ell$-ordering indexed as $s = \ell(\ell+1) + m$, with $-\ell \leq m \leq \ell$.
Results are shown for two observer positions: one  on the reflection plane(s) of the manifold and another at a distance from the reflection plane(s).
The selected topological scales ensure no detectable matched circle pairs in the CMB for these observers, consistent with existing constraints from \rcite{Cornish:1997ab,Cornish:2003db,ShapiroKey:2006hm,Vaudrevange:2012da,Planck:2013okc,Planck:2015gmu}. 

\subsection{KL divergence}

To assess the detectability of the Universe having a non-trivial topology, we evaluate whether the probability distributions of CMB fluctuations, denoted $p({a_{\ell m}})$ for non-trivial topologies and $q({a_{\ell m}})$ for the covering space, are distinguishable.
The KL divergence, or the average cross-entropy, quantifies the information lost when approximating the true distribution ($p$) with a different distribution ($q$) and is defined as \cite{kullback1951, kullback1959information}
\begin{equation}
    D_{\mathrm{KL}}(p || q) = \int \mathrm{d}\{a_{\ell m}\} \,\, p(\{a_{\ell m}\}) \ln\! \left[\frac{p(\{a_{\ell m}\})}{q(\{a_{\ell m}\})} \right] .
\end{equation}

The KL divergence can also be understood as the expected value of the log-Bayes factor between models $p$ and $q$ assuming that data follow model $p$.
A commonly used convention is to regard $D_{\mathrm{KL}}\geq 1$ as a threshold of distinguishability. This value provides a quantitative measure of the detectability of non-trivial topology in an ideal experiment with no noise, foreground emission, or masking.

For the CMB coefficients $a_{\ell m}$, which follow zero-mean Gaussian distributions, the KL divergence simplifies to\footnote{This corrects an error in \cite{COMPACT:2023rkp,Samandar:2025kuf} which incorrectly used $1/\lambda_j$ instead of $\lambda_j$ in $D_{\mathrm{KL}}(p || q)$.}
\begin{equation}
\label{eq:DKLqp}
    D_{\mathrm{KL}}(p || q) = \frac{1}{2} \sum_j \left(\lambda_j - \ln \lambda_j - 1 \right), 
\end{equation}
where the $\lambda_j$ are the eigenvalues of the matrix $ \Xi^{\E{i}}_{\ell m\ell'm'}$.
Reversing the question, i.e., reversing the roles of $p$ and $q$, produces the KL divergence $D_{\mathrm{KL}}(q || p)$ which is computed as in \eqref{eq:DKLqp} using the inverse of the eigenvalues: $\lambda_j \to 1/\lambda_j$.
In this study, we focus on $D_{\mathrm{KL}}(p || q)$ to analyze the detectability of non-trivial topologies against the $\Lambda$CDM model.
For further details, see \rcite{COMPACT:2023rkp,Samandar:2025kuf}.

\subsection{Results and discussion}
\label{subsecn:numerical_results}

To interpret the results more intuitively, we choose the observer of the CMB to be located at the origin of the coordinate system.
For convenience, we have also chosen to consider the same pair of $\vec{x}_0$ for all four compact topologies.
As noted above, only the components of $\vec{x}_0$ off the reflections plane(s) have observable consequences: affecting the symmetries, clone patterns, etc.
This means that only the $y$-components affects \E{7} and \E{9}, whereas both the $x$- and $y$-components affect \E{8} and \E{10}.
In none of these topologies is the value of the $z$-component significant so it has been set to zero.

\begin{figure}[htbp]
\caption*{\E{7}}
    \centering
    \begin{subfigure}[t]{0.5\textwidth}
        \centering
        \includegraphics[width=\textwidth]{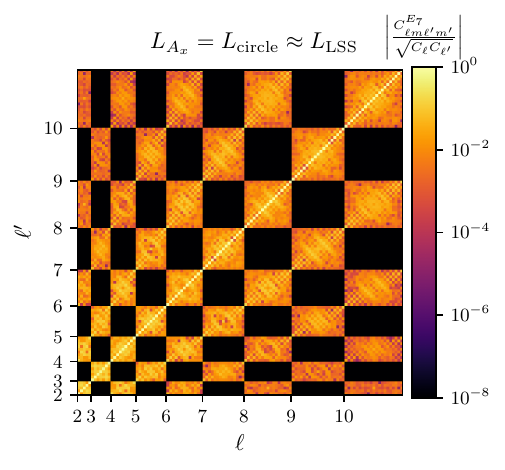}
        \caption{$\vec{x}_0 = (0, 0, 0)$}
    \end{subfigure}%
    \begin{subfigure}[t]{0.5\textwidth}
        \centering
        \includegraphics[width=\textwidth]{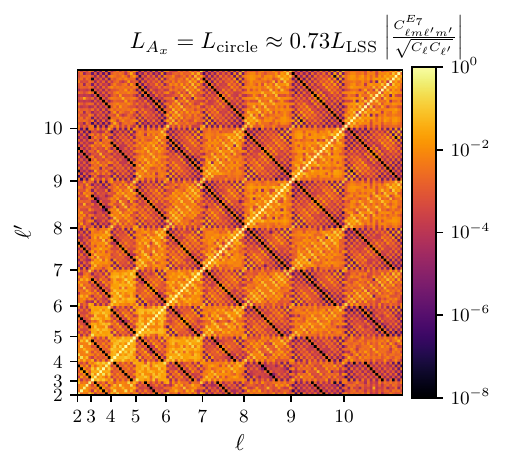}
        \caption{$\vec{x}_0 = (-0.1, 0.34, 0)$}
    \end{subfigure}
    \begin{subfigure}[t]{0.9\textwidth}
        \centering
        \includegraphics[width=\textwidth]{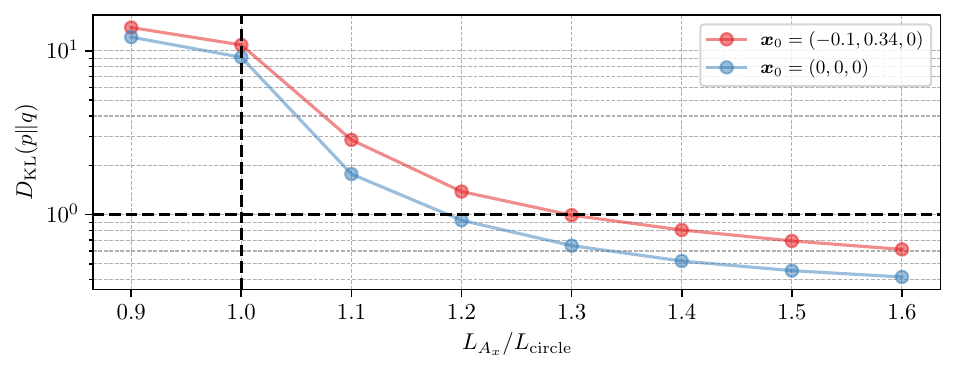}
        \caption{KL divergence vs $L_{Ax}$}
    \end{subfigure}
    \caption{Upper panels: Absolute values of the rescaled $TT$ CMB covariance matrix, $\Xi^{\E{7}}_{\ell m \ell' m'}$, for the $\E{7}$ topology at low multipoles ($\ell_{\mathrm{max}} = 10$).
    The manifold parameters are set to $L_{Ay} = 0$, $L_{2x} = 0.7\LLSS$, $L_{2z} = \LLSS$, $L_{1y} = 1.4\LLSS$, and $L_{Ax} = \Lcircle$.
    The left panel corresponds to an on-plane observer at $\vec{x}_0 = (0, 0, 0)$, with $\Lcircle = \LLSS$, while the right panel shows results for an off-plane observer at $\vec{x}_0 = (-0.1, 0.34, 0)\LLSS$, where $\Lcircle \approx 0.73\LLSS$.
    Lower panels: KL divergence for the \E{7} topology as a function of $L_{Ax} / \Lcircle$, computed up to $\ell_{\mathrm{max}} = 30$.
    Circles indicate calculated data points, and solid lines connect them for visual guidance.}
    \label{fig:E7_corr_kl}
\end{figure}

\begin{figure}[htbp]
\caption*{\E{9}}
    \centering
    \begin{subfigure}[t]{0.5\textwidth}
        \centering
        \includegraphics[width=\textwidth]{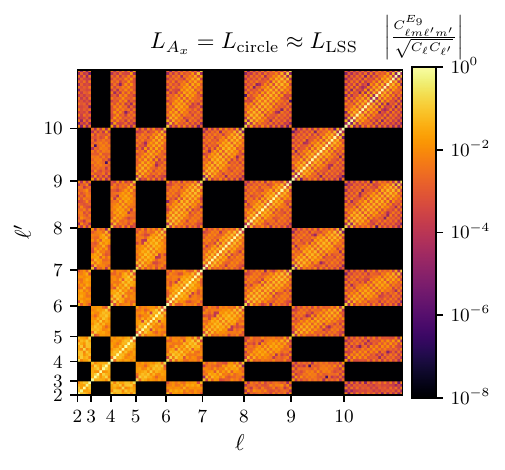}
        \caption{$\vec{x}_0 = (0, 0, 0)$}
    \end{subfigure}%
    \begin{subfigure}[t]{0.5\textwidth}
        \centering
        \includegraphics[width=\textwidth]{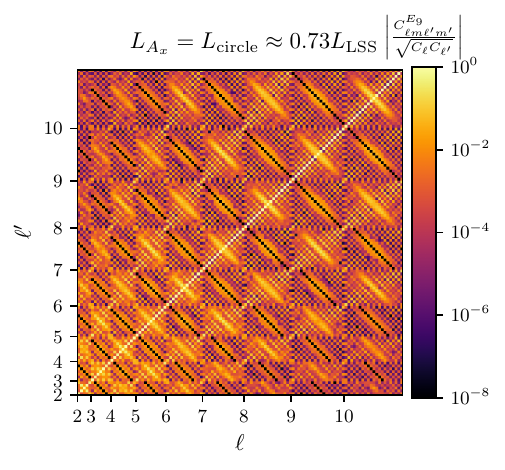}
        \caption{$\vec{x}_0 = (-0.1, 0.34, 0)$}
    \end{subfigure}
    \begin{subfigure}[t]{0.9\textwidth}
        \centering
        \includegraphics[width=\textwidth]{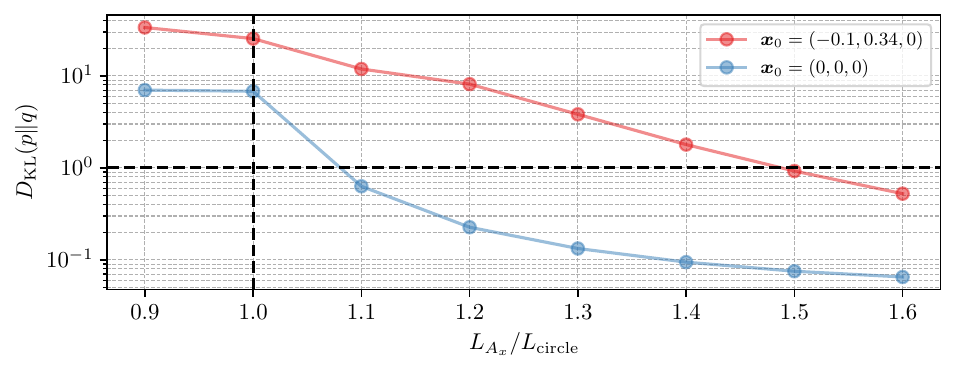}
        \caption{KL divergence vs $L_{Ax}$}
    \end{subfigure}
        \caption{Same as \cref{fig:E7_corr_kl}, but for the \E{9} topology.
        The only difference in manifold parameters compared to \E{7} is that the $y$-component of the generator $\vec{T}^{\E{9}}_{B_2}$ is $L_{1y}/2$ instead of zero.
        All other parameters, observer positions remain unchanged. Unlike the other cases, for the off-axis point, the KL divergence is computed up to $\ell_{\mathrm{max}} = 60$, since it saturates at this value.}
    \label{fig:E9_corr_kl}
\end{figure}

For the two topologies \E{7} and \E{9} (upper panels of \cref{fig:E7_corr_kl,fig:E9_corr_kl}), we set the manifold parameters as follows: $L_{Ay} = 0$, $L_{2x} = 0.7\LLSS$, $L_{2z} = \LLSS$, $L_{1y} = 1.4\LLSS$, and $L_{Ax} = \Lcircle$, where $\Lcircle$ is defined as the minimum value of $L_{Ax}$ that ensures, for this set of parameters and the given $\vec{x}_0$, that no pairs of matched circles appear on the CMB sky as observed from the coordinate origin. 
Our choice of manifold parameters is not intended to capture generic properties but to highlight distinctive features of these manifolds and the rich phenomenology of non-orientable topologies through illustrative, \emph{but not necessarily representative}, examples. 
To have a consistent comparison across the \E{7}--\E{10} topologies, we adopt similar parameter choices for both on-plane and off-plane observers, keeping the fundamental domain volume fixed for the same choice of $L_{A_x}$ in all cases.
Specifically, for all cases we set $\Lcircle = \LLSS$ for the on-plane observer at $\vec{x}_0 = (0, 0, 0)$ (left panel), and $\Lcircle \approx 0.73 \LLSS$ for the off-plane observer at $\vec{x}_0 = (-0.1, 0.34, 0)\LLSS$ (right panel).
In addition, we focus on configurations in which the off-plane observer has a shorter distance to the nearest clone than in the corresponding on-plane case for the same $L_{A_x}/\Lcircle$.

The first row of \cref{fig:E7_corr_kl,fig:E9_corr_kl} shows the rescaled temperature covariance matrices for \E{7} and \E{9}.
As extensively discussed in \rcite{COMPACT:2024cud,Samandar:2025kuf} regarding symmetries in the CMB correlations, isotropy violation is required to induce non-zero elements in the off-diagonal elements of the $TT$ covariance matrix. Additionally, non-zero elements in $\ell + \ell'$-odd blocks can only be produced by parity violation (combined with a violation of isotropy). 
When the observer is located on the reflection plane (left panels), the $TT$ correlations conserve parity, even though the manifold itself is parity-violating.
In this case, correlations vanish for odd values of $(\ell + \ell')$, while non-zero correlations appear for even $(\ell + \ell')$.
This results from the symmetric nature of the clone pattern relative to the reflection plane for this observer.
However, when the observer is moved from the reflection plane, the correlations become parity-violating, resulting in non-zero values for all $(\ell + \ell')$ combinations.

To quantify the information content of these correlations, we compute the KL divergence $D_{\mathrm{KL}}(p || q)$, shown in the lower panel of \cref{fig:E7_corr_kl,fig:E9_corr_kl}.
The KL divergence is evaluated up to $\ell_{\mathrm{max}} = 30$, using the same manifold parameters and observer positions as in the upper panels (which show covariance matrices for $\ell = 2-10$).
In this analysis, we vary the manifold parameter $L_{Ax}$ in units of $\Lcircle$, while keeping all other parameters fixed.
The two statistics, $D_{\mathrm{KL}}(p || q)$ and $D_{\mathrm{KL}}(q || p)$, where $p$ is the probability distribution for the non-trivial topology and $q$ corresponds to the probability distribution for the covering space, exhibit similar behavior and convey comparable information.
Consequently, we present only $D_{\mathrm{KL}}(p || q)$.

As evident in the lower panels of \cref{fig:E7_corr_kl,fig:E9_corr_kl},
when $L_{Ax} < \Lcircle$ matched circles appear in the CMB sky, resulting in a large KL divergence.
As $L_{Ax}$ increases, the KL divergence gradually decreases.
Once $L_{Ax} \geq \Lcircle$, the amount of information is still significant but matched circles are no longer present.
With further increases in $L_{Ax}$, the KL divergence eventually falls below the standard threshold for model distinguishability (i.e., $D_{\mathrm{KL}}(p || q) = 1$), for both on-plane and off-plane observers in \E{7} and \E{9}.

For this parameter configuration, the KL divergence is generally higher for the off-plane observer than for the on-plane observer in both \E{7} and \E{9}.
Furthermore, in \E{9}, the KL divergence for the off-plane observer is significantly larger than in \E{7}.
As discussed in \cref{secn:topologyE7,secn:topologyE9}, the only structural difference between \E{9} and \E{7} is that the $y$-component of the generator $\vec{T}^{\E{9}}_{2}$ is $L_{1y}/2$, whereas in \E{7} it is zero.
Despite the overall similarity between the two topologies, this seemingly minor change in a single translation vector causes the nearest-clone distance in \E{9} to increase much more slowly with varying $L_{A_x}$ for our specific choice of $\vec{x}_0$.
As a result, the KL divergence is significantly enhanced for any given $L_{A_x}$ in the off-plane observer case of \E{9}.
Notably, the KL divergence in this case remains above the detectability threshold ($D_{\mathrm{KL}}> 1$) up to $L_{Ax}/\Lcircle = 1.6$, which is remarkably high even compared to all other cases studied in this work and in \rcite{COMPACT:2023rkp} for orientable Euclidean topologies.

\begin{figure}[htbp]
\caption*{\E{8}}
    \centering
    \begin{subfigure}[t]{0.5\textwidth}
        \centering
        \includegraphics[width=\textwidth]{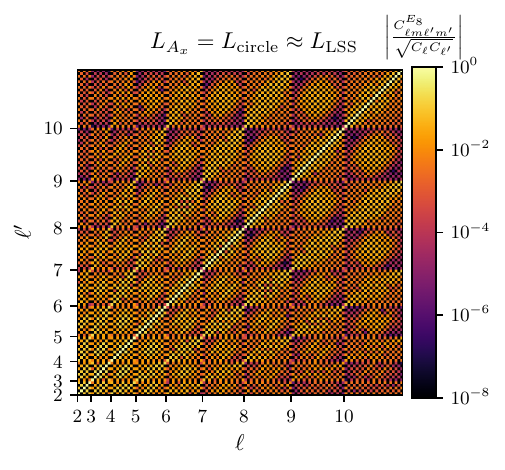}
        \caption{$\vec{x}_0 = (0, 0, 0)$}
    \end{subfigure}%
    \begin{subfigure}[t]{0.5\textwidth}
        \centering
        \includegraphics[width=\textwidth]{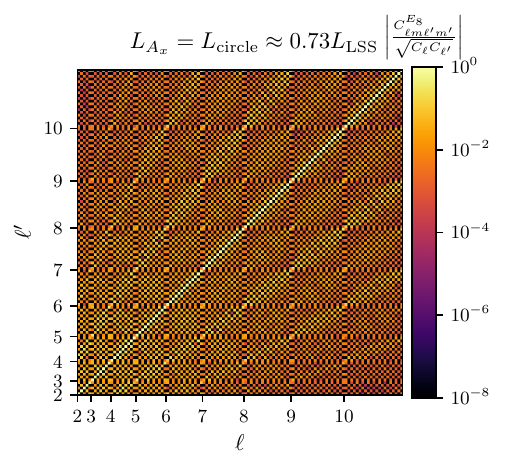}
        \caption{$\vec{x}_0 = (-0.1, 0.34, 0)$}
    \end{subfigure}
    \begin{subfigure}[t]{0.9\textwidth}
        \centering
        \includegraphics[width=\textwidth]{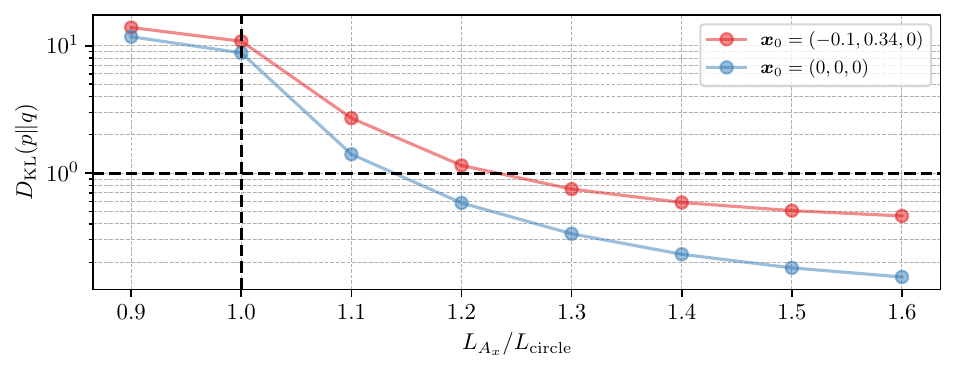}
        \caption{KL divergence vs $L_{Ax}$}
    \end{subfigure}
        \caption{Upper panels: Absolute values of the rescaled $TT$ CMB covariance matrix, $\Xi^{\E{8}}_{\ell m \ell' m'}$, for the $\E{8}$ topology at low multipoles ($\ell_{\mathrm{max}} = 10$).
        The manifold parameters are set to $L_{Ay} = 0$, $L_{Bx} = 0.7\LLSS$, $L_{Bz} = \LLSS$, $L_{Cy} = 1.4\LLSS$, and $L_{Ax} = \Lcircle$.
        The left panel corresponds to an on-plane observer located at $\vec{x}_0 = (0, 0, 0)$, with $\Lcircle = \LLSS$, while the right panel shows results for an off-plane observer at $\vec{x}_0 = (-0.1, 0.34, 0)\LLSS$, where $\Lcircle \approx 0.73\LLSS$.
        Lower panels: KL divergence for the $\E{8}$ topology as a function of $L_{Ax} / \Lcircle$, computed up to $\ell_{\mathrm{max}} = 30$.
        Circles indicate calculated data points, and solid lines connect them for visual guidance.}
    \label{fig:E8_corr_kl}
\end{figure}

\begin{figure}[htbp]
\caption*{\E{10}}
    \centering
    \begin{subfigure}[t]{0.5\textwidth}
        \centering
        \includegraphics[width=\textwidth]{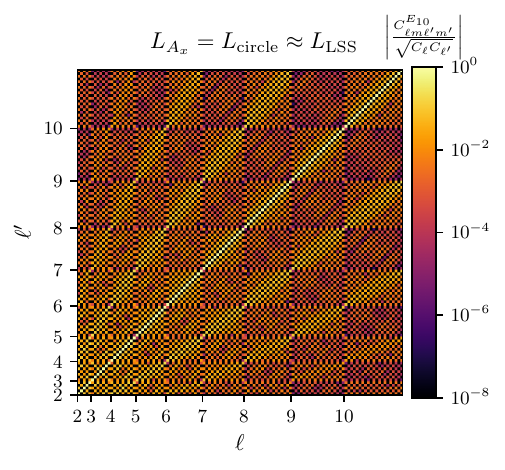}
        \caption{$\vec{x}_0 = (0, 0, 0)$}
    \end{subfigure}%
    \begin{subfigure}[t]{0.5\textwidth}
        \centering
        \includegraphics[width=\textwidth]{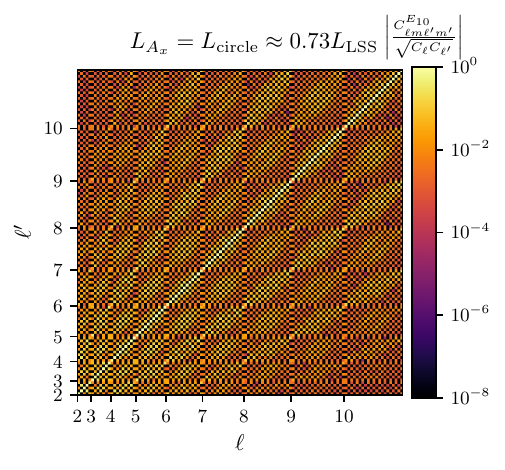}
        \caption{$\vec{x}_0 = (-0.1, 0.34, 0)\LLSS$}
    \end{subfigure}
    \begin{subfigure}[t]{0.9\textwidth}
        \centering
        \includegraphics[width=\textwidth]{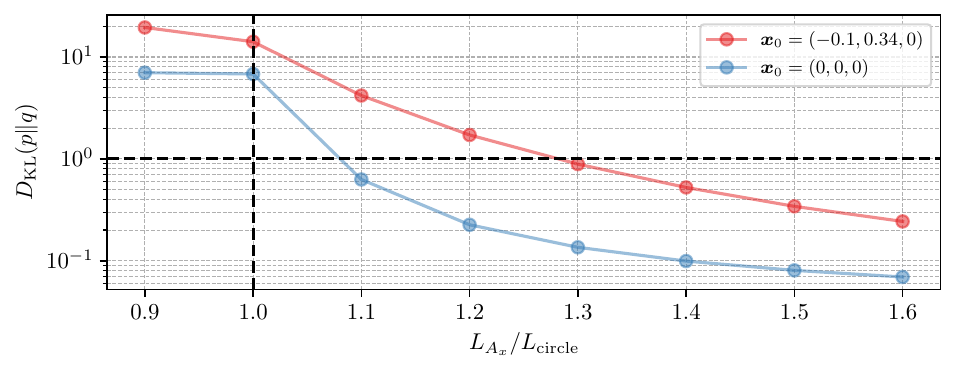}
        \caption{KL divergence vs $L_{Ax}$}
    \end{subfigure}
    \caption{Same as \cref{fig:E8_corr_kl}, but for the \E{10} topology.
    The only difference in manifold parameters compared to \E{8} is that the $y$-component of the generator $\vec{T}^{\E{10}}_{B}$ is $L_{Cy}/2$ instead of zero.
    All other parameters, observer positions, remain unchanged.}
    \label{fig:E10_corr_kl}
\end{figure}

For \E{8} and \E{10} (upper panels of \cref{fig:E8_corr_kl,fig:E10_corr_kl}), we adopt a parameter configuration analogous to that used for \E{7} and \E{9}.
Specifically, we set $L_{Ay} = 0$, $L_{Bx} = 0.7\LLSS$, $L_{B_z} = \LLSS$, $L_{Cy} = 1.4\LLSS$, and $L_{Ax} = \Lcircle$, with $\Lcircle = \LLSS$ for an on-plane observer at $\vec{x}_0 = (0, 0, 0)$ (left panel), and $\Lcircle \approx 0.73\LLSS$ for an off-plane observer, again at $\vec{x}_0 = (-0.1, 0.34, 0)\LLSS$ (right panel).

The first row of \cref{fig:E8_corr_kl,fig:E10_corr_kl} shows the rescaled temperature covariance matrices for \E{8} and \E{10}.
Unlike \E{7} and \E{9}, parity violation is evident even for an observer located on the reflection planes (left panels).
This arises from the non-zero translation along the $x$-direction ($L_{B_x}$) in $g^{\E{8}}_B$ and $g^{\E{10}}_B$, which is perpendicular to the $yz$ reflection plane of these generators. Consequently, when $g^{\E{8}}_B$ or $g^{\E{10}}_B$ is applied to the on-plane observer, the non-zero $L_{B_x}$ moves the observer away from the $yz$ reflection plane, thereby breaking the parity symmetry in the correlations.\footnote{A similar effect would occur with a non-zero $L_{A_y}$ in all four topologies: when $g^{\E{i}}_A$ is applied to the on-plane observer, the non-zero $L_{A_y}$ shifts the observer off of the $xz$ reflection plane, again making the correlations parity violating.}
When the observer is displaced from the reflection planes, the correlations remain parity-violating, leading to non-zero entries across all $(\ell + \ell')$ blocks, similar to the on-plane case.

The KL divergence $D_{\mathrm{KL}}(p || q)$ for \E{8} and \E{10} is shown in the lower panels of \cref{fig:E8_corr_kl,fig:E10_corr_kl}.
As in the previous cases, it is computed up to $\ell_{\mathrm{max}} = 30$, with $L_{Ax}$ varied in units of $\Lcircle$, while all other parameters are held fixed.
Again for this parameter configuration and consistent with the behavior observed in \E{7} and \E{9}, the KL divergence is generally higher for the off-plane observer than for the on-plane observer in both \E{8} and \E{10}.
Moreover, the KL divergence for the off-plane observer in \E{10} is larger than in \E{8}.
As discussed in \cref{secn:topologyE8,secn:topologyE10}, the only structural difference between these two topologies is that the $y$-component of the generator $\vec{T}^{\E{10}}_B$ is $L_{Cy}/2$, while in \E{8} it is zero.
Despite the similarity between the two manifolds, this seemingly minor difference in a single translation vector substantially boosts the KL divergence for the off-plane observer in \E{10}.

\begin{figure}[ht]
    \centering
    \begin{subfigure}[b]{\linewidth}
        \centering
        \includegraphics[width=\linewidth]{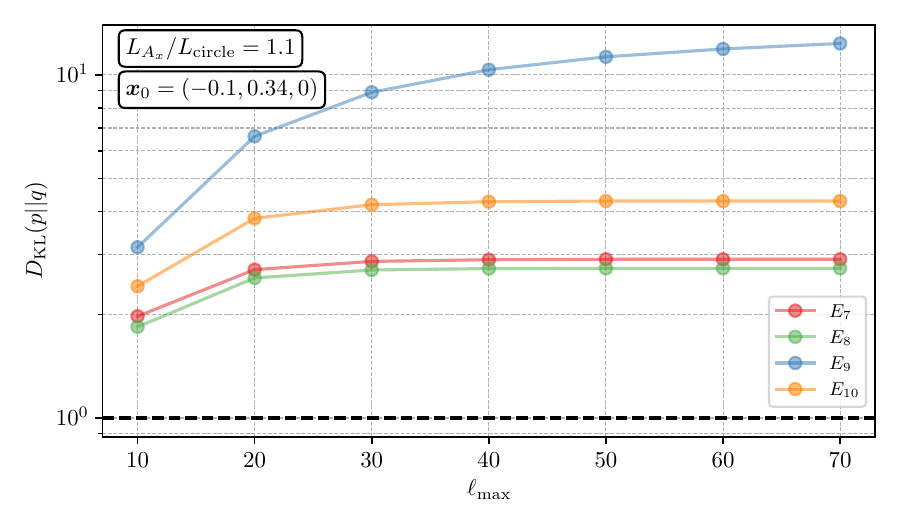}
    \end{subfigure}
 
    \caption{Comparison of $D_{\mathrm{KL}}(p|| q)$ as a function of $\ell_{\mathrm{max}}$ (from 10 to 70) for off-axis observers in \E{7}--\E{10} with $L_{A_x}=1.1L_{\mathrm{circle}}$. Circles represent computed data points. Notably, for the \E{9} topology, unlike the other cases where convergence occurs around $\ell_{\mathrm{max}} = 30$, it begins to converge at $\ell_{\mathrm{max}} \gtrsim 60$, due to the shorter distance to the nearest clone compared to the other cases.}
    \label{fig:E7_E8_E9_E10_kl}
\end{figure}

In \cref{fig:E7_E8_E9_E10_kl}, we compare the KL divergence as a function of $\ell_{\mathrm{max}}$ for the off-plane observers introduced in \cref{fig:E7_corr_kl,fig:E9_corr_kl,fig:E8_corr_kl,fig:E10_corr_kl}. These observers share the same fundamental domain volume, $L_{\mathrm{circle}} \approx 0.73$, which makes them an interesting case study. We extend the analysis up to $\ell_{\mathrm{max}} = 70$ to test the convergence behavior. Our results confirm that, in most cases, the KL divergence saturates around $\ell_{\mathrm{max}} = 30$, except for the \E{9} off-plane configuration, where convergence occurs only at $\ell_{\mathrm{max}} \gtrsim 60$.

This discrepancy originates from the distance to the nearest clone in each topology. When this distance is smaller than $L_{\mathrm{LSS}}$, no convergence occurs with increasing $\ell_{\mathrm{max}}$, since the observer can see circles in the sky and higher multipoles continue to provide more information. If the distance is close to $L_{\mathrm{LSS}}$, slightly smaller or slightly larger, the KL divergence eventually converges, but at a higher value of $\ell_{\mathrm{max}}$. For example, in the off-axis case of \E{9} with $L_{A_x} = 1.1 L_{\mathrm{circle}}$, the nearest-clone distance is $\approx 1.006$, and convergence is reached only at $\ell_{\mathrm{max}} \gtrsim 60$. By contrast, for the corresponding off-axis points of \E{7}, \E{8}, and \E{10}, where the nearest-clone distance is $\approx 1.05$, convergence occurs already at $\ell_{\mathrm{max}} = 30$.
Overall, our findings suggest that when the nearest-clone distance exceeds roughly $\approx 1.03$, the KL divergence reliably converges by $\ell_{\mathrm{max}} = 30$.

\begin{figure}[ht]
    \centering
    \begin{subfigure}[b]{\linewidth}
        \centering
        \includegraphics[width=\linewidth]{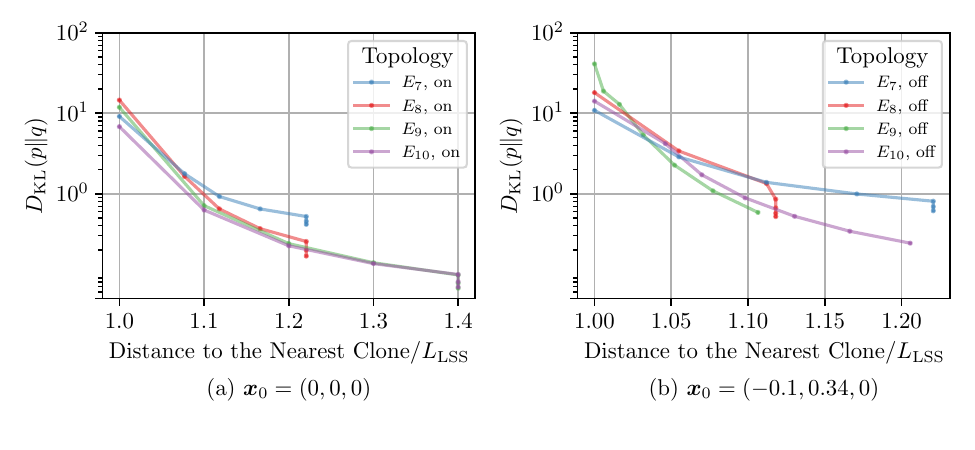}
    \end{subfigure}
 
    \caption{KL divergence $D_{\mathrm{KL}}(p || q)$ versus distance to the nearest clone (in units of $\LLSS$) for all manifolds of non-orientable Euclidean topologies \E{7}--\E{10} studied in this paper. Panel (a) corresponds to on-plane observers, and panel (b) corresponds to off-plane observers.}
    \label{fig:E7_E8_E9_E10_kl_nearest_clone_dist}
\end{figure}

In \cref{fig:E7_E8_E9_E10_kl_nearest_clone_dist}, we plot the KL divergence $D_{\mathrm{KL}}(p || q)$ versus the distance to the nearest clone (in units of $\LLSS$) for all example configurations of \E{7}--\E{10} considered in this paper, including both on-plane (panel a) and off-plane (panel b) observers.
The results show a clear trend across all topologies: the KL divergence decreases as the distance to the nearest clone increases. Moreover, in all topologies, there are cases with the same distance to the nearest clone but different manifold parameters, specifically here, different $L_{A_x}$. These points demonstrate that factors beyond the distance to the nearest clone can affect the information content of the topological correlations. In this analysis, we only vary $L_{A_x}$ for each curve, which serves as a representative for how parameters can alter the information content.
For instance in \E{8} and \E{10}, for points with the same nearest-clone distance along the same curve, increasing $L_{A_x}$ eventually changes the generator corresponding to the nearest clone from $g_A$ to $g_B$. 

Our limited study of the non-orientable Euclidean topologies indicates that the distance to the nearest clone is the principal factor in the information content of the temperature correlations. However, other factors, such as the number of nearest clones, the generators corresponding to the nearest clone, the distance to the second-nearest clone, and the pattern of the nearest clones, could potentially also be important for the information content of the correlations.
These factors will be studied in follow-up work performing Bayesian statistical inference for the topology of the Universe using the {\it Planck} data.

\section{Conclusion}
\label{secn:conclusion}

The local spatial geometry of the Universe may be consistent with Euclidean flatness when averaged over domains exceeding the scale of the largest cosmic structures but small relative to the Hubble volume or the surface of last scattering.
Nevertheless, this does not imply that its topology corresponds to the infinitely extended covering space of Euclidean geometry \Espace.
There are eighteen possible topologies (labeled \E{1}--\E{18}) for three-manifolds that have homogeneous flat local geometry; the covering space is just one of them -- \E{18}.
In previous papers we studied scalar and spin-2 modes in the ten orientable Euclidean topologies \cite{COMPACT:2023rkp,Samandar:2025kuf,COMPACT:2024cud}.
In this work we have studied the scalar modes in the remaining eight non-orientable Euclidean topologies.
These eight topologies of non-orientable manifolds fall into three classes: four with compact manifolds, i.e., all three spatial dimensions are compact, (\E{7}--\E{10}); three that have two compact dimensions (\E{13}--\E{15}); and one that has one compact dimension (\E{17}).

For each of these topologies (cf.\ \cref{secn:topologiesmanifolds-general}) we have, for the first time, provided a completely general parametrization of all the possible manifolds.
This builds on previous work (see especially \rcite{Riazuelo2004:prd}), which included all these topologies, but not, or not explicitly, with their most general parametrization.
To do so, we have allowed the full range of possibilities for the translation vectors associated with the generators of the manifolds, whereas in \rcite{Riazuelo2004:prd}, specific special cases were chosen for certain manifolds.

In deriving a general parametrization of the manifolds, it becomes clear that there are many choices possible for a minimal set of generators of each manifold (i.e., one with the mininum number of generators -- $3$ for \E{7}--\E{10}, $2$ for \E{13}--\E{15}, and $1$ for \E{17}).  
In general there is some freedom in how to choose the $O(3)$ matrices in the generators.
While this freedom is present in the orientable topologies, e.g., \E{2}--\E{6}, the standard choices seem natural so it was not discussed in \cite{COMPACT:2023rkp}).
In this work, we adopt choices of $O(3)$ elements for the generators that differ from those in \rcite{Riazuelo2004:prd}, while also presenting 
alternatives, including direct generalizations of their conventions in \cref{app:derivations}.
The standard naming of spaces, however, follows the mathematical conventions and those used in \rcite{Riazuelo2004:prd}, which often reflect particular symmetries tied to their selected translation vectors.
For consistency, we retain these standard names.

It is important for many purposes to identify the physically distinct values of parameters of manifolds for each topology.
For example, for \E{7}--\E{10}, it would appear, naively, that the Laplacian eigenmodes, correlation matrices, \ldots, and thus observables, for each manifold could depend on the nine components of the three vectors $\vec{T}^{\E{i}}_a$  associated with a minimal set of three generators, plus the vector ($-\vec{x}_0$) giving the location of the observer/origin relative to the planes of reflection or the axes of rotation associated with the generators. 
However, three of these twelve degrees of freedom represent the relative orientation of observer and  topology frames --
 we use them to simplify the $\vec{T}^{\E{i}}_a$ in each \E{i} and do not display the Euler angles of the rotations.
Of the six remaining components of $\vec{T}^{\E{i}}_a$ in each compact topology, one or more may be zero or may have a specific relationship with other components.
The number of independent parameters of the manifold is thus at most six.

The vector $\vec{x}_0$ parametrizes the choice of origin.
Certain components of $\vec{x}_0$ are redundant physically with components of the translation vectors -- changes in one cause changes in the other; the other components of $\vec{x}_0$ are non-physical.
Certain choices of the redundant components can therefore be used to ``simplify'' the translation vectors; we have avoided this simplification and included $\vec{x}_0$ in our expressions for eigenmodes and correlation matrices, allowing the reader to make the choices they prefer. 
We also provide a specific choice of the range of physically distinguishable values of the manifold parameters.

For each of these manifolds presented in \cref{secn:topologiesmanifolds}, we have, in \cref{secn:eigenmodes}, presented an analytic formula for: the eigenmodes of the scalar Laplacian,  the Fourier-mode correlation matrix $C^{\E{i};XY}_{\vec{k}_{\vec{n}}\vec{k}_{\vec{n}}'}$, 
where $X$ and $Y$ represent any scalar fields (e.g., temperature or scalar $E$-mode polarization in the context of the CMB), and the topology factors $\xi^{\E{i};\unitvec{k}_{\vec{n}}*}_{k_{\vec{n}} \ell' m'}$ that are needed for calculation of the harmonic component correlation matrix $C^{\E{i};XY}_{\ell m \ell' m'}$(under the assumption that the background metric and all relevant microphysics is isotropic).
These should fully equip the reader to make their own calculations of the statistical properties of observables sourced by the scalar perturbations in these topologies.

In \cref{secn:numerical_results} we begin presenting, topology by topology, numerical harmonic component correlation matrices $C^{\E{i};TT}_{\ell m \ell' m'}$ for small $\ell$ and $\ell'$ for the compact non-orientable topologies \E{7}--\E{10}. Our choices of manifold parameters are meant to highlight interesting features of these manifolds rather than to be particularly representative.
In doing so, we selected one set of manifold parameters for each topology, but made two choices for $\vec{x}_0$. In the first case, the observer is placed on the plane(s) of reflection associated with the isometry group elements to increase the symmetry of the correlation matrix. The corresponding results are shown in the left-hand panels of \cref{fig:E7_corr_kl,fig:E9_corr_kl,fig:E8_corr_kl,fig:E10_corr_kl}. In the second case, the observer is displaced from this point of symmetry to highlight these various interesting features of the correlation matrices. These results are shown in the right-hand panels of the same figures.

These examples demonstrate that the violation of statistical isotropy inherent in all non-trivial topologies fundamentally changes the $TT$ correlation matrices. In contrast to the diagonal matrices of the covering space, all elements of the correlation matrix can in principle be non-zero in non-orientable topologies.
We emphasize that these parity-violating elements are non-zero despite the absence of microphysical parity violation -- topology alone causes parity violation -- except in some cases where the observer is positioned on the reflection plane(s), where parity is conserved due to the symmetric clone pattern.

While the patterns in these correlation matrices are informative, the key question is whether they enable detection of cosmic topology. To address this, the second row of \cref{fig:E7_corr_kl,fig:E9_corr_kl,fig:E8_corr_kl,fig:E10_corr_kl} shows the KL divergence $D_{\mathrm{KL}}(p || q)$, where $p$ is the probability distribution for the non-trivial topology \E{i} ($i\in \{7,8,9,10\}$) and $q$ corresponds to the covering space \E{18} (assuming the same cosmological parameters). This represents the information potentially available from measured CMB temperature fluctuations to measure the probability of the topology of the Universe to be \E{18} if we assume the Universe has non-trivial topology \E{i}.
The KL divergence $D_{\mathrm{KL}}(p || q)$ is plotted against the value $L_{A_x}$, one of the several components of the translations in the generators $\vec{T}^{\E{i}}_a$, in units of $\Lcircle$.
Here $\Lcircle$ is the minimum value of $L_{A_x}$ would need for which the shortest closed path around the manifold through the origin has length equal to $\LLSS$, the diameter of the last scattering surface of CMB photons.
The so-called ``circles-in-the-sky'' limits on cosmic topology \rcite{Cornish:1997ab,Cornish:2003db,ShapiroKey:2006hm,Vaudrevange:2012da,Planck:2013okc,Planck:2015gmu,COMPACT:2022nsu,COMPACT:2024qni} from WMAP and {\it Planck} require that this distance is greater than $0.985\LLSS$.

The KL divergence $D_{\mathrm{KL}}$ is a widely used measure of distinguishability of models, and $D_{\mathrm{KL}}>1$ is a commonly employed threshold of distinguishability.  
Our results indicate that when $L_{Ax} < \Lcircle$, where the matched circles appear in the CMB sky, yielding large KL divergences.
As $L_{Ax}$ increases beyond $\Lcircle$, the KL divergence decreases, but remains significant initially. For the topology cases studied in this work, the KL divergence is higher for off-plane observers than for on-plane observers in all topologies \E{7}--\E{10}.
Moreover, the KL divergence is notably enhanced in \E{9} and \E{10} compared to \E{7} and \E{8} for off-plane observers, due to subtle differences in translation vectors that affect the nearest-clone distance.
In all cases, however, the KL divergence falls below the distinguishability threshold of 1 once $L_{Ax}/\Lcircle$ reaches between 1.1 and 1.6, with non-orientable topologies exhibiting richer phenomenology through parity violation and position-dependent effects.
These correlation matrices and KL divergences are meant to be illustrative and not necessarily representative of the full range of behaviors one might encounter across the \E{i} parameter spaces.

In \cref{fig:E7_E8_E9_E10_kl}, we compared the KL divergence as a function of $\ell_{\mathrm{max}}$ for off-plane observers across topologies, with a fixed fundamental domain volume at $L_{A_x} = 1.1 L_{\mathrm{circle}}$. Convergence typically occurs around $\ell_{\mathrm{max}} = 30$, except for the \E{9} configuration, which requires $\ell_{\mathrm{max}} = 60$ due to its nearest-clone distance ($\approx 1.006\LLSS$) being closer to $\LLSS$ than in \E{7}, \E{8}, and \E{10} ($\approx 1.05\LLSS$). This highlights that convergence depends on the nearest-clone distance relative to $\LLSS$; when the distance is smaller or only marginally larger than $\LLSS$, higher multipoles are needed for the KL divergence to converge.

This study provides key insights into the information content of topological temperature correlations in non-orientable Euclidean topologies \E{7} through \E{10}. As shown in \cref{fig:E7_E8_E9_E10_kl_nearest_clone_dist}, which plots the KL divergence $D_{\mathrm{KL}}(p || q)$ against the distance to the nearest clone (in units of $\LLSS$) for various configurations, including both on-plane (panel a) and off-plane (panel b) observers, the KL divergence consistently decreases as the distance to the nearest clone increases across all examined topologies.

When the distance to the nearest clone is smaller than the diameter of the LSS, the topological information in the CMB is substantial, as expected. Notably, given existing CMB constraints indicating that the shortest distance around the Universe through us exceeds $\approx 0.985\LLSS$ 
(at $\sim 95\%$ confidence \rcite{Cornish:2003db,Cornish:2011ys}), the KL divergence remains above 1 for nearest-clone distances slightly larger than the LSS diameter. Depending on the topology and observer position, distinguishability from the trivial topology persists up to nearest-clone distances of 1.2 in units of $\LLSS$.

In summary, our analysis confirms that the distance to the nearest clone is the primary factor influencing the information content of topological temperature correlations. However, other factors, such as the number of nearest clones, their associated generators, the distance to the second-nearest clone, and the pattern of clones, may also play significant roles. These findings set the stage for future investigations, including comprehensive Bayesian statistical inference of the topology of the Universe using {\it Planck} data, extending prior work \rcite{Planck:2015gmu,Planck:2013okc} beyond cubic \E{1} and \slabh(unrotated slab space) topologies to all topologies with their full parameter spaces. Such studies will systematically evaluate these secondary factors, potentially reducing the parameter space and enabling more efficient topological inference for the Universe.

\acknowledgments
We thank Jeffrey Weeks for valuable conversations.
G.D.S. and Y.A. thank Miguel Montero for valuable conversations.
J.C.D. is supported by the Spanish Research Agency (Agencia Estatal de Investigaci\'on), the Ministerio de Ciencia, Inovaci\'on y Universidades, and the European Social Funds through grant JDC2023-052152-I, as part of the Juan de la Cierva program.
Y.A. acknowledges support by the Spanish Research Agency (Agencia Estatal de Investigaci\'on)'s grant RYC2020-030193-I/AEI/10.13039/501100011033, by the European Social Fund (Fondo Social Europeo) through the  Ram\'{o}n y Cajal program within the State Plan for Scientific and Technical Research and Innovation (Plan Estatal de Investigaci\'on Cient\'ifica y T\'ecnica y de Innovaci\'on) 2017-2020, by the Spanish Research Agency through the grant IFT Centro de Excelencia Severo Ochoa No CEX2020-001007-S funded by MCIN/AEI/10.13039/501100011033, by the Spanish National Research Council (CSIC) through the Talent Attraction grant 20225AT025, and by the Spanish Research Agency's Consolidaci\'on Investigadora
2024 grant CNS2024-154430.
C.J.C., G.D.S., A.K., and D.P.M.\ acknowledge partial support from NASA ATP grant RES240737 and from NASA ADAP grant 24-ADAP24-0018; G.D.S. and Y.A.\ from the Simons Foundation; A.S. and G.D.S.\ from DOE grant DESC0009946; G.D.S., Y.A., and A.H.J.\ from the Royal Society (UK); and A.H.J.\ from STFC in the UK\@.
F.C.G\ is supported by the Presidential Society of STEM Postdoctoral Fellowship at Case Western Reserve University and by Ministerio de Ciencia, Innovaci\'on y Universidades, Spain, through a Beatriz Galindo Junior grant BG23/00061.
J.R.E. acknowledges support from the European Research Council under the Horizon 2020 Research and Innovation Programme (Grant Agreement No.~819478).
A.N.\ is supported by the Richard S.\ Morrison Fellowship.
T.S.P is supported by Funda\c{c}\~ao Arauc\'aria (NAPI Fen\^omenos Extremos do Universo, grant 347/2024 PD\&I). J.C.D. and A.T. are supported by CSIC through grant No. 20225AT025. A.T.\ was also supported by the Richard S.\ Morrison Fellowship. 
This work made use of the High-Performance Computing Resource in the Core Facility for Advanced Research Computing at Case Western Reserve University and the facilities of the Ohio Supercomputing Center.

\appendix

\section{Appendix: Construction of general generators}
\label{app:derivations}

In this work, the most general allowed set of three generators $g_{a_j}$ for each of the non-orientable \E{i} has been provided.
Here we describe a constructive, algebraic approach for determining the parametrization of these generators.

In contrast to the orientable manifolds, the topology of each non-orientable manifold cannot be set just by the choice of elements of $O(3)$, i.e., by the matrices $\mat{M}^{\E{i}}_a$.
Identical choices of the $\mat{M}^{\E{i}}_a \in O(3)$ can lead to distinct sets of generators that embody distinct symmetries.
This is in contrast to the orientable manifolds \cite{COMPACT:2023rkp} where the choice of elements of $SO(3)$ uniquely determine the topology.\footnote{
    This does not preclude other choices.
    For example, the Hantzsche-Wendt space (\E{6}), can be represented by generators with half-turns around each of the orthogonal coordinate axes, or as a two generators with a half-turn around one axis and one generator with a half-turn about one of the orthogonal axes.
    Regardless, either of these choices is distinct from the other five compact, orientable topologies, \E{1}--\E{5}, in contrast to what is found for the non-orientable spaces.
}
Once a choice of matrices is made, there is limited freedom to adjust the $\mat{M}^{\E{i}}_a$: only the freedom to rotate the coordinate system to choose the orientations of the axes of rotation or the planes of reflection.
The remainder and essence of our task will then be to determine the most general allowed vectors $\vec{T}^{\E{i}}_{a_j}$ associated with the $\mat{M}^{\E{i}}_a$ of each \E{i}.
After the orientation of the coordinate system has been set, any remaining freedom will be used to simplify the three vectors $\vec{T}^{\E{i}}_{a_j}$.
There are two principal tools available to constrain the components of $\vec{T}^{\E{i}}_{a_j}$:
\begin{enumerate}
    \item Any finite sequence of generators and their inverses are group elements.
    Functionally, this is enforced by checking that any such sequence that is a pure translation is an integer linear combination of some ``basis set'' of three linearly independent translations, so that the group is a discrete group, $\Gamma^{\E{i}}$.
    In \cref{secn:topologiesmanifolds} we call this basis set (the generators of) ``the associated \E{1}''.
    Part of this process is the determination of the set of such pure translations that can be chosen as the basis set.
    
    For example, consider two generators $g^{\E{n}}_{ai}$ and $g^{\E{n}}_{bj}$ of $\Gamma^{\E{n}}$ for one of the orientable Euclidean manifolds $\E{n}$, associated with the $SO(3)$ elements $\mat{M}^{\E{n}}_a$ and $\mat{M}^{\E{n}}_b$, respectively, and let $\vec{T}^{\E{n}}_k$ ($k=1,2,3$) represent the pure translations of the associated \E{1}.
    Since $(\mat{M}^{\E{n}}_a)^{-1} (\mat{M}^{\E{n}}_b)^{-1} \mat{M}^{\E{n}}_a \mat{M}^{\E{n}}_b=\identity$ for all $a$ and $b$ in all such $\Gamma^{\E{n}}$,\footnote{
        Note that this is not true for arbitrary elements of $O(3)$ since it is a non-abelian group.
        However, it is true for the particular elements of $O(3)$ used in the generators for each of the \E{n}.}
    we must insist that 
    \begin{equation}
        (g^{\E{n}}_{ai})^{-1} (g^{\E{n}}_{bj})^{-1} g^{\E{n}}_{ai} g^{\E{n}}_{bj}: \vec{x} \to \vec{x} + \sum_{k=1}^3 m_k \vec{T}^{\E{n}}_k
        \label{eqn:topologycondition}
    \end{equation}
    for some triplet of integers $m_k$.
    
    This may not be sufficient.
    Consider the case where $\mat{M}^{\E{n}}_a\neq \identity$ but $(\mat{M}^{\E{n}}_a)^2 = \identity$ and where there are (at least) two generators associated with this matrix.
    Then the application of any two of these generators should result in a pure translation.
    In other words, all combinations of the form
    \begin{equation}
        g^{\E{n}}_{ai} g^{\E{n}}_{aj}, \quad g^{\E{n}}_{a{i}} (g^{\E{n}}_{a{j}})^{-1}, \quad (g^{\E{n}}_{a{i}})^{-1} g^{\E{n}}_{a{j}}, \quad (g^{\E{n}}_{a{i}})^{-1} (g^{\E{n}}_{a{j}})^{-1}
    \end{equation}
    for each of $i,j\in \{1, 2\}$ must lead to pure translations that are integer linear combinations of the basis set.
    If this is true, then \eqref{eqn:topologycondition} is trivially satisfied.
    For some topologies two of these combinations will be used to define the translation vectors and the rest will lead to constraints that must be satisfied.
    \item We must also ensure that the set of transformations (the group elements) consists only of freely acting transformations, i.e., no transformation (other than the identity transformation) has a fixed point.
\end{enumerate}
    
Once these conditions have been enforced, we may find that there appear to be distinct sets of ``solutions'', i.e., parametrizations of the $\vec{T}^{\E{i}}_{a_j}$ that cannot be transformed into one another by rotations, reorderings, or rescalings.
We must still prove that two such sets do not generate the same lattice of clones for a given starting point.
A simple way that this can happen is if the $\vec{T}^{\E{i}}_{a_j}$ of one set are just integer linear combinations of the vectors of the other set.

We will that we are able to bring this program to a successful conclusion for each of the non-orientable manifolds.

\subsection{\E{7} and \E{9}: Klein spaces without and with a vertical flip}
\label{app:E7E9}

As noted above, the Klein spaces are not uniquely determined by their choice of $O(3)$ matrices.
The $O(3)$ structure of the two Klein spaces \E{7} and \E{9} is the same.
Conventionally (c.f.\ \cite{Riazuelo2004:prd}) for \E{7} the generators are written using the matrices
\begin{equation}
    \mat{M}^{\E{7}}_A = \diag(1, -1, 1) \mbox{ and } \mat{M}^{\E{7}}_B = \identity,
\end{equation}
with two generators $g^{\E{7}}_{\A{j}}$ and one generator $g^{\E{7}}_B$; whereas for \E{9} the single matrix
\begin{equation}
    \mat{M}^{\E{9}}_A = \diag(1, -1, 1)
\end{equation}
has three generators $g^{\E{9}}_{\A{j}}$.
However, there are other choices.

Notice that we can define another generator associated with $\mat{M}^{\E{7}}_A$ as
\begin{equation}
    g^{\E{7}}_{\A{3}} \equiv g^{\E{7}}_{\A{1}} g^{\E{7}}_B
\end{equation}
and use this in place of $g^{\E{7}}_B$.
The set $\{ g^{\E{7}}_{\A{j}}| j=1, 2, \mbox{ or } 3 \} \in \Gamma^{\E{7}}$.
Equivalently, for \E{9} we note that $\mat{M}^{\E{9}}_B \equiv (\mat{M}^{\E{9}}_A)^2 = \identity$, define
\begin{equation}
    g^{\E{9}}_B \equiv g^{\E{9}}_{\A{2}} g^{\E{9}}_{\A{3}} ,
\end{equation}
and use this in place of $g^{\E{9}}_{\A{3}}$.
The set $\{ g^{\E{9}}_{\A{1}}, g^{\E{9}}_{\A{2}}, g^{\E{9}}_B \} \in \Gamma^{\E{9}}$.

There are even more possibilities.
One of particular use is to have one generator a glide reflection and two generators pure translations.
To proceed in this manner we use the rotational freedom to choose the flip to be across the $xz$-plane: this fixes the $y$-axis.
We thus choose as our starting form\footnote{
  Here we drop topology labels for convenience and since the results will apply to two different topologies.
  The labels will be restored when the results are related to the standard naming conventions.
  Further, we will use $\mat{M}_A$ to represent the flip and $\mat{M}_B$ to represent the translations, as is conventional in the Klein spaces.
}
\begin{align}
    & \mat{M}_A = \diag(1, -1, 1), \quad \mat{M}_B = \identity \quad \mbox{with } \nonumber \\
    &  \vec{T}'_A = \begin{pmatrix} L'_{Ax} \\ L'_{Ay} \\ L'_{Az} \end{pmatrix}, \quad
    \vec{T}'_{\B{1}} = \begin{pmatrix} L'_{1x} \\ L'_{1y} \\ L'_{1z} \end{pmatrix}, \quad
    \vec{T}'_{\B{2}} = \begin{pmatrix} L'_{2x} \\ L'_{2y} \\ L'_{2z} \end{pmatrix} .
\end{align}
The remaining rotational freedom around the $y$-axis allows us to remove the $z$-component of one of the translation vectors.
We will employ this freedom below.
From these a set of pure translations, an associated \E{1}, can be constructed.
The actions of both the $g_{\B{i}}$ are already pure translations, so two of the needed translation vectors can be chosen as $\vec{T}'_1 \equiv \vec{T}'_{\B{1}}$ and $\vec{T}'_2 \equiv \vec{T}'_{\B{2}}$.
A third translation vector can be defined based on the fact that $(\mat{M}_A)^2=\identity$.
Note that
\begin{equation}
    g'_3 \equiv (g'_A)^2:\vec{x} \to \vec{x} + (\identity + \mat{M}_A) \vec{T}'_A
\end{equation}
leads to the definition 
\begin{equation}
    \vec{T}'_3 \equiv (\identity + \mat{M}_A) \vec{T}'_A = \transpose{(2 L'_{Ax}, 0, 2 L'_{Az})}.
\end{equation}

The remaining conditions to impose (all other combinations of generators will either already be pure translations or reduce to pure translations of these) is that for $i$ and $j \in \{1, 2\}$
\begin{equation}
    \inverse{(g'_A)} \inverse{(g'_{\B{i}})} g'_A g'_{\B{j}} : \vec{x} \to \vec x + \vec{T}'_{\B{j}} - \mat{M}_A \vec{T}'_{\B{i}} = \vec{x} + \sum_{k} m^{(i)}_k \vec{T}'_k,
\end{equation}
for some set of $m^{(i)}_k \in \integers$.
Since $\vec{T}'_{\B{j}}$ is already one of the pure translations, $\vec{T}'_j$, this is equivalent to requiring
\begin{equation}
    \label{eqn:E7E9groupcondition}
    (\identity - \mat{M}_A) \vec{T}'_{\B{i}} =  \sum_{k} m^{(i)}_k \vec{T}'_k, \quad \mbox{for some } m^{(i)}_k \in \integers,
\end{equation}
leading to the set of conditions
\begin{align}
    \label{eqn:E7E9groupcondition1}
    \begin{pmatrix}
    0 \\
    2 L'_{1y} \\
    0
    \end{pmatrix}
    &= 
    \begin{pmatrix}
        m^{(1)}_1 L'_{1x} + m^{(1)}_2 L'_{2x} + m^{(1)}_3 2 L'_{Ax}  \\
        m^{(1)}_1 L'_{1y} + m^{(1)}_2 L'_{2y}  \\
        m^{(1)}_1 L'_{1z} + m^{(1)}_2 L'_{2z} + m^{(1)}_3 2 L'_{Az}
    \end{pmatrix},
    \\
    \label{eqn:E7E9groupcondition2}
    \begin{pmatrix}
        0 \\
        2 L'_{2y} \\
        0
    \end{pmatrix}
    &= 
    \begin{pmatrix}
        m^{(2)}_1 L'_{1x} + m^{(2)}_2 L'_{2x} + m^{(2)}_3 2 L'_{Ax}  \\
        m^{(2)}_1 L'_{1y} + m^{(2)}_2 L'_{2y}  \\
        m^{(2)}_1 L'_{1z} + m^{(2)}_2 L'_{2z} + m^{(2)}_3 2 L'_{Az}
    \end{pmatrix}.
\end{align}
For the translation vectors $\vec{T}'_i$ to span \Espace\ we must have at least one of $L'_{1y}$, and $L'_{2y}$ non-zero and similarly at least one of $L'_{1z}$, $L'_{2z}$, and $L'_{Az}$ non-zero.
Finally, the remaining rotational freedom allows us to always choose one of the $z$-components to be zero, thus, we only need consider the cases when one or two of the $z$-components are non-zero.

To determine the form of the generators, we begin by rotating such that $L'_{1z} = 0$ and consider the special case where $L'_{Az} = 0$, meaning that $L'_{2z} \neq 0$.
This further requires that $L'_{Ax} \neq 0$ otherwise $\vec{x} = \transpose{(x, L'_{Ay}/2, z)}$ would be a fixed point, i.e., $g'_A$ would not be freely acting.
For this case, from the $z$-components of \cref{eqn:E7E9groupcondition1} and \cref{eqn:E7E9groupcondition2} we immediately see that $m^{(1)}_2 = m^{(2)}_2 = 0$.
The $y$-components of \cref{eqn:E7E9groupcondition1} and \cref{eqn:E7E9groupcondition2} then require that $m^{(1)}_1 = 2$ and $L'_{2y} = m^{(2)}_1 L'_{1y} / 2$, and then 
the $x$-components of \cref{eqn:E7E9groupcondition1} and \cref{eqn:E7E9groupcondition2} require $L'_{1x} = -m^{(1)}_3 L'_{Ax}$.
Proceeding, we note that we can always shift $\vec{T}'_{\B{1}}$ by integer multiples of $\vec{T}'_3$, thus we only need consider the cases where $m^{(1)}_3 \in \{0, -1 \}$.
However, when  $m^{(1)}_3 = -1$ then $\vec{x} = \transpose{(x, (L'_{Ay} + L'_{1y})/2, z)}$ will be a fixed point of $g'_{\B{1}} g'_A$, meaning that we must have $m^{(1)}_3 = 0$.
Relabeling $p \equiv m^{(2)}_2$ we thus arrive at the solution
\begin{equation}
    \label{eqn:E7E9generalTa}
    \vec{T}'_A = \begin{pmatrix} L'_{Ax} \\ L'_{Ay} \\ 0 \end{pmatrix}, \quad
    \vec{T}'_{\B{1}} = \begin{pmatrix} 0 \\ L'_{1y} \\ 0 \end{pmatrix}, \quad
    \vec{T}'_{\B{2}} = \begin{pmatrix} L'_{2x} \\ \frac{p}{2} L'_{1y} \\ L'_{2z} \end{pmatrix}.
\end{equation}
Since $\vec{T}'_{\B{2}}$ can always be shifted by integer multiples of $\vec{T}'_{\B{1}}$, we can always choose $p \in \{0, 1\}$.
Thus we see there are two distinct solutions labeled by the integer $p \in \{0, 1\}$.

The analysis of this special case is the template for the analysis of other cases.
In fact, it turns out that this special case is sufficient: it has found all the allowed solutions.
The next step is to consider the other special case where we begin with $L'_{1z} = L'_{2z} = 0$ and $L'_{Az} \neq 0$.
However, by rotating around the $y$-axis this becomes the case with $L'_{Az} = 0$, $L'_{1z} \neq 0$, and $L'_{2z} \neq 0$.

There is only one remaining case to consider to complete the search for the general form of the generators:  $L'_{Az} = 0$, $L'_{1z} \neq 0$, and $L'_{2z} \neq 0$.
As noted above, $L'_{Ax} \neq 0$ to ensure that $g'_A$ is freely acting.
Again, we will start with a special case.
Suppose $m^{(2)}_1 = 0$.
From \cref{eqn:E7E9groupcondition2} this implies that also $m^{(2)}_2 = m^{(2)}_3 = 0$ and $L'_{2y} = 0$.
The $y$-component of \cref{eqn:E7E9groupcondition1} then requires that $m^{(1)}_2 = 2$.
We are then left with just two conditions that must be satisfied
\begin{align}
    0 &= 2 L'_{1x} + m^{(1)}_2 L'_{2x} + 2 m^{(1)}_3 L'_{Ax} , \nonumber \\
    0 &= 2 L'_{1z} + m^{(1)}_2 L'_{2z} .
\end{align}
Again, using the freedom to shift by integer multiples of $\vec{T}'_i$, we can restrict to the cases $m^{(1)}_2 \in \{0, 1\}$ and $m^{(1)}_3 \in \{0, 1\}$.
Further using this shift freedom, we can show these cases reduce as follows:
\begin{description}
    \setlength{\itemsep}{0pt}
    \setlength{\baselineskip}{0pt}
    \item[$m^{(1)}_2 = 0$, $m^{(1)}_3 = 0$: ] This is the $p=0$ solution from \cref{eqn:E7E9generalTa}.
    \item[$m^{(1)}_2 = 0$, $m^{(1)}_3 = 1$: ] This is invalid as it would require $L'_{1z} = 0$, in contradiction to our starting assumption.
    \item[$m^{(1)}_2 = 1$, $m^{(1)}_3 = 0$: ] This is the $p=1$ solution from \cref{eqn:E7E9generalTa} which can be seen by replacing $\vec{T}'_{\B{2}}$ with $\vec{T}'_{\B{2}} + 2 \vec{T}'_1$ and relabeling.
    \item[$m^{(1)}_2 = 1$, $m^{(1)}_3 = 1$: ] This is the $p=1$ solution from \cref{eqn:E7E9generalTa} which can be seen by replacing $\vec{T}'_{\B{2}}$ with $\vec{T}'_{\B{2}} + 2 \vec{T}'_1 + \vec{T}'_3$ and relabeling.
\end{description}
Thus this case reproduces the previous solutions without introducing new ones.
Though we began with the assumption $m^{(2)}_1 = 0$, the same argument holds if $m^{(1)}_1 = 0$, $m^{(1)}_2 = 0$, or $m^{(2)}_2 = 0$.

We are left to consider the case where none of $m^{(1)}_1$, $m^{(1)}_2$, $m^{(2)}_1$, and $m^{(2)}_2$ are zero.
The conditions \cref{eqn:E7E9groupcondition1} and \cref{eqn:E7E9groupcondition2} then lead to the requirements

\begin{align}
    \label{eqn:E7E9conditions_a}
    & \frac{m^{(2)}_2} {m^{(2)}_1} = \frac{m^{(1)}_2} {m^{(1)}_1} , \quad
    \frac{m^{(2)}_3} {m^{(2)}_2}= \frac{m^{(1)}_3} {m^{(1)}_2}, \quad
    \frac{m^{(2)}_1} {m^{(2)}_3}= \frac{m^{(1)}_1} {m^{(1)}_3},
     \\
    \label{eqn:E7E9conditions_b}
    &  m^{(1)}_1 =   2 - m^{(2)}_2 .
\end{align}
\cref{eqn:E7E9conditions_a} is the statement that $m^{(2)}_j = q m^{(1)}_j$ for $q \in \integers$.
The remaining constraints come from the $x$ and $z$-components of \cref{eqn:E7E9groupcondition1}.
By replacing $\vec{T}'_{\B{2}}$ with $m^{(1)}_2 \vec{T}'_{\B{2}} + (2 - q m^{(1)}_2) \vec{T}'_{\B{1}} + m^{(1)}_3 \vec{T}'_3$ and relabeling, we can show that this case reduces to the $p=1$ solution of \cref{eqn:E7E9generalTa}.

The general solution is thus of the form \cref{eqn:E7E9generalTa}.
This solution can be converted to the more conventional forms by looking at alternative generators.
Notice that
\begin{align}
    g'_A g'_{\B{1}}: \vec{x} &\to \mat{M}_A \vec{x} + \begin{pmatrix} L'_{Ax} \\ L'_{Ay} - L'_{1y} \\ 0 \end{pmatrix} , \nonumber \\
    g'_A g'_{\B{2}}: \vec{x} &\to \mat{M}_A \vec{x} + \begin{pmatrix} L'_{Ax} + L'_{2x} \\ L'_{Ay} - \frac{p}{2} L'_{1y} \\ L'_{2z} \end{pmatrix} .
\end{align}
With these the conventional choice of $O(3)$ elements that correspond to \E{7} involves the set of generators $\{g'_A, g'_A g'_{\B{1}}, g'_{\B{2}}\}$, while the conventional choice of $O(3)$ elements that correspond to \E{9} involves the set of generators $\{g'_A, g'_A g'_{\B{1}}, g'_A g'_{\B{2}}\}$.

More explicitly, the $p=0$ solution corresponds to the conventional set of generators for \E{7}.
To see this, let $L_{1x} \equiv L'_{Ax}$, $L_{1y} \equiv L'_{Ay}$, $L_{2y} \equiv L'_{Ay} - L'_{1y}$, $L_{Bx} \equiv L'_{2x}$, and $L_{Bz} \equiv L'_{2z}$.
Plugging these in we can identify $g^{\E{7}}_{\A{1}} \equiv g'_A$, $g^{\E{7}}_{\A{2}} \equiv g'_A g'_{\B{1}}$, and $g^{\E{7}}_B \equiv g'_{\B{2}}$ with
\begin{equation}
    \vec{T}^{\E{7}}_{\A{1}} = \begin{pmatrix} L_{1x} \\ L_{1y} \\ 0 \end{pmatrix} , \quad
    \vec{T}^{\E{7}}_{\A{2}} = \begin{pmatrix} L_{1x} \\ L_{2y} \\ 0 \end{pmatrix} , \quad
    \vec{T}^{\E{7}}_B = \begin{pmatrix} L_{Bx} \\ 0 \\ L_{Bz} \end{pmatrix} .
\end{equation}

Similarly, the $p=1$ solution corresponds to the conventional set of generators for \E{9}.
To see this, let $L_{1x} \equiv L'_{Ax}$, $L_{1y} \equiv L'_{Ay}$, $L_{2y} \equiv L'_{Ay} - L'_{1y}$ (so that $L'_{1y} = L_{1y} - L_{2y}$), $L_{3x} \equiv L'_{Ax} + L'_{2x}$, and $L_{3z} \equiv L'_{2z}$.
Plugging these in we can identify $g^{\E{9}}_{\A{1}} \equiv g'_A$, $g^{\E{9}}_{\A{2}} \equiv g'_A g'_{\B{1}}$, and $g^{\E{9}}_{\A{3}} \equiv g'_A g'_{\B{2}}$ with
\begin{equation}
    \vec{T}^{\E{9}}_{\A{1}} = \begin{pmatrix} L_{1x} \\ L_{1y} \\ 0 \end{pmatrix} , \quad
    \vec{T}^{\E{9}}_{\A{2}} = \begin{pmatrix} L_{1x} \\ L_{2y} \\ 0 \end{pmatrix} , \quad
    \vec{T}^{\E{9}}_{\A{3}} = \begin{pmatrix} L_{3x} \\ \frac{1}{2}(L_{1y} + L_{2y}) \\ L_{3z} \end{pmatrix} .
\end{equation}

There is no requirement to restrict to the conventional choice.
As the derivation above shows, other choices are possible.
Considering the $O(3)$ matrices listed above, there are multiple choices that can be made for the set of generators.
Different choices may ease computations in some cases.
The natural choices are listed here, where we have dropped the primes and relabeled the components of the translation vectors.
\begin{description}
    \item[I. 1 glide reflection ($A$), 2 translations ($B$)] (the case used in the derivation):
    \begin{align}
        \label{eqn:E7E9-caseI}
        & \mat{M}_A = \diag(1,-1,1), \quad \mat{M}_B = \identity , \nonumber \\
        & \vec{T}_A = \begin{pmatrix} L_{Ax} \\ L_{Ay} \\ 0 \end{pmatrix} , \quad
        \vec{T}_{\B{1}} = \begin{pmatrix} 0 \\ L_{1y} \\ 0 \end{pmatrix} , \quad
        \vec{T}_{\B{2}} = \begin{pmatrix} L_{2x} \\ \frac{p}{2} L_{1y} \\ L_{2z} \end{pmatrix} ,
    \end{align}
    with associated \E{1} given by
    \begin{equation}
        \vec{T}_1 = \begin{pmatrix} 2 L_{Ax} \\ 0 \\ 0 \end{pmatrix} , \quad
        \vec{T}_2 = \begin{pmatrix} 0 \\ L_{1y} \\ 0 \end{pmatrix} , \quad
        \vec{T}_3 = \begin{pmatrix} L_{2x} \\ \frac{p}{2} L_{1y} \\ L_{2z} \end{pmatrix} ,
    \end{equation}
    \item[II. 2 glide reflections ($A$), 1 translation ($B$)] 
        (the conventional choice for \E{7} with $p=0$):
    \begin{align}
        \label{eqn:E7E9-caseII}
        & \mat{M}_A = \diag(1,-1,1), \quad \mat{M}_B = \identity , \nonumber \\
        & \vec{T}_{\A{1}} = \begin{pmatrix} L_{1x} \\ L_{1y} \\ 0 \end{pmatrix} , \quad
        \vec{T}_{\A{2}} = \begin{pmatrix} L_{1x} \\ L_{2y} \\ 0 \end{pmatrix} , \quad
        \vec{T}_B = \begin{pmatrix} L_{Bx} \\ \frac{p}{2} (L_{1y} - L_{2y}) \\ L_{Bz} \end{pmatrix} ,
    \end{align}
    with associated \E{1} given by
    \begin{equation}
        \label{eqn:E7E9-caseII-E1}
        \vec{T}_1 = \begin{pmatrix} 2 L_{1x} \\ 0 \\ 0 \end{pmatrix} , \quad
        \vec{T}_2 = \begin{pmatrix} 0 \\ L_{1y} - L_{2y} \\ 0 \end{pmatrix} , \quad
        \vec{T}_3 = \begin{pmatrix} L_{Bx} \\ \frac{p}{2} (L_{1y} - L_{2y}) \\ L_{Bz} \end{pmatrix} ,
    \end{equation}
    \item[III. 3 glide reflections ($A$)]
        (the conventional choice for \E{9} with $p=1$):
    \begin{align}
        \label{eqn:E7E9-caseIII}
        & \mat{M}_A = \diag(1,-1,1) , \nonumber \\
        & \vec{T}_{\A{1}} = \begin{pmatrix} L_{1x} \\ L_{1y} \\ 0 \end{pmatrix} , \quad
        \vec{T}_{\A{2}} = \begin{pmatrix} L_{1x} \\ L_{2y} \\ 0 \end{pmatrix} , \quad
        \vec{T}_{\A{3}} = \begin{pmatrix} L_{3x} \\ L_{1y} - \frac{p}{2}(L_{1y} - L_{2y}) \\ L_{3z} \end{pmatrix} ,
    \end{align}
    with associated \E{1} given by
    \begin{equation}
        \label{eqn:E7E9-caseIII-E1}
        \vec{T}_1 = \begin{pmatrix} 2 L_{1x} \\ 0 \\ 0 \end{pmatrix} , \quad
        \vec{T}_2 = \begin{pmatrix} 0 \\ L_{1y} - L_{2y} \\ 0 \end{pmatrix} , \quad
        \vec{T}_3 = \begin{pmatrix} L_{3x} - L_{1x} \\ \frac{p}{2} (L_{1y} - L_{2y}) \\ L_{3z} \end{pmatrix} .
    \end{equation}
\end{description}
Up to redefinition of the parameters, the associated \E{1} is the same for all choices of the set of generators, as it must be.
Regardless of the form chosen, the $p=0$ solutions are the generators of the Klein space, \E{7}, and the $p=1$ solutions are the generators of the Klein space with a vertical flip, \E{9}.

\subsection{\E{8} and \E{10}: Klein spaces with horizontal flip or half-turn}
\label{app:E8E10}

Similar to \E{7} and \E{9}, the $O(3)$ structures of the two Klein spaces \E{8} and \E{10} are the same.
Conventionally (c.f.\ \cite{Riazuelo2004:prd}) their sets of generators are written with two generators associated with a vertical flip,
\begin{equation}
    \mat{M}^{\E{8}}_A = \mat{M}^{\E{10}}_A = \diag(1, -1, 1).
\end{equation}
For \E{8} the third generator is conventionally associated with a horizontal flip,
\begin{equation}
    \mat{M}^{\E{8}}_B = \diag(-1, 1, 1),
\end{equation}
whereas for \E{10} the third generator is conventionally associated with a half-turn,
\begin{equation}
    \mat{M}^{\E{10}}_B = \diag(-1, -1, 1).
\end{equation}
However, since $\mat{M}^{\E{8}}_A \mat{M}^{\E{8}}_B = \diag(-1, -1, 1) = \mat{M}^{\E{10}}_B$, the same sets of three matrices could be used to construct a set of generators for either of the topologies.
As above, for a given choice of $\mat{M}_A$ and $\mat{M}_B$, there will be two independent sets of generators, with one identified as \E{8} and the other as \E{10}.

To derive the general form of the generators we will use an alternative set of matrices.
Choose $\mat{M}_A$ a vertical flip, $\mat{M}_B$ a half-turn, and $\mat{M}_C = \identity$, with one generator associated with each of these three matrices.
Starting with this choice we can use two rotational degrees of freedom to fix the vertical flip to be across the $xz$-plane (which fixes the $y$-axis) and the remaining rotational degree of freedom to fix the half-turn to be around the $z$-axis.
We thus choose as our starting form
\begin{align}
    & \mat{M}_A = \diag(1, -1, 1), \quad \mat{M}_B = \diag(-1, -1, 1), \quad \mat{M}_C = \identity, \nonumber \\
    & \vec{T}'_A = \begin{pmatrix} L'_{Ax} \\ L'_{Ay} \\ L'_{Az} \end{pmatrix} , \quad
     \vec{T}'_B = \begin{pmatrix} L'_{Bx} \\ L'_{By} \\ L'_{Bz} \end{pmatrix} , \quad
     \vec{T}'_C = \begin{pmatrix} L'_{Cx} \\ L'_{Cy} \\ L'_{Cz} \end{pmatrix} ,
\end{align}
with the corresponding pure translations coming from $(g_A)^2$, $(g_B)^2$, and $g_C$, respectively, to be
\begin{equation}
    \vec{T}'_1 = \begin{pmatrix} 2 L'_{Ax} \\ 0 \\ 2 L'_{Az} \end{pmatrix} , \quad
    \vec{T}'_2 = \begin{pmatrix} 0 \\ 0 \\ 2 L'_{Bz} \end{pmatrix} , \quad
    \vec{T}'_3 = \begin{pmatrix} L'_{Cx} \\ L'_{Cy} \\ L'_{Cz} \end{pmatrix} .
\end{equation}

The remaining conditions to impose (all other combinations of generators will either already be pure translations or reduce to pure translations of these) is that the following combinations of generators must be pure translations:\footnote{The list of combinations of generators that must be pure translations is not unique.
    The fact that the provided list is sufficient is also not generic, it depends on the matrices chosen, i.e., on $\mat{M}_A$, $\mat{M}_B$, and $\mat{M}_C$.
    Finally, these particular forms were chosen to make the derivation of the generators simpler.
    This is most notably true for $(g'_A)^2 \inverse{(g'_B)} (g'_A)^2 g'_B$ where the prepended $(g'_A)^2$ is redundant.
}
\begin{equation}
    \left\{ \inverse{(g'_A)} \inverse{(g'_B)} g'_A g'_B , \quad
    (g'_A)^2 \inverse{(g'_B)} (g'_A)^2 g'_B , \quad
    \inverse{(g'_A)} g'_C g'_A  , \quad
    \inverse{(g'_B)} g'_C g'_B  \right\} .
\end{equation}
This leads to the system of equations with $m^{(i)}_a \in \integers$ for $i\in\{1, 2, 3, 4\}$ and $a\in\{A, B, C\}$:
\begin{align}
    \label{eqn:E8E10groupcondition1}
    \begin{pmatrix}
    -2 L'_{Ax} \\
    2(L'_{Ay} - L'_{By}) \\
    0
    \end{pmatrix}
    &= 
    \begin{pmatrix}
        m^{(1)}_A 2 L'_{Ax} + m^{(1)}_C L'_{Cx}  \\
        m^{(1)}_C L'_{Cy}  \\
        m^{(1)}_A 2 L'_{Az} + m^{(1)}_B 2 L'_{Bz} + m^{(1)}_C L'_{Cz}
    \end{pmatrix},
    \\
    \label{eqn:E8E10groupcondition2}
    \begin{pmatrix}
        0 \\
        0 \\
        4 L'_{Az}
    \end{pmatrix}
    &= 
    \begin{pmatrix}
        m^{(2)}_A 2 L'_{Ax} + m^{(2)}_C L'_{Cx}  \\
        m^{(2)}_C L'_{Cy}  \\
        m^{(2)}_A 2 L'_{Az} + m^{(2)}_B 2 L'_{Bz} + m^{(2)}_C L'_{Cz}
    \end{pmatrix},
    \\
    \label{eqn:E8E10groupcondition3}
    \begin{pmatrix}
        \hphantom{-} L'_{Cx} \\
        -L'_{Cy} \\
        \hphantom{-} L'_{Cz}
    \end{pmatrix}
    &= 
    \begin{pmatrix}
        m^{(3)}_A 2 L'_{Ax} + m^{(3)}_C L'_{Cx}  \\
        m^{(3)}_C L'_{Cy}  \\
        m^{(3)}_A 2 L'_{Az} + m^{(3)}_B 2 L'_{Bz} + m^{(3)}_C L'_{Cz}
    \end{pmatrix} ,
    \\
    \label{eqn:E8E10groupcondition4}
    \begin{pmatrix}
        \hphantom{-} L'_{Cx} \\
        \hphantom{-} L'_{Cy} \\
        -L'_{Cz}
    \end{pmatrix}
    &= 
    \begin{pmatrix}
        m^{(4)}_A 2 L'_{Ax} + m^{(4)}_C L'_{Cx}  \\
        m^{(4)}_C L'_{Cy}  \\
        m^{(4)}_A 2 L'_{Az} + m^{(4)}_B 2 L'_{Bz} + m^{(4)}_C L'_{Cz}
    \end{pmatrix} .
\end{align}
For the translation vectors $\vec{T}'_i$ to be non-trivial and span \Espace\ we require $L'_{Ax} \neq 0$, $L'_{Bz} \neq 0$, and $L'_{Cy} \neq 0$.

While this appears to be a complicated set of relations that must be satisfied for many lengths and many integers, it can be reduced in a straightforward manner.
\begin{enumerate}
    \item Beginning with the $y$-components, since $L'_{Cy} \neq 0$ we immediately have that $m^{(2)}_C = 0$, $m^{(3)}_C = -1$, and $m^{(4)}_C = 1$, along with $L'_{By} = L'_{Ay} - m^{(1)}_C L'_{Cy} / 2$.
    \item Next, since $L'_{Ax} \neq 0$ the $x$-component of \cref{eqn:E8E10groupcondition4} requires $m^{(4)}_A = 0$, the $x$-component of \cref{eqn:E8E10groupcondition2} requires $m^{(2)}_A = 0$, and the $x$-component of \cref{eqn:E8E10groupcondition3} gives $L'_{Cx} = m^{(3)}_A L'_{Ax}$.
    \item To continue, we note that the $z$-component of \cref{eqn:E8E10groupcondition4} gives $L'_{Cz} = -m^{(4)}_B L'_{Bz}$.
    However, if $m^{(4)}_B$ is even we can always replace $\vec{T}'_C$ with $\vec{T}'_C - \frac{m^{(4)}_B}{2} \vec{T}'_2$, making $L'_{Cz} = 0$.
    On the other hand, if $m^{(4)}_B$ is odd then $\inverse{(g'_C)} (g'_B)^{m^{(4)}_B}$ will have a fixed point for $\vec{x} = \transpose{(m^{(4)}_B L'_{Bx} - L'_{Cx}/2, m^{(4)}_B L'_{By} - L'_{Cy}/2, z)}$.
    Thus we must have $L'_{Cz} = 0$.
    \item Finally, the $z$-component of \cref{eqn:E8E10groupcondition2} gives $L'_{Az} = m^{(2)}_B L'_{Bz} / 2$.
    \item We are left with three integers.
    For convenience we redefine them as $m_A \equiv m^{(3)}_A$, $m_B \equiv m^{(2)}_B$, and $m_C \equiv m^{(1)}_C$.
    The remaining conditions lead to the constraints that if $m_A$ is odd then $m_B$ and $m_C$ must be even, or that if either $m_B$ or $m_C$ is odd then $m_A$ must be even.
\end{enumerate}
This exhausts the useful knowledge from the sets of conditions.

The general solution now has the form
\begin{equation}
    \vec{T}'_A = \begin{pmatrix} L'_{Ax} \\ L'_{Ay} \\ m_B L'_{Bz} / 2 \end{pmatrix} , \quad
    \vec{T}'_B = \begin{pmatrix} L'_{Bx} \\ L'_{Ay} - m_C L'_{Cy} / 2 \\ L'_{Bz} \end{pmatrix} , \quad
    \vec{T}'_C = \begin{pmatrix} m_A L'_{Ax} \\ L'_{Cy} \\ 0 \end{pmatrix} , 
\end{equation}
and the associated pure translation vectors
\begin{equation}
    \vec{T}'_1 = \begin{pmatrix} 2 L'_{Ax} \\ 0 \\ m_B L'_{Bz} \end{pmatrix} , \quad
    \vec{T}'_2 = \begin{pmatrix} 0 \\ 0 \\ 2 L'_{Bz} \end{pmatrix} , \quad
    \vec{T}'_3 = \begin{pmatrix} m_A L'_{Ax} \\ L'_{Cy} \\ 0 \end{pmatrix} .
\end{equation}
The range of these integers can be restricted and many choices lead to generators that are not freely acting.

Beginning with $\vec{T}'_A$ we see that shifting by integer multiples of $\vec{T}'_2$ allows us to restrict $m_B\in\{-1, 0, 1, 2\}$.
To analyze the remaining cases we start with the special one where $m_A = m_B = 0$.
We can then shift $\vec{T}'_B$ by integer multiples of $\vec{T}'_C$ to restrict $m_C \in \{0, 1\}$.
This gives us two solutions.
Again redefining $p \equiv m_C$ we have the general solutions
\begin{equation}
    \label{eqn:E8E10generalTa}
    \vec{T}'_A = \begin{pmatrix} L'_{Ax} \\ L'_{Ay} \\ 0 \end{pmatrix} , \quad
    \vec{T}'_B = \begin{pmatrix} L'_{Bx} \\ L'_{Ay} - p L'_{Cy} / 2 \\ L'_{Bz} \end{pmatrix} , \quad
    \vec{T}'_C = \begin{pmatrix} 0 \\ L'_{Cy} \\ 0 \end{pmatrix} , 
\end{equation}
for $p \in \{0, 1\}$.
Despite being a special case, this is the the general solution.
All other cases will be invalid or reduce to \cref{eqn:E8E10generalTa}.

For $m_A = 0$ we can also have $m_B = 2$ where we can again restrict $m_C \in \{0, 1\}$.
In this case, when $m_C = 0$, then $\vec{x} = \transpose{((L'_{Bx} - L'_{Ax})/2, y, z)}$ will be a fixed point of $\inverse{(g'_A)} g'_B$, meaning that there is only a solution when $m_C = 1$.
At first glance the resulting solution does not look the same as \cref{eqn:E8E10generalTa} since the $z$-component of $\vec{T'}_A$ is $L'_{Bz} \neq 0$.
However, by rotating around the $y$-axis we can set the $z$-component of $\vec{T}'_A$ to zero without changing $\vec{T}'_C$ and, through redefinition of $L'_{Bx}$ and $L'_{Bz}$ we arrive at the $p=1$ solution in \cref{eqn:E8E10generalTa}, hence this is not a new solution.

Finally for $m_A = 0$ we can have $m_B = \pm 1$.
However, the set of generators will not be freely acting for either choice of $m_B$.
When $m_B = +1$, $(g'_A)^{-2} g_B:\vec{x} \to \vec{x}$ for $\vec{x} = \transpose{(-L'_{Ax} + L'_{Bx}/2, L'_{Ay}/2 - m_C L'_{Cy} / 4, z)}$,
whereas when $m_B = -1$, $(g'_A)^{2} g_B:\vec{x} \to \vec{x}$ for $\vec{x} = \transpose{(L'_{Ax} + L'_{Bx}/2, L'_{Ay}/2 - m_C L'_{Cy} / 4, z)}$.
Thus for $m_A = 0$ the valid solutions are those in \cref{eqn:E8E10generalTa}.

To complete the derivation we return to the possibilities for $m_B$.
When $m_B = 0$ then $\vec{T}'_C$ can be shifted by integer multiples of $\vec{T}'_1$ so that we can always restrict $m_A \in \{ 0, 1 \}$ with $m_A = 0$ having already been studied.
When $m_A = 1$ then $\inverse{(g'_C)} g'_A:\vec{x} \to \vec{x}$ for $\vec{x} = \transpose{(x, (L'_{Ay}-L'_{Cy})/2, z)}$ so is not freely acting.
Thus there are no new solutions when $m_B = 0$.

Next, when $m_B = 2$ then we again can restrict to $m_A \in \{ 0, 1 \}$ with $m_A = 0$ having already been studied.
As in the $m_A = 0$ and $m_B = 2$ case we next can restrict to $m_C \in \{ 0, 1 \}$ with $m_C = 0$ invalid.
When $m_C = 1$ we must have $m_A = 0$ (since it must be even), so this case has already been studied.
Thus there are no new solutions when $m_B = 2$.

When $m_B = \pm 1$ then $m_A$ must be even.
Shifting $\vec{T}'_C$ by integer multiples of $\vec{T}'_1$ and $\vec{T}'_2$ we can restrict $m_A \in \{0, 2\}$ with the $m_A = 0$ case already studied.
However, when $m_A = 2$ then this reduces to the same calculation as the $m_A = 0$ where we can show that the generators will not be freely acting.
Thus there are no new solutions when $m_B = \pm 1$.

The general solution is thus of the form \cref{eqn:E8E10generalTa}.
This solution can be converted to the more conventional forms by looking at alternative generators.
Notice that
\begin{align}
    g'_A g'_C: \vec{x} & \to \mat{M}_A \vec{x} + \begin{pmatrix} L'_{Ax} \\ L'_{Ay} - L'_{Cy} \\ 0 \end{pmatrix} , \nonumber \\
    g'_A g'_B: \vec{x} & \to \mat{M}^{\E{8}} \vec{x} + \begin{pmatrix} L'_{Ax} + L'_{Bx} \\ \frac{p}{2} L'_{Cy} \\ L'_{Bz} \end{pmatrix} .
\end{align}
Recall that our solution is based on $\mat{M}_B \equiv \mat{M}^{\E{10}}_B = \diag(-1, -1, 1)$ and that $\mat{M}^{\E{8}} = \diag(-1, 1, 1)$.
With these the conventional choice of $O(3)$ elements that correspond to \E{8} involves the set of generators $\{ g'_A, g'_A g'_C, g'_A g'_B \}$, while the conventional choice of $O(3)$ elements that correspond to \E{10} involves the set of generators $\{ g'_A, g'_A g'_C,  g'_B \}$.

More explicitly, the $p=0$ solution corresponds to the set of generators for \E{8}.
To see this, let $L_{1x} \equiv L'_{Ax}$, $L_{1y} \equiv L'_{Ay}$, $L_{2y} \equiv L'_{Ay} - L'_{Cy}$, $L_{Bx} \equiv L'_{Ax} + L'_{Bx}$, and $L_{Bz} \equiv L'_{Bz}$.
Plugging these in we can identify $g^{\E{8}}_{\A{1}} \equiv g'_A$, $g^{\E{8}}_{\A{2}} \equiv g'_A g'_C$, and $g^{\E{8}}_B \equiv g'_A g'_B$ with
\begin{equation}
    \vec{T}^{\E{8}}_{\A{1}} = \begin{pmatrix} L_{1x} \\ L_{1y} \\ 0 \end{pmatrix} , \quad
    \vec{T}^{\E{8}}_{\A{2}} = \begin{pmatrix} L_{1x} \\ L_{2y} \\ 0 \end{pmatrix} , \quad
    \vec{T}^{\E{8}}_B = \begin{pmatrix} L_{Bx} \\ 0 \\ L_{Bz} \end{pmatrix} .
\end{equation}

Similarly, the $p=1$ solution corresponds to the conventional set of generators for \E{10}.
To see this let $L_{1x} \equiv L'_{Ax}$, $L_{1y} \equiv L'_{Ay}$, $L_{2y} \equiv L'_{Ay} - L'_{Cy}$ (so that $L'_{Cy} = L_{1y} - L_{2y}$), $L_{Bx} \equiv L'_{Bx}$, $L_{By} \equiv L'_{Ay} - L'_{Cy}/2 = (L_{1y} + L_{2y})/2$, and $L_{Bz} \equiv L'_{Bz}$.
Plugging these in we can identify $g^{\E{10}}_{\A{1}} \equiv g'_A$, $g^{\E{10}}_{\A{2}} \equiv g'_A g'_C$, and $g^{\E{10}}_B \equiv g'_B$ with
\begin{equation}
    \vec{T}^{\E{10}}_{\A{1}} = \begin{pmatrix} L_{1x} \\ L_{1y} \\ 0 \end{pmatrix} , \quad
    \vec{T}^{\E{10}}_{\A{2}} = \begin{pmatrix} L_{1x} \\ L_{2y} \\ 0 \end{pmatrix} , \quad
    \vec{T}^{\E{10}}_B = \begin{pmatrix} L_{Bx} \\ \frac{1}{2} (L_{1y} + L_{2y}) \\ L_{Bz} \end{pmatrix} .
\end{equation}

As with \E{7} and \E{9}, there is no requirement to restrict to the conventional choice.
Here the natural choices are listed where we have dropped the primes, relabeled the components in the translation vectors, and, in some cases, reordered the associated \E{1} vectors.
\begin{description}
    \item[I. 1 glide reflection ($A$), 1 half-turn ($B$), and 1 translation ($C$)] (the case used in the derivation):
    \begin{align}
        \label{eqn:E8E10-caseI}
        & \mat{M}_A = \diag(1,-1,1), \quad \mat{M}_B = \diag(-1,-1,1), \quad \mat{M}_C = \identity , \nonumber \\
        & \vec{T}_A = \begin{pmatrix} L_{Ax} \\ L_{Ay} \\ 0 \end{pmatrix} , \quad
        \vec{T}_B = \begin{pmatrix} L_{Bx} \\ L_{Ay} - \frac{p}{2} L_{Cy} \\ L_{Bz} \end{pmatrix} , \quad
        \vec{T}_C = \begin{pmatrix} 0 \\ L_{Cy} \\ 0 \end{pmatrix} ,
    \end{align}
    with associated \E{1} given by
    \begin{equation}
        \vec{T}_1 = \begin{pmatrix} 2 L_{Ax} \\ 0 \\ 0 \end{pmatrix} , \quad
        \vec{T}_2 = \begin{pmatrix} 0 \\ L_{Cy} \\ 0 \end{pmatrix} , \quad
        \vec{T}_3 = \begin{pmatrix} 0 \\ 0 \\ 2 L_{Bz} \end{pmatrix} ,
    \end{equation}
    \item[II. 1 glide reflection ($A$), 1 orthogonal glide reflection ($B$), and 1 translation ($C$)]:
    \begin{align}
        & \mat{M}_A = \diag(1,-1,1), \quad \mat{M}_B = \diag(-1,1,1), \quad \mat{M}_C = \identity , \nonumber \\
        & \vec{T}_A = \begin{pmatrix} L_{Ax} \\ L_{Ay} \\ 0 \end{pmatrix} , \quad
        \vec{T}_B = \begin{pmatrix} L_{Bx} \\ \frac{p}{2} L_{Cy} \\ L_{Bz} \end{pmatrix} , \quad
        \vec{T}_C = \begin{pmatrix} 0 \\ L_{Cy} \\ 0 \end{pmatrix} ,
    \end{align}
    with associated \E{1} given by
    \begin{equation}
        \vec{T}_1 = \begin{pmatrix} 2 L_{Ax} \\ 0 \\ 0 \end{pmatrix} , \quad
        \vec{T}_2 = \begin{pmatrix} 0 \\ L_{Cy} \\ 0 \end{pmatrix} , \quad
        \vec{T}_3 = \begin{pmatrix} 0 \\ 0 \\ 2 L_{Bz} \end{pmatrix} ,
    \end{equation}
    \item[III. 2 glide reflections ($A$), 1 orthogonal glide reflection ($B$)] (the conventional choice for \E{8} with $p=0$):
    \begin{align}
        & \mat{M}_A = \diag(1,-1,1), \quad \mat{M}_B = \diag(-1,1,1) , \nonumber \\
        & \vec{T}_{\A{1}} = \begin{pmatrix} L_{1x} \\ L_{1y} \\ 0 \end{pmatrix} , \quad
        \vec{T}_{\A{2}} = \begin{pmatrix} L_{1x} \\ L_{2y} \\ 0 \end{pmatrix} , \quad
        \vec{T}_B = \begin{pmatrix} L_{Bx} \\ \frac{p}{2} (L_{1y} - L_{2y}) \\ L_{Bz} \end{pmatrix} ,
    \end{align}
    with associated \E{1} given by
    \begin{equation}
        \vec{T}_1 = \begin{pmatrix} 2 L_{Ax} \\ 0 \\ 0 \end{pmatrix} , \quad
        \vec{T}_2 = \begin{pmatrix} 0 \\ L_{1y} - L_{2y} \\ 0 \end{pmatrix} , \quad
        \vec{T}_3 = \begin{pmatrix} 0 \\ 0 \\ 2 L_{Bz} \end{pmatrix} ,
    \end{equation}
    \item[IV. 2 glide reflections ($A$), 1 half-turn ($B$)] (the conventional choice for \E{10} with $p=1$):
    \begin{align}
        & \mat{M}_A = \diag(1,-1,1), \quad \mat{M}_B = \diag(-1,-1,1) , \nonumber \\
        & \vec{T}_{\A{1}} = \begin{pmatrix} L_{1x} \\ L_{1y} \\ 0 \end{pmatrix} , \quad
        \vec{T}_{\A{2}} = \begin{pmatrix} L_{1x} \\ L_{2y} \\ 0 \end{pmatrix} , \quad
        \vec{T}_B = \begin{pmatrix} L_{Bx} \\ L_{1y} - \frac{p}{2} (L_{1y} - L_{2y}) \\ L_{Bz} \end{pmatrix} ,
    \end{align}
    with associated \E{1} given by
    \begin{equation}
        \vec{T}_1 = \begin{pmatrix} 2 L_{Ax} \\ 0 \\ 0 \end{pmatrix} , \quad
        \vec{T}_2 = \begin{pmatrix} 0 \\ L_{1y} - L_{2y} \\ 0 \end{pmatrix} , \quad
        \vec{T}_3 = \begin{pmatrix} 0 \\ 0 \\ 2 L_{Bz} \end{pmatrix} .
    \end{equation}
\end{description}
Up to redefinition of the parameters, the associated \E{1} is the same for all choices of the set of generators, as it must be.
Regardless of the form chosen, the $p=0$ solutions are the generators of the Klein space with a horizontal flip, \E{8}, and the $p=1$ solutions are the generators of the Klein space with a half-turn, \E{10}.

\subsection{\E{13} and \E{14}: Chimney spaces with a vertical or horizontal flip}
\label{app:E13E14}

As with the other spaces involving a flip there are multiple choices for the $O(3)$ structure of the generators.
Conventionally (c.f.\ \cite{Riazuelo2004:prd}) the set of generators are written with one generator associated with a translation, $\mat{M}_A^{\E{13}} = \mat{M}_A^{\E{14}} = \identity$, and the other associated with a flip, $\mat{M}_B^{\E{13}} = \diag(1, 1, -1)$ or $\mat{M}_B^{\E{14}} = \diag(-1, 1, 1)$.
We will deviate from this choice by keeping the flip associated with $\mat{M}_A$ and the translation associated with $\mat{M}_B$.
Further, we will choose the flip to be consistent with that from the Klein space, \E{7}, so that $\mat{M}_A = \diag(1,-1,1)$.
A horizontal flip is the same as the quarter-turn of a vertical flip.
The distinction between these two types of flips need not be made.
Finally, both generators could be associated with the same flip.
For derivation purposes it is convenient to work with this form.

Using two rotational degrees of freedom to fix the vertical flip to be across the $xz$-plane (which fixes the $y$-axis) and the remaining rotational degree of freedom to rotate around the $y$ axis to set the $z$-component of one of the vectors to zero, we choose as our starting form
\begin{equation}
    \mat{M}_A = \diag(1, -1, 1) , \qquad
    \vec{T}'_{\A{1}} = \begin{pmatrix} L'_{1x} \\ L'_{1y} \\ 0 \end{pmatrix} , \quad
    \vec{T}'_{\A{2}} = \begin{pmatrix} L'_{2x} \\ L'_{2y} \\ L'_{2z} \end{pmatrix} .
\end{equation}
The corresponding pure translations come from $(g'_{\A{1}})^2$ and $\inverse{(g'_{\A{1}})} g'_{\A{2}}$ and are given by
\begin{equation}
    \vec{T}'_1 = \begin{pmatrix} 2 L'_{1x} \\ 0 \\  0 \end{pmatrix} \quad \mbox{and} \quad
    \vec{T}'_2 = \begin{pmatrix} -L'_{1x} + L'_{2x} \\ L'_{1y} - L'_{2y} \\ L'_{2z} \end{pmatrix} .
\end{equation}

The remaining condition to impose is that $(g'_{\A{2}})^2$ is a pure translation.
This leads to the system of equations with $m_1$ and $m_2$ integers:
\begin{equation}
    \label{eqn:E13E14groupcondition}
    \begin{pmatrix} 2 L'_{2x} \\ 0 \\ 2 L'_{2z} \end{pmatrix}
    =
    \begin{pmatrix} 2 m_1 L'_{1x} + m_2 ( L'_{2x} - L'_{1x} ) \\ m_2 ( L'_{1y} - L'_{2y} ) \\ m_2 L'_{2z} \end{pmatrix} .
\end{equation}
For the translation vectors $\vec{T}'_i$ to be non-trivial we require $L'_{1x} \neq 0$ and at least one of $L'_{2z} \neq 0$ or $L'_{1y} \neq L'_{2y}$.

To find the general solutions we begin with the $z$-component of \cref{eqn:E13E14groupcondition} which has two solutions: $m_2 = 2$ or $L'_{2z} = 0$.
We first consider $m_2 = 2$.
With this the $y$-component requires $L'_{2y} = L'_{1y}$ and the $x$-component requires $m_1 = 1$ (since $L'_{1x} \neq 0$).
We thus arrive at the solution
\begin{equation}
    \label{eqn:E13E14-1}
    \vec{T}'_{\A{1}} = \begin{pmatrix} L'_{1x} \\ L'_{1y} \\ 0 \end{pmatrix} , \quad
    \vec{T}'_{\A{2}} = \begin{pmatrix} L'_{2x} \\ L'_{1y} \\ L'_{2z} \end{pmatrix} .
\end{equation}

Alternatively we can consider $L'_{2z} = 0$.
This requires that we must have $L'_{1y} \neq L'_{2y}$, thus the $y$-component of \cref{eqn:E13E14groupcondition} requires $m_2 = 0$.
With this, the $x$-component of \cref{eqn:E13E14groupcondition} then requires $L'_{2x} = m_1 L'_{1x}$.
Shifting by integer multiples of $\vec{T}'_1$ allows us to limit $m_1 \in \{ 0, 1 \}$.
However, if $m_1 = 0$ then $\inverse{(g'_{\A{1}})} g'_{\A{2}} g'_{\A{1}}: \vec{x} \to \vec{x}$ for $\vec{x} = \transpose{(x, L'_{1y} - L'_{2y} / 2, z)}$ so is not freely acting.
Thus we must have $m_1 = 1$ which leads to the solution
\begin{equation}
    \label{eqn:E13E14-2}
    \vec{T}'_{\A{1}} = \begin{pmatrix} L'_{1x} \\ L'_{1y} \\ 0 \end{pmatrix} , \quad
    \vec{T}'_{\A{2}} = \begin{pmatrix} L'_{1x} \\ L'_{2y} \\ 0 \end{pmatrix} .
\end{equation}

The two solutions correspond to the two topologies named \E{13} and \E{14}.
As with \E{7}--\E{10} there are multiple forms in which the generators can be written.
The natural choices are listed below where we have dropped the primes and relabeled the components in the translation vectors.
\begin{description}
    \item[I. 2 glide reflections ($A$)] (first solution in the derivation):
    \begin{align}
        & \mat{M}_A = \diag(1,-1,1), \nonumber \\
        & \vec{T}_{\A{1}} = \begin{pmatrix} L_{1x} \\ L_{1y} \\ 0 \end{pmatrix} , \quad
        \vec{T}_{\A{1}} = \begin{pmatrix} L_{2x} \\ L_{1y} \\ L_{2z} \end{pmatrix} , \quad
    \end{align}
    with associated \E{11} given by
    \begin{equation}
        \label{eqn:E13E14-caseI}
        \vec{T}_1 = \begin{pmatrix} 2 L_{1x} \\ 0 \\ 0 \end{pmatrix} , \quad
        \vec{T}_2 = \begin{pmatrix} -L_{1x} + L_{2x} \\ 0 \\ L_{2z} \end{pmatrix} ,
    \end{equation}
    \item[II. 2 glide reflections ($A$)] (second solution in the derivation):
    \begin{align}
        \label{eqn:E13E14-caseII}
        & \mat{M}_A = \diag(1,-1,1), \nonumber \\
        & \vec{T}_{\A{1}} = \begin{pmatrix} L_{1x} \\ L_{1y} \\ 0 \end{pmatrix} , \quad
        \vec{T}_{\A{2}} = \begin{pmatrix} L_{1x} \\ L_{2y} \\ 0 \end{pmatrix} , \quad
    \end{align}
    with associated \E{11} given by
    \begin{equation}
        \vec{T}_1 = \begin{pmatrix} 2 L_{1x} \\ 0 \\ 0 \end{pmatrix} , \quad
        \vec{T}_2 = \begin{pmatrix} 0 \\ L_{1y} - L_{2y} \\ 0 \end{pmatrix} ,
    \end{equation}
    \item[III. 1 glide reflection ($A$), 1 translation ($B)$] (based on the first solution in the derivation):
    \begin{align}
        \label{eqn:E13E14-caseIII}
        & \mat{M}_A = \diag(1,-1,1), \quad \mat{M}_B = \identity , \nonumber \\
        & \vec{T}_A = \begin{pmatrix} L_{Ax} \\ L_{Ay} \\ 0 \end{pmatrix} , \quad
        \vec{T}_B = \begin{pmatrix} L_{Bx} \\ 0 \\ L_{Bz} \end{pmatrix} , \quad
    \end{align}
    with associated \E{11} given by
    \begin{equation}
        \vec{T}_1 = \begin{pmatrix} 2 L_{Ax} \\ 0 \\ 0 \end{pmatrix} , \quad
        \vec{T}_2 = \begin{pmatrix} L_{Bx} \\ 0 \\ L_{Bz} \end{pmatrix} ,
    \end{equation}
    \item[IV. 1 glide reflection ($A$), 1 translation ($B)$] (based on the second solution in the derivation):
    \begin{align}
        \label{eqn:E13E14-caseIV}
        & \mat{M}_A = \diag(1,-1,1), \quad \mat{M}_B = \identity , \nonumber \\
        & \vec{T}_A = \begin{pmatrix} L_{Ax} \\ L_{Ay} \\ 0 \end{pmatrix} , \quad
        \vec{T}_B = \begin{pmatrix} 0 \\ L_{By} \\ 0 \end{pmatrix} , \quad
    \end{align}
    with associated \E{11} given by
    \begin{equation}
        \vec{T}_1 = \begin{pmatrix} 2 L_{Ax} \\ 0 \\ 0 \end{pmatrix} , \quad
        \vec{T}_2 = \begin{pmatrix} 0 \\ L_{By} \\ 0 \end{pmatrix} .
    \end{equation}
\end{description}
These can be related to the conventional choices (c.f.\ \cite{Riazuelo2004:prd}).
Conventionally \E{13} can be identified as case III in \cref{eqn:E13E14-caseIII} though $L_{Bx} \neq 0$ generically.
Conventionally \E{14} can be identified as case IV in \cref{eqn:E13E14-caseIV} when rotated around the $y$-axis by $\pi/2$.
In general, either of cases I and III describe the chimney space \E{13} and either of cases II and IV describe the chimney space \E{14}.
Further, we see that \E{13}, case III in \cref{eqn:E13E14-caseIII}, is the limit of \E{7}, $p=0$ in \cref{eqn:E7E9-caseI}, with $\vert \vec{T}_{\B{1}} \vert \to \infty$ and \E{14}, case IV in \cref{eqn:E13E14-caseIV}, is the limit of \E{7}, $p=0$ in \cref{eqn:E7E9-caseI}, with $\vert \vec{T}_{\B{2}} \vert \to \infty$.

\subsection{\E{15}: Chimney space with half-turn and flip}
\label{app:E15}

As with the other spaces involving a flip there are multiple choices for the $O(3)$ structure of the generators.
Conventionally (c.f.\ \cite{Riazuelo2004:prd}) the set of generators are written with one generator associated with a vertical flip and the other associated with a half-turn.
Alternatively we could consider one generator associated with the vertical flip and the other with a horizontal flip.
To derive the general form of the generators we will consider a vertical flip and a half-turn but break with the orientation conventions of \cite{Riazuelo2004:prd}.

Using two rotational degrees of freedom we align the rotation axis with the $z$-axis and the remaining rotational degree of freedom to associate the flip with the $y$-axis.
This uses all the rotational degrees of freedom.
We thus choose our general starting case to be
\begin{align}
    & \mat{M}_A = \diag(1, -1, 1), \quad \mat{M}_B = \diag(-1, -1, 1) , \nonumber \\
    & \vec{T}'_A = \begin{pmatrix} L'_{Ax} \\ L'_{Ay} \\ L'_{Az} \end{pmatrix}, \quad
    \vec{T}'_B = \begin{pmatrix} L'_{Bx} \\ L'_{By} \\ L'_{Bz} \end{pmatrix} .
\end{align}
From these, a set of pure translations, an associated \E{11}, can be constructed from $(g'_A)^2$ and $(g'_B)^2$ to be
\begin{equation}
   \vec{T}'_1 = \begin{pmatrix} 2 L'_{Ax} \\ 0 \\ 2 L'_{Az} \end{pmatrix} \quad \mbox{and} \quad
   \vec{T}'_2 = \begin{pmatrix} 0 \\ 0 \\ 2 L'_{Bz} \end{pmatrix} .
\end{equation}

The remaining conditions to impose are that the following two generators are pure translations (where the prefactor $(g'_A)^2$ is included to make the subsequent derivation simpler)
\begin{equation}
    g'_A g'_B \inverse{(g'_A)} \inverse{(g'_B)} \quad \mbox{and} \quad
    (g'_A)^2 g'_B (g'_A)^2 \inverse{(g'_B)} .
\end{equation}
This leads to the set of conditions
\begin{align}
    \label{eqn:E15groupcondition1}
    \begin{pmatrix} 2 L'_{Ax} \\  2 (L'_{Ay} - L'_{By}) \\ 0 \end{pmatrix}
    & = \begin{pmatrix} 2 m^{(1)}_1 L'_{Ax} \\ 0 \\ 2 m^{(1)}_1 L'_{Az} + 2 m^{(1)}_2 L'_{Bz} \end{pmatrix} , \\
    \label{eqn:E15groupcondition2}
    \begin{pmatrix} 0 \\ 0 \\ 4 L'_{Az} \end{pmatrix}
    & = \begin{pmatrix} 2 m^{(2)}_1 L'_{Ax} \\ 0 \\ 2 m^{(2)}_1 L'_{Az} + 2 m^{(2)}_2 L'_{Bz} \end{pmatrix} .
\end{align}
For the translation vectors $\vec{T}'_i$ to be non-trivial we require $L'_{Ax} \neq 0$ and $L'_{Bz} \neq 0$.

To find the general solution we note that the $x$-components of \cref{eqn:E15groupcondition1} and \cref{eqn:E15groupcondition2} require $m^{(1)}_1 = 1$ and $m^{(2)}_1 = 0$, while the $y$-component of \cref{eqn:E15groupcondition1} requires $L'_{By} = L'_{Ay}$.
With this the $z$-component of \cref{eqn:E15groupcondition1} requires $L'_{Az} = -m^{(1)}_2 L'_{Bz}$.
Shifting $\vec{T}'_A$ by integer multiples of $\vec{T}'_2$ we can restrict $m^{(1)}_2 \in \{ 0, 1 \}$.
However, when $m^{(1)}_2 = 1$, then $g'_B g'_A: \vec{x} \to \vec{x}$ for $\vec{x} = \transpose{((-L'_{Ax} + L'_{Bx})/2, y, z)}$, so the generators are not freely acting.
We are thus left with the solution when $m^{(1)}_2 = 0$ given by
\begin{equation}
    \vec{T}'_A = \begin{pmatrix} L'_{Ax} \\ L'_{Ay} \\ 0 \end{pmatrix}, \quad
    \vec{T}'_B = \begin{pmatrix} L'_{Bx} \\ L'_{Ay} \\ L'_{Bz} \end{pmatrix} .
\end{equation}

As noted above there are multiple forms in which the generators may be written.
Here the natural choices are listed where we have dropped the primes and relabeled the components in the translation vectors.
\begin{description}
    \item[I. 1 glide reflection ($A$), 1 half-turn ($B$)] (the case used in the derivation):
    \begin{align}
        \label{eqn:E15-caseI}
        & \mat{M}_A = \diag(1,-1,1), \quad \mat{M}_B = \diag(-1,-1,1) , \nonumber \\
        & \vec{T}_A = \begin{pmatrix} L_{Ax} \\ L_{Ay} \\ 0 \end{pmatrix} , \quad
        \vec{T}_B = \begin{pmatrix} L_{Bx} \\ L_{Ay} \\ L_{Bz} \end{pmatrix} ,
    \end{align}
    with associated \E{11} given by
    \begin{equation}
        \vec{T}_1 = \begin{pmatrix} 2 L_{Ax} \\ 0 \\ 0 \end{pmatrix} , \quad
        \vec{T}_2 = \begin{pmatrix} 0 \\ 0 \\ 2 L_{Bz} \end{pmatrix} , \quad
    \end{equation}
    \item[II. 1 glide reflection ($A$), 1 orthogonal glide reflection ($B$)]:
    \begin{align}
        & \mat{M}_A = \diag(1,-1,1), \quad \mat{M}_B = \diag(-1,1,1) , \nonumber \\
        & \vec{T}_A = \begin{pmatrix} L_{Ax} \\ L_{Ay} \\ 0 \end{pmatrix} , \quad
        \vec{T}_B = \begin{pmatrix} L_{Bx} \\ 0 \\ L_{Bz} \end{pmatrix} ,
    \end{align}
    with associated \E{11} given by
    \begin{equation}
        \vec{T}_1 = \begin{pmatrix} 2 L_{Ax} \\ 0 \\ 0 \end{pmatrix} , \quad
        \vec{T}_2 = \begin{pmatrix} 0 \\ 0 \\ 2 L_{Bz} \end{pmatrix} . \quad
    \end{equation}
\end{description}
In both cases we see that the associated \E{11} is the same, as must be the case.
Either choice can be used to describe \E{15}.
Further, we see that \E{15}, case I in \cref{eqn:E15-caseI}, is the limit of \E{8}, $p=0$ in \cref{eqn:E8E10-caseI}, with $\vert \vec{T}_C \vert \to \infty$.

\subsection{\E{17}: Slab space with flip}
\label{app:E17}

The slab space slab space with a flip has one generator.
Conventionally (c.f.\ \cite{Riazuelo2004:prd}) this was chosen as a horizontal flip, here we will employ a vertical flip.

Using two rotational degrees of freedom we align the flip to be across the $xz$-plane.
The remaining rotational degree of freedom is used to set the $x$-component of the translation to zero.
This immediately determines the general form of the generator to be
\begin{equation}
    \mat{M}_A = \diag(1,-1,1) \quad \mbox{with} \quad
    \vec{T}_A = \begin{pmatrix} 0 \\ L_y \\ L_z \end{pmatrix} ,
\end{equation}
with the pure translation vector, an associated \slabh, constructed from $(g_A)^2$ so that
\begin{equation}
    \vec{T}_1 = \begin{pmatrix} 0 \\ 0 \\ 2 L_z \end{pmatrix}.
\end{equation}

We note that in contrast to \E{16} there is not a rotated slab space with a flip, i.e., there is no root of \slabi, as this is equivalent to \E{17} viewed in a rotated coordinate system.
Explicitly, consider the horizontal flip $\mat{M}_A$ and let $\mat{R}_{\unitvec{z}}(\psi)$ be a rotation by $\psi$ about the $z$-axis (or any axis in the reflection plane).
Define $\mat{M} \equiv \mat{R}_{\unitvec{z}}(\psi) \mat{M}_A$ and the action of a generator as $g:\vec{x} \to \mat{M} \vec{x} + \vec{T}$.
It can be shown that $\transpose{\mat{M}} = \mat{M}$ so that $\mat{M}^2 = \identity$.
From this
\begin{equation}
    g^2:\vec{x} \to \vec{x} + (\identity + \mat{M}) \vec{T}
\end{equation}
so is a pure translation (with the requirement that $(\identity + \mat{M}) \vec{T} \neq \vec{0}$).
Further, the eigenvalues of $\mat{M}$ can be shown to be $\{-1, 1, 1\}$, thus, $\mat{M}$ can be rotated to a coordinate system such that $\mat{M} \to \mat{M}_A$.
In other words, the pattern of clones generated by $g$ is identical to that generated by $g_A$ as viewed in a rotated coordinate system.

\subsection{Desiderata}

As noted above, the non-orientable topologies are not uniquely defined by their $O(3)$ structure.
Above we have provided a number of different sets of generators for each of the non-orientable topologies: all are equally valid choices.
In this work we have chosen to not follow the conventional choice as defined in \cite{Riazuelo2004:prd}.
Instead we have chosen to maximize the number of generators that are pure translations and written all expressions with this choice (c.f.\ \cref{secn:topologiesmanifolds-general} and \cref{secn:topologiesmanifolds}).

For \E{7} and \E{9}, there are more significant differences between the general form of the generators and those provided in \cite{Riazuelo2004:prd}.
Using a $(c)$ superscript for the conventional choice, their form of the generators for \E{7} contain the translations
\begin{equation}
    \vec{T}^{(c)\E{7}}_{\A{1}} = \begin{pmatrix} L_x/2 \\ L_y/2 \\ 0 \end{pmatrix}, \quad
    \vec{T}^{(c)\E{7}}_{\A{2}} = \begin{pmatrix} \hphantom{-} L_x/2 \\ -L_y/2 \\ \hphantom{-} 0 \end{pmatrix}, \quad
    \vec{T}^{(c)\E{7}}_{B} = \begin{pmatrix} 0 \\ 0 \\ L_z \end{pmatrix}.
\end{equation}
Comparing to \eqref{eqn:E7E9-caseII} with $p=0$ we see these are equivalent when
\begin{equation}
    L_{1x} = L_x/2, \; L_{1y} = -L_{2y} = -L_y/2, \; L_{Bx} = 0, \; L_{Bz} = L_z.
\end{equation}
These choices have imposed two special restrictions on the general form.
Firstly, the choice $L_{Bx} = 0$ is a special case.
Secondly, as seen from the associated \E{1} in \eqref{eqn:E7E9-caseII-E1}, they have written $L_{1y} - L_{2y} = L_y$ and then made the further restriction that $L_{1y} = -L_{2y}$, a special case achieved by using the freedom to shift the origin.
Due to the first choice their associated \E{1} is rectangular.

The choices in \E{9} are similar and even more restrictive.
Their form of the generators for \E{9} contain the translations
\begin{equation}
    \vec{T}^{(c)\E{9}}_{\A{1}} = \begin{pmatrix} L_x/2 \\ L_y/2 \\ 0 \end{pmatrix}, \quad
    \vec{T}^{(c)\E{9}}_{\A{2}} = \begin{pmatrix} \hphantom{-} L_x/2 \\ -L_y/2 \\ \hphantom{-} 0 \end{pmatrix}, \quad
    \vec{T}^{(c)\E{9}}_{\A{3}} = \begin{pmatrix} 0 \\ 0 \\ L_z/2 \end{pmatrix}.
\end{equation}
Comparing to \eqref{eqn:E7E9-caseIII} with $p=1$ we see these are equivalent when
\begin{equation}
    L_{1x} = L_x/2, \; L_{1y} = -L_{2y} = -L_y/2, \; L_{3x} = 0, \; L_{3z} = L_z/2.
\end{equation}
These choices are similar to those for \E{7} and have thus imposed the same two special restrictions on the general form.
Firstly, the choice $L_{3x} = 0$ is a special case.
Secondly, as seen from the associated \E{1} in \eqref{eqn:E7E9-caseIII-E1}, they have written $L_{1y} - L_{2y} = L_y$ and then made the further restriction that $L_{1y} = -L_{2y}$, a special case achieved by using the freedom to shift the origin.

For \E{9} the conventional choice can be misleading.
Notice that in the generic form of the associated \E{1} there are two translations that contain a component in the direction normal to the plane of the flip; the $y$-direction in our case \eqref{eqn:E7E9-caseIII-E1}.
However, the simplifications made to write the eigenmodes in \cite{Riazuelo2004:prd} (see their equation (67)) appear to show that the associated \E{1} of \E{9} is rectangular.
This is not the case and the authors \emph{do not claim} that this is the case.
Explicitly, they have the equivalent of our $N=2$ mode (see \eqref{eqn:E9_eigenmode_N2} and the discussion thereafter) written as
\begin{equation}
    \frac{1}{\sqrt{2}} \left[ \Upsilon_{2\pi(n_x/L_x, n_y/L_y,n_z/L_z)} + (-1)^{n_x + n_y} \Upsilon_{2\pi(n_x/L_x, -n_y/L_y,n_z/L_z)} \right],
\end{equation}
with the conditions that $n_x \in \integers$, $n_y \in \integers^{>0}$, $n_z \in \integers$, and $n_x + n_y \equiv n_z \Mod{2}$.
From \eqref{eqn:associatedhomogeneousdiscretization} this is of the form expected for a rectangular associated \E{1}.
Letting $\vec{n} = (n_1, n_2, n_3)$ we can easily identify $n_1 = n_x$ and $n_2 = n_y$.
The third pure translation comes from $g^{(c)\E{9}}_3 = g^{(c)\E{9}}_{\A{1}} g^{(c)\E{9}}_{\A{3}}$ which leads to
\begin{equation}
    \vec{T}^{(c)\E{9}}_3 = \begin{pmatrix} L_x/2 \\ L_y/2 \\ L_z/2 \end{pmatrix}.
\end{equation}
Thus $n_3$ comes from
\begin{equation}
    \vec{k}_{\vec{n}} \cdot \vec{T}^{(c)\E{9}}_3 = 2\pi n_3 = \frac{2\pi n_x}{2} + \frac{2\pi n_y}{2} + \frac{(k_{\vec{n}})_z L_z}{2}
\end{equation}
which we solve to find
\begin{equation}
    (k_{\vec{n}})_z = \frac{2\pi}{L_z}[2 n_3 - (n_x + n_y)] \equiv \frac{2\pi n_z}{L_z}.
\end{equation}
We see that we \emph{can define} the integer $n_z$ with the quoted relationship between $n_x + n_y$ and $n_z$.
This was possible due to the restrictive choices made.
It is not generically possible.

\bibliographystyle{utphys}
\bibliography{topology,additional}
\end{document}